\documentclass[11pt, a4paper]{article}
\usepackage[dvipsnames]{xcolor}
\usepackage[utf8]{inputenc}
\usepackage{url}
\usepackage{amsmath}
\usepackage{amsfonts}
\usepackage{amssymb} 
\usepackage{bm}
\usepackage{graphicx}
\usepackage{array}
\usepackage{lipsum}
\usepackage{float}
\usepackage{multirow}
\usepackage{hhline}
\usepackage{tabularx}
\usepackage{subfigure}
\usepackage{afterpage}
\usepackage{epstopdf}
\usepackage{MnSymbol}
\usepackage[utf8]{inputenc}
\usepackage{nicefrac}

\usepackage[pdftitle={One Loop to Rule Them All: Eight and Nine Dimensional String Vacua from Junctions},
  pdfauthor={Mirjam Cvetic, Markus Dierigl, Ling Lin, Hao Y. Zhang},
  pdfsubject={string landscape, swampland, supergravity, 1-form symmetry},
  bookmarksopen, bookmarksnumbered, bookmarksopenlevel=2, colorlinks=false, linkcolor=blue, citecolor=blue, urlcolor=blue]{hyperref}

\usepackage{comment}
\usepackage{rotating}
\usepackage{setspace}
\usepackage{longtable}
\usepackage{pdflscape}
\usepackage{caption}

\usepackage{bbold}

\newcommand{\tabincell}[2]{\begin{tabular}{@{}#1@{}}#2\end{tabular}}

\usepackage{tikz}
\usepackage{tikz-3dplot}
\usetikzlibrary{calc}
\usetikzlibrary{decorations} 
\usepackage{tikz-cd} 

\usepackage[all]{xy}

\numberwithin{equation}{section}

\usepackage[nosort]{cite}

\def\fkg{\mathfrak{g}}

\def\bbZ{\mathbb{Z}}
\def\cN{\mathcal{N}}
\def\bA{\mathbf{A}}
\def\bB{\mathbf{B}}
\def\bC{\mathbf{C}}
\def\bN{\mathbf{N}}
\def\bX{\mathbf{X}}
\def\ba{\mathbf{a}}
\def\bb{\mathbf{b}}
\def\bc{\mathbf{c}}
\def\bn{\mathbf{n}}
\def\bx{\mathbf{x}}

\newcommand{\colpq}[2]{\left( \begin{smallmatrix} #1 \\ #2 \end{smallmatrix} \right)}

\begin{document}

\baselineskip=18pt

\vspace*{-2cm}
\begin{flushright}
	\texttt{CERN-TH-2022-032} \\
  \texttt{LMU-ASC 10/22}\\
	\texttt{UPR-1316-T}
\end{flushright}

\vspace*{0.6cm} 
\begin{center}
{\LARGE{\textbf{One Loop to Rule Them All:}}} \\
\vspace*{.3cm}
{\Large{\textbf{Eight and Nine Dimensional String Vacua from Junctions}}} \\
 \vspace*{1.5cm}
Mirjam Cveti{\v c}$^{1,2,3}$, Markus Dierigl$^4$, Ling Lin$^{5}$, Hao Y.~Zhang$^{1}$\\

{
 \vspace*{1.0cm} 
{\it ${}^1$ Department of Physics and Astronomy, \\University of Pennsylvania,
Philadelphia, PA 19104, USA\\ \vspace{.2cm}
${}^2$ Department of Mathematics, \\University of Pennsylvania,
Philadelphia, PA 19104, USA\\ \vspace{.2cm}
${}^3$ Center for Applied Mathematics and Theoretical Physics,\\
University of Maribor, SI20000 Maribor, Slovenia\\ \vspace{.2cm}
${}^4$ Arnold-Sommerfeld-Center for Theoretical Physics,\\
Ludwig-Maximilians-Universit\"at, 80333 M\"unchen, Germany\\ \vspace{.2cm}
${}^5$CERN Theory Department, CH-1211 Geneva, Switzerland}
}

\vspace*{0.8cm}
\end{center}
\vspace*{.5cm}

\noindent String and 5-brane junctions are shown to succinctly classify all known 8d ${\cal N}=1$ string vacua.
This requires an extension of the description for ordinary $[p,q]$-7-branes to consistently include O7$^+$-planes, which then naturally encodes the dynamics of $\mathfrak{sp}_n$ gauge algebras, including their $p$-form
center symmetries.
Central to this analysis are loop junctions, i.e., strings/5-branes which encircle stacks of 7-branes and O7$^+$'s.
Loop junctions further signal the appearance of affine symmetries of emergent 9d descriptions at the 8d moduli space's boundaries.
Such limits reproduce all 9d string vacua, including the two disconnected rank $(1,1)$ moduli components.

\thispagestyle{empty}
\clearpage

\setcounter{page}{1}

\newpage
\tableofcontents

\newpage

\section{Introduction}

Supergravity theories in a large number of dimensions form an ideal laboratory to investigate the manifestation of quantum gravitational consistency conditions in the low-energy limit of string theory.
In particular, they provide a concrete class of models which corroborates the conjecture of ``string universality'', or ``string lamppost principle'', stating that all consistent (super)gravity theories arise from string theory.

Formulating and sharpening the relevant conditions on consistent effective theories of quantum gravity are at the heart of the Swampland Program \cite{Vafa:2005ui,Ooguri:2006in}.
Arguably, among the best motivated of these constraints is that quantum gravity theories should have no exact global symmetries. This statement can also be applied to higher-form global symmetries, such as 1-form center symmetries of non-Abelian gauge sectors. The condition demands that these symmetries are either gauged or broken. However, if they are to be gauged, one needs to demand the absence of obstructions/anomalies to turning on the gauge fields of these generalized symmetries.
This absence of anomalies of center 1-form symmetries can lead to severe restrictions on the global topology of the allowed gauge groups in supersymmetric theories \cite{Apruzzi:2020zot,Cvetic:2020kuw}. Similarly, the absence of global symmetries requires certain topological invariants called bordism groups to be trivial \cite{McNamara:2019rup}, once more leading to powerful constraints, in particular on the total rank of the gauge symmetry, of consistent supergravity theories in more than six dimensions \cite{Montero:2020icj,Bedroya:2021fbu}.

Being confronted with the set of supergravity models that pass the above consistency tests, the remaining question is whether all of these can be realized in string theory. To answer this we therefore need good control of the realization of the global form of the gauge groups, i.e., the fate of the center 1-form symmetries in string theory constructions. In the present work we focus on compactifications to eight and nine dimensions (8d and 9d) with 16 supercharges (i.e., ${\cal N}=1$).

A powerful approach that successfully utilizes the machinery of geometry is F-theory \cite{Vafa:1996xn}, which ties the algebraic and arithmetic properties of elliptic K3-surfaces to 8d gauge theories with ADE gauge algebras of total rank $(2,18)$.\footnote{Throughout this work, we collect the number $a$ of independent gravi-photons, and the maximal non-Abelian gauge rank $r$ into a pair $(a, r)$, which we often refer to as the (total) gauge rank.
At generic values of moduli, the gauge algebra is hence $\mathfrak{u}(1)^{a+r}$.}
In this context, the global gauge group structure is encoded in the Mordell--Weil group of rational sections of the elliptic fibration \cite{Aspinwall:1998xj,Mayrhofer:2014opa,Cvetic:2017epq}. 
Under M-/F-theory duality, this can be phrased in terms of gauging and breaking higher-form symmetries \cite{Gaiotto:2014kfa}, that is reflected geometrically in the gluing of torsional homology cycles in local patches containing the non-Abelian gauge dynamics \cite{Cvetic:2021sxm}.\footnote{The investigation of generalized symmetries within the geometric engineering framework has received broad attention in recent literature \cite{DelZotto:2015isa,Morrison:2020ool,Albertini:2020mdx,Dierigl:2020myk,Apruzzi:2020zot,Cvetic:2020kuw,Bhardwaj:2020phs,DelZotto:2020sop,Bhardwaj:2021pfz,Hosseini:2021ged,Cvetic:2021sxm,Bhardwaj:2021wif,Apruzzi:2021mlh,Apruzzi:2021nmk,Tian:2021cif,DelZotto:2022fnw}.}
Moreover, through suitable deformations that correspond to infinite distance points in the 8d moduli space, the F-theory geometry also classifies 9d ${\cal N}=1$ string vacua with gauge rank $(1,17)$ \cite{Lee:2021qkx,Lee:2021usk}.
This is consistent with the dual heterotic description, where the 9d moduli space --- described via the rank $(1,17)$ Narain lattice --- is contained in that of the 8d moduli space, with a rank $(2,18)$ Narain lattice description (see \cite{Font:2020rsk} for a recent comprehensive study).

However, the dictionary between geometry and physics is less understood in the presence of so-called frozen singularities \cite{Witten:1997bs,deBoer:2001wca,Tachikawa:2015wka,Bhardwaj:2018jgp}.
While these are known to be the necessary ingredient for an F-theory description of $\mathfrak{sp}$ gauge algebras on the 8d ${\cal N}=1$ moduli branches of gauge ranks $(2,10)$ and $(2,2)$, the characterization of, e.g., the gauge group topology is no longer purely geometric (i.e., given by the Mordell--Weil group) \cite{Cvetic:2021sjm}.
Likewise, it is not immediately clear how to identify decompactification limits on these moduli spaces.
On the other hand, advances in the Swampland program \cite{Hamada:2021bbz} strongly suggest, that all 8d ${\cal N}=1$ vacua should have a characterization in terms of an elliptically-fibered K3.

As we will demonstrate in this work, string junctions provide a \emph{unified framework} that encompasses all these features.
In this description, the underlying elliptic K3 is encoded in the configuration of $[p,q]$-7-branes of type IIB string theory, whose $[p,q]$-type are in one-to-one correspondence to elliptic singularities characterized by an $SL(2,\bbZ)$-monodromy $M_{[p,q]}$.
The junctions are then $(p,q)$-strings or -5-branes stretched between the 7-branes.
In their original formulation \cite{Gaberdiel:1997ud,Gaberdiel:1998mv,DeWolfe:1998zf} that is equivalent to F-theory without frozen singularities, junctions describe the 8d gauge dynamics with ADE gauge algebras,\footnote{String junctions have been also used to construct lower-dimensional theories, see \cite{Bonora:2010bu,Cvetic:2011gp,Garcia-Etxebarria:2013tba,Grassi:2014ffa,Agarwal:2016rvx,Grassi:2018wfy,Hassler:2019eso,Heckman:2020svr,Grassi:2021ptc}.} as well as their higher-form symmetries \cite{Cvetic:2021sxm}.
To account for a junction description of all 8d ${\cal N}=1$ vacua, we extend the discussion to include O7$^+$-planes, which are the IIB avatars of frozen singularities, and have the same monodromy as an elliptic $\text{D}_8$ singularity \cite{Witten:1997bs,deBoer:2001wca,Tachikawa:2015wka}.

A concept that will be key to this work are so-called fractional null junctions, which are certain fractional (and hence, unphysical) $\colpq{p}{q}$-charges encircling all 7-branes, i.e., \emph{loop junctions}.
In the absence of O7$^+$-planes, these are known to be equivalent to Mordell--Weil torsion of the underlying elliptic K3 \cite{Fukae:1999zs,Guralnik:2001jh}.
To correctly account for the electric and magnetic center symmetries and the gauge group topology for the $\mathfrak{sp}$ gauge symmetries that arise in the presence of O7$^+$, it turns out to be instrumental to study separately $(p,q)$-strings and -5-branes.
The key difference is, while any integer number of 5-branes can end on an O7$^+$, the number of \emph{string}-prongs there must be even.
Indeed, with this modification, we find that the junction description of 8d vacua with one O7$^+$ is equivalent to so-called CHL vacua \cite{Chaudhuri:1995fk,Chaudhuri:1995bf} of rank $(2,10)$, including the characterization of the global gauge group structure \cite{Cvetic:2021sjm,Font:2021uyw}.
Moreover, it is straightforward to include two O7$^+$-planes, thereby giving a junction-esque classification of 8d string vacua with gauge rank $(2,2)$ including their gauge group topology, for which there is no known heterotic or CHL string description.

In addition, we also propose a junction description for decompactification limits to 9d including O7$^+$-planes.
Parallel to the 9d uplifts of the rank $(2,18)$ setting \cite{Lee:2021qkx,Lee:2021usk} (see also \cite{Collazuol:2022jiy} for a related discussion of 10d uplifts of 9d heterotic vacua), we identify the corresponding infinite distance limits with O7$^+$-planes by the emergence of loop junctions that affinize the 8d gauge algebra.
Again, the subtle differences from having modified boundary conditions for strings and 5-branes can be cross-checked with the momentum lattice description of the CHL string for uplifting 8d rank $(2,10)$ theories to 9d rank $(1,9)$ theories.
Like in 8d, the junction description naturally encodes the gauge group topology of 9d vacua.
For the rank $(2,2)$ theories without a momentum lattice analog, the 9d theories with rank $(1,1)$ that result from the junction description live on two disconnected moduli branches that are only connected through an $S^1$-reduction to 8d, which matches other stringy constructions \cite{Aharony:2007du}.
This further establishes junctions as a complimentary framework to sharpen aspects of the Swampland Distance conjecture \cite{Ooguri:2006in} in string compactifications.

The rest of the paper is organized as follows.
After reviewing the junction framework with ordinary $[p,q]$-7-branes in Section \ref{sec:local_analysis}, we discuss, in Section \ref{subsec:junctions_on_O7}, the modified boundary conditions for strings and 5-branes on an O7$^+$-plane that give rise to the correct higher-form symmetries of $\mathfrak{sp}$ gauge algebras in 8d.
In Section \ref{sec:global}, we then describe global 8d models by ``gluing'' together local patches with 7-brane stacks involving O7$^+$-planes.
A particular focus will be on the gauge group topology that is encoded in the fractional null junctions.
We then examine, in Section \ref{sec:9d}, the infinite distance limits described via 7-branes and junctions that correspond to 9d ${\cal N}=1$ vacua, for which we will also determine the global gauge group structure.
The appendices contain some technical aspects, as well as the full list of gauge group topologies for all 8d vacua with maximally-enhanced non-Abelian symmetries in Appendix \ref{app:results}.

\section{String and 5-brane junctions}
\label{sec:local_analysis}

String junctions provide an efficient way to classify electrically charged states with respect to gauge symmetries localized on 7-brane stacks in type IIB string theory. Therefore, they also contain information about the electric 1-form center symmetries and the global realization of the 8d gauge group \cite{Fukae:1999zs,Guralnik:2001jh,Cvetic:2021sxm}. 
The magnetically dual perspective is provided by 5-brane webs, which can also be described by junctions \cite{Aharony:1997bh,Leung:1997tw,Kol:1998cf,DeWolfe:1999hj}.

In this section we recall some properties of string and 5-brane junctions in the presence of a general $[p,q]$-7-brane stack. This provides a local construction of the charged states. Importantly, the charge under the center symmetry is related to the appearance of certain \emph{fractional junctions}, the extended weight junctions, that determine the global properties of the model \cite{Cvetic:2021sxm}. We then generalize the discussion of string and 5-brane junctions to backgrounds containing O7$^+$-planes, whose geometric interpretation in F-theory is more challenging \cite{Tachikawa:2015wka,Bhardwaj:2018jgp}.
With the help of string junctions we can successfully extract the correct properties of these configurations and identify the electric center symmetries also for symplectic gauge groups.
This analysis is repeated with 5-brane junctions, which, as opposed to the ADE algebras realized without O7$^+$'s, have a subtle distinction from string junctions that is precisely needed to correctly account for the magnetic center symmetry.

\subsection[Basics of \texorpdfstring{$[p,q]$}{[p,q])}-7-branes and string junctions]{Basics of \boldmath{$[p,q]$}-7-branes and junctions}

In this section we will recall the basics of the junction description for 8d ${\cal N}=1$ dynamics from type IIB compactifications \cite{Gaberdiel:1997ud,DeWolfe:1998zf}.
The key players are spacetime filling $[p,q]$-7-branes $\mathbf{X}_{[p,q]}$, which we will denote with square brackets.
A single 7-brane must have coprime $p$ and $q$.
In the plane perpendicular to its worldvolume, $\mathbf{X}_{[p,q]}$ induces a singularity in the axio-dilaton profile $\tau = C_0 + i e^{- \phi}$, composed of the RR 0-form $C_0$ and the dilaton field $\phi$, which is characterized by an $SL(2,\bbZ)$ monodromy
\begin{align}
M_{[p,q]} = \begin{pmatrix} 1 + p q & - p^2 \\ q^2 & 1 - p q \end{pmatrix} \in SL(2,\mathbb{Z}) \, ,
\end{align}
which acts on $\tau$ by a M{\" o}bius transformation, and in the doublet representation,
\begin{align}
    \colpq{B_2}{C_2} \mapsto M_{[p,q]} \colpq{B_2}{C_2} \, ,
\end{align}
on the NSNS- and RR-2-form fields $(B_2, C_2)$.
This monodromy can be captured by a branch cut in the perpendicular plane that emanates form the 7-brane.
In the following it will prove useful to introduce conventions for some special 7-branes, that will later appear in the construction of non-Abelian gauge algebras
\begin{equation}
\begin{split}
&\mathbf{A} = \bX_{[1,0]}: \quad M_{[1,0]} = \begin{pmatrix} 1 & -1 \\ 0 & 1 \end{pmatrix} \,, \\
&\mathbf{B} = \bX_{[1,-1]}: \quad M_{[1,-1]} = \begin{pmatrix} 0 & -1 \\ 1 & 2 \end{pmatrix} \,, \\
& \mathbf{C} = \bX_{[1,1]}: \quad M_{[1,1]} = \begin{pmatrix} 2 & -1 \\ 1 & 0 \end{pmatrix} \,, \\
& \mathbf{N} = \bX_{[0,1]}: \quad M_{[0,1]} = \begin{pmatrix} 1 & 0 \\ 1 & 1 \end{pmatrix} \,.
\end{split}
\label{eq:branespecies}
\end{equation}

In a local model, where the perpendicular plane is non-compact (i.e., is $\mathbb{R}^2 = \mathbb{C} \ni z$), it is customary to extend the branch cuts all downwards (meeting at $z = -i \infty$, without crossing each other before).
Starting from a configuration describing a certain 8d vacuum, we can obtain another one on the same 8d ${\cal N}=1$ moduli space by moving the 7-branes.
When $\bX_{[p_1, q_1]}$ crosses the branch cut of $\bX_{[p_2,q_2]}$ from the left to right, the $[p,q]$-type changes according to:
\begin{align}\label{eq:brane_move_left-to-right}
    \begin{split}
        \underrightarrow{{\bf X}_{[p_1, q_1]}} {\bf X}_{[p_2, q_2]} \rightarrow {\bf X}_{[p_2, q_2]} {\bf X}_{[p_1 + D \cdot p_2, \, q_1 + D \cdot q_2]} \, ,
    \end{split}
\end{align}
where $D \equiv \det\left( \begin{smallmatrix} p_1 & p_2 \\ q_1 & q_2 \end{smallmatrix} \right)$, and the arrow indicating the branch-cut-crossing 7-brane.
Likewise, when it crosses from right to left, one has
\begin{align}\label{eq:brane_move_right-to-left}
    {\bf X}_{[p_2, q_2]} \underleftarrow{{\bf X}_{[p_1, q_1]}} \rightarrow & {\bf X}_{[p_1 + D \cdot p_2, \, q_1 + D \cdot q_2]} {\bf X}_{[p_2, q_2]} \, .
\end{align}
For any concrete configuration, we can arrange the 7-branes along a horizontal axis, and denote them as $\bX_{[p_1, q_1]} \bX_{[p_2,q_2]} \, ...$ by labelling from left to right.
The $SL(2,\bbZ)$ monodromy around any (connected) part of this chain is the product of the individual monodromies of the encircled branes \emph{from right to left}.
To obtain a valid global configuration (i.e., where the perpendicular plane is $\mathbb{P}^1$) describing an 8d ${\cal N}=1$ supergravity model, tadpole cancellation requires exactly 24 $[p,q]$-7-branes, whose overall monodromy must be the identity.
Note that, while their \emph{relative} $[p,q]$-types are pivotal for distinguishing different physical configurations, an \emph{overall} $SL(2,\bbZ) \ni g$ transformation,
\begin{align}
    \left[ \begin{smallmatrix} p_i \\ q_i \end{smallmatrix} \right] \mapsto g \left[ \begin{smallmatrix} p_i \\ q_i \end{smallmatrix} \right] \, , \quad M_{[p_i, q_i]} \mapsto g M_{[p_i, q_i]} g^{-1} \quad \text{for all } \, i \, ,
\end{align}
does not matter physically for a model (local or global) described by a collection of 7-branes $\bX_{[p_i, q_i]}$.

\begin{figure}[ht]
    \centering
    \includegraphics[width = \textwidth]{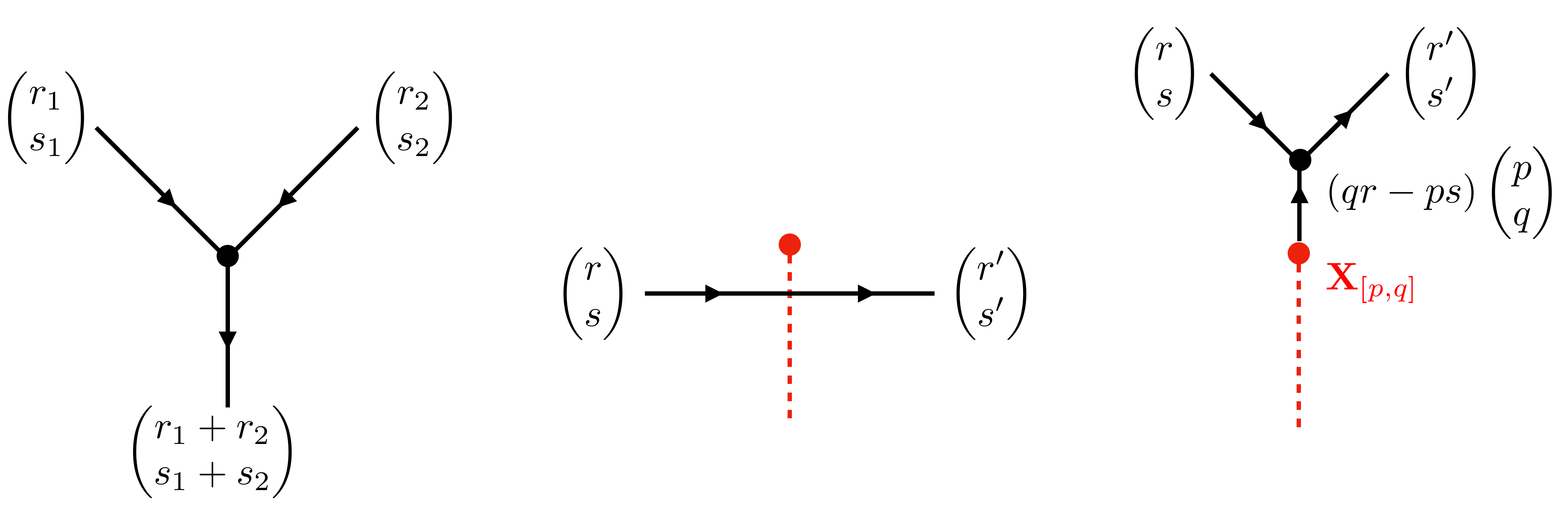}
    \caption[matrix in text]{Strings and 5-branes, which are represented as lines in the perpendicular plane, form junctions, where the $\colpq{p}{q}$-charge at each vertex is conserved (left). In the presence of 7-branes, they undergo monodromy transformations \eqref{eq:monodromy_on_prongs} when they cross a branch cut (middle). By a Hanany--Witten transition, the same junction can be represented as having a prong on the 7-brane (right).}
    \label{fig:Juncrules}
\end{figure}

BPS-particles in 8d arise from $(p,q)$-strings --- a bound state of $p$ fundamental and $q$ D-strings with electric charge $\colpq{p}{q}$ under $\colpq{B_2}{C_2}$ --- anchored on the 7-branes of the same $[p,q]$-type, and extending as a directed line into the perpendicular plane.
Their magnetically dual objects, which are four-dimensional in 8d, are given by $(p,q)$-5-branes --- a bound state of $p$ NS5- and $q$ D5-branes with magnetic charge $\colpq{p}{q}$ under $\colpq{B_2}{C_2}$ --- that share four common spatial directions as the 7-brane and also project to lines in the perpendicular plane.
Since strings and 5-branes can fuse and split, so long as the overall $\colpq{p}{q}$-charge is conserved at every vertex, they form junctions, see left of Figure \ref{fig:Juncrules}.
Note that, by flipping the direction on any prong, its $\colpq{p}{q}$-charge acquires a minus sign.\footnote{To make contact with the F-theory description of type IIB, note that (directed) junctions can be interpreted as (oriented) 2-cycles in an elliptic K3, on which M2- and M5-branes can be wrapped, which are the objects dual to strings and 5-branes under M-/F-theory duality. See, e.g., \cite{Cvetic:2021sxm} for details of this correspondence.}

As charged objects of the 2-form fields $(B_2, C_2)$, strings and 5-branes also experience $SL(2,\bbZ)$ monodromies as they are transported around 7-branes.
This action can be represented in the perpendicular plane, after choosing the branch cuts, by an analogous transformation
\begin{align}\label{eq:monodromy_on_prongs}
\colpq{r}{s} \rightarrow \colpq{r'}{s'} =  M_{[p,q]} \colpq{r}{s} =  \colpq{r}{s} + (q r - p s) \colpq{p}{q} \,
\end{align}
on the $\colpq{r}{s}$-charges of a junction-prong as it crosses the branch cut of a $[p,q]$-7-brane, see middle of Figure \ref{fig:Juncrules}.
Finally, in analogy to Hanany--Witten transitions \cite{Hanany:1996ie}, the same junction can be expressed, by moving the branch-cut-crossing prong across the 7-brane, as a junction with an additional prong on the 7-brane, see right of Figure \ref{fig:Juncrules}.

To have non-Abelian gauge dynamics in 8d, we have to collide 7-branes to form stacks.
Strings that stretch between different constituents of a stack then become light and form massless W-bosons of the enhanced gauge symmetry.
In terms of the 7-brane types \eqref{eq:branespecies}, ADE-gauge algebras are realized when the following stacks form\footnote{We have chosen a particular $SL(2,\bbZ)$-frame that is common in the literature, but any $SL(2,\bbZ)$-conjugated configuration would obviously give the same gauge algebra.}:
\begin{align}\label{eq:ADE_brane_stacks}
\begin{array}{| c | c | c |}
\hline
\text{Lie algebra} & \text{brane constituents} & \text{monodromy} \\ \hline \hline
\mathfrak{su}_{n} & \mathbf{A}^{n} & \begin{pmatrix} 1 & - n \\ 0 & 1 \end{pmatrix} \\ \hline
\mathfrak{so}_{2n} & \mathbf{A}^n \mathbf{B} \mathbf{C} & \begin{pmatrix} -1 & n-4 \\ 0 & -1 \end{pmatrix} \\ \hline
\mathfrak{e}_{n \geq 1} & {\bf A}^{n-1} {\bf B} {\bf C}^2 & \begin{pmatrix} -2 & 2n-9 \\ -1 & n-5 \end{pmatrix} \\ \hline
\tilde{\mathfrak{e}}_{n \geq 0} & {\bf A}^n {\bf X}_{[2,-1]} {\bf C} & \begin{pmatrix} -3 & 3n-11 \\ -1 & n-4 \end{pmatrix} \\ \hline
\end{array}
\end{align}
where we have used exponents to group the same type of branes that are appear consecutively.
The overall monodromy of a 7-brane stack is the product of the individual branes from right to left; e.g., $M_{\mathfrak{so}_{2n}} = M_{[1,1]} M_{[1,-1]} M_{[1,0]}^n$.
The realizations of the exceptional algebras are physically equivalent, i.e., equal up to 7-brane moves inside the stack and $SL(2,\bbZ)$ conjugations, for $n\geq 2$,\footnote{We use the standard identifications $\mathfrak{e}_2 \cong \mathfrak{su}_2 \oplus \mathfrak{u}(1)$, $\mathfrak{e}_3 \cong \mathfrak{su}_3 \oplus \mathfrak{su}_2$, $\mathfrak{e}_4 \cong \mathfrak{su}_5$, $\mathfrak{e}_5 \cong \mathfrak{so}_{10}$.} while $\mathfrak{e}_1 \cong \mathfrak{su}_2$ and $\tilde{\mathfrak{e}}_1 \cong \mathfrak{u}(1)$; finally, the $\tilde{\mathfrak{e}}_0$ configuration corresponds to a trivial gauge algebra.
There are additional strongly coupled versions of the Lie algebra $\mathfrak{su}_n$ with $n \in \{2,3\}$ of the form $\mathbf{A}^{n+1} \mathbf{C}$.
In the remaining part of this section, we will focus mainly on the ``standard'' cases $\mathfrak{su}_n$, $\mathfrak{so}_{2n}$ and $\mathfrak{e}_{n\geq 6}$, while $\tilde{\mathfrak{e}}_{n}$ will be relevant in Section \ref{sec:9d}.
Of course there is a beautiful relation between the 7-branes stacks above with their induced $SL(2,\mathbb{Z})$ monodromies, and the classification of singularities in elliptic fibrations by Kodaira, which is central in F-theory (see \cite{Weigand:2018rez,Cvetic:2018bni} for recent reviews and additional references). 
In the following, we will focus solely on the junction perspective.

\subsection{The junction lattice}\label{subsec:junction_lattice}

In the following, we give an abstract definition of junctions as lines in the plane perpendicular to the 7-branes satisfying the axioms above.
In principle, one has to specify if they represent $(p,q)$-strings or 5-brane webs to attach physical meaning to them.

Consider the junctions formed by a single prong extending from one 7-brane $\bX_{[p,q]}$, which we denote with a lower case letter as $\bx_{[p,q]}$, and sometimes call a unit junction.
In analogy to the different types defined in \eqref{eq:branespecies}, there are then also junctions
\begin{align}
\mathbf{a}\,, \enspace \mathbf{b} \,, \enspace \mathbf{c} \,, \enspace \mathbf{n} \,.
\label{eq:prongs}
\end{align}
Since a general string or 5-brane junction takes the form of a linear combination of the individual prongs, the set of all physical junctions (strings or 5-branes) on a 7-brane configuration, $\bX_{[p_1, q_1]} \bX_{[p_2,q_2]} \cdots \bX_{[p_i,q_i]} \cdots$, form a $\bbZ$-module,
\begin{align}
J_{\text{phys}} =  \big\{ {\bf j} = \sum_i a^i \mathbf{x}_{[p_i,q_i]} \, | \, a^i \in \mathbb{Z} \big\} \, .
\label{eq:Jphys}
\end{align}
One important physical invariant is the net, or \emph{asymptotic} $\colpq{p}{q}$-charge of a junction ${\bf j}$, given by $\colpq{p}{q}_\text{asymp} = \sum_i a^{i} \colpq{p_i}{q_i}$.

One further defines a symmetric bi-linear pairing $(.,.)$ on this module as follows.
For the basis junction $\bx_{[p_i, q_i]}$ (note that the ordering of the 7-branes is important), one defines\footnote{Here, we simply present the rules as stated in \cite{DeWolfe:1998zf}. It can be shown that they agree with the geometric intersection pairing for the elliptic K3 of the dual F-theory description.}
\begin{align}
\begin{split}
& \big( \mathbf{x}_{[p_i,q_i]}, \mathbf{x}_{[p_j,q_j]} \big) = \big( \mathbf{x}_{[p_j,q_j]}, \mathbf{x}_{[p_i,q_i]} \big) = \begin{cases}
    -1 \, , & \text{if } \, i=j \\
    \tfrac12 \det \left( \begin{smallmatrix} p_i & p_j \\ q_i & q_j \end{smallmatrix} \right) \, , & \text{if } \, \bX_{[p_i,q_i]} \, \text{ is on the left of } \, \bX_{[p_j, q_j]} \, .
\end{cases}
\end{split}
\label{eq:selfintersec}
\end{align}
By linearly extending to the module $J_\text{phys}$, we endow it with a lattice structure, which will be called the (physical) junction lattice.
For example, consider an arrangement of only $\bA$, $\bB$ and $\bC$ branes which are ordered ``alphabetically'',
\begin{align}
    \bA_1 \, \cdots \, \bA_\alpha \, \cdots \, \bB_1 \, \cdots \, \bB_\beta \, \cdots \, \bC_1 \, \cdots \, \bC_\gamma \, .
\end{align}
For this 7-brane configuration, we have
\begin{align}
\begin{split}
    & (\ba_\alpha, \ba_{\alpha'}) = -\delta_{\alpha, \alpha'} \, , \quad (\bb_\beta, \bb_{\beta'}) = - \delta_{\beta,\beta'} \, , \quad (\bc_{\gamma}, \bc_{\gamma'}) = -\delta_{\gamma,\gamma'} \, , \\
    & (\ba_\alpha, \bb_\beta) = -\tfrac12 \, , \quad (\ba_\alpha, \bc_\gamma) = \tfrac12 \, , \quad (\bb_\beta, \bc_\gamma) = 1 \, .
\end{split}
\end{align}

An important property of the pairing \eqref{eq:selfintersec} is that it is invariant under 7-brane motions.
That is, given a fixed set of 7-branes $\bX_{[p_i,q_i]}$, the lattice $(J_\text{phys}, ( \cdot, \cdot))$ changes only up to a unimodular transformation (i.e., change of basis) when we move the 7-branes.
To see this it suffices to consider a two-branes configuration $\bX_{[p_1, q_1]} \bX_{[p_2,q_2]}$ with $J_\text{phys} = \{a^1 \bx_{[p_1,q_1]} + a^2 \bx_{[p_2,q_2]} \}$, for which the pairing matrix is $\left( \begin{smallmatrix} -1 & D/2 \\ D/2 & -1 \end{smallmatrix} \right)$, with $D = \det\left( \begin{smallmatrix} p_1 & p_2 \\ q_1 & q_2 \end{smallmatrix} \right)$.
After moving $\bX_{[p_1,q_1]}$ across the branch cut to the right, as in \eqref{eq:brane_move_left-to-right} (the other direction, \eqref{eq:brane_move_right-to-left}, works analogously), the configuration $\bX_{[p_2,q_2]} \bX_{[p_1 + D p_2, q_1 + D q_2]} \equiv \bX_{l} \bX_{r}$ has the lattice
\begin{align}
    J_\text{phys} = \left\{ a_l \bx_{l} + a_r \bx_{r}  \right\} \, , \ \text{with} \ (\bx_{i} ,\bx_{j}) = \left( \begin{smallmatrix}
        -1 & -\tfrac{D}{2} \\
        -\tfrac{D}{2} & -1
    \end{smallmatrix} \right) = \left( \left( \begin{smallmatrix}
        -D & 1 \\
        1 & 0
    \end{smallmatrix} \right)^{-1} \right)^T \left( \begin{smallmatrix}
        -1 & \tfrac{D}{2} \\
        \tfrac{D}{2} & -1
    \end{smallmatrix} \right) \left( \begin{smallmatrix}
        -D & 1 \\
        1 & 0
    \end{smallmatrix} \right)^{-1} .
\end{align}
The unimodular transformation $\left( \begin{smallmatrix} -D & 1 \\ 1 & 0 \end{smallmatrix} \right)$ precisely traces how the original unit prongs $\{\bx_{[p_i,q_i]}\}$ are expressed in terms of the new basis $\{\bx_l, \bx_r\}$ after the 7-brane transition \eqref{eq:brane_move_left-to-right},
\begin{align}
    \bx_{[p_1,q_1]} \rightarrow -D \bx_{l} + \bx_{r} \, , \quad \bx_{[p_2,q_2]} \rightarrow \bx_{l} \, .
\end{align}

\subsubsection*{Loop junctions and their self-pairings}

\begin{figure}[ht]
    \centering
    \includegraphics[width = .35 \textwidth]{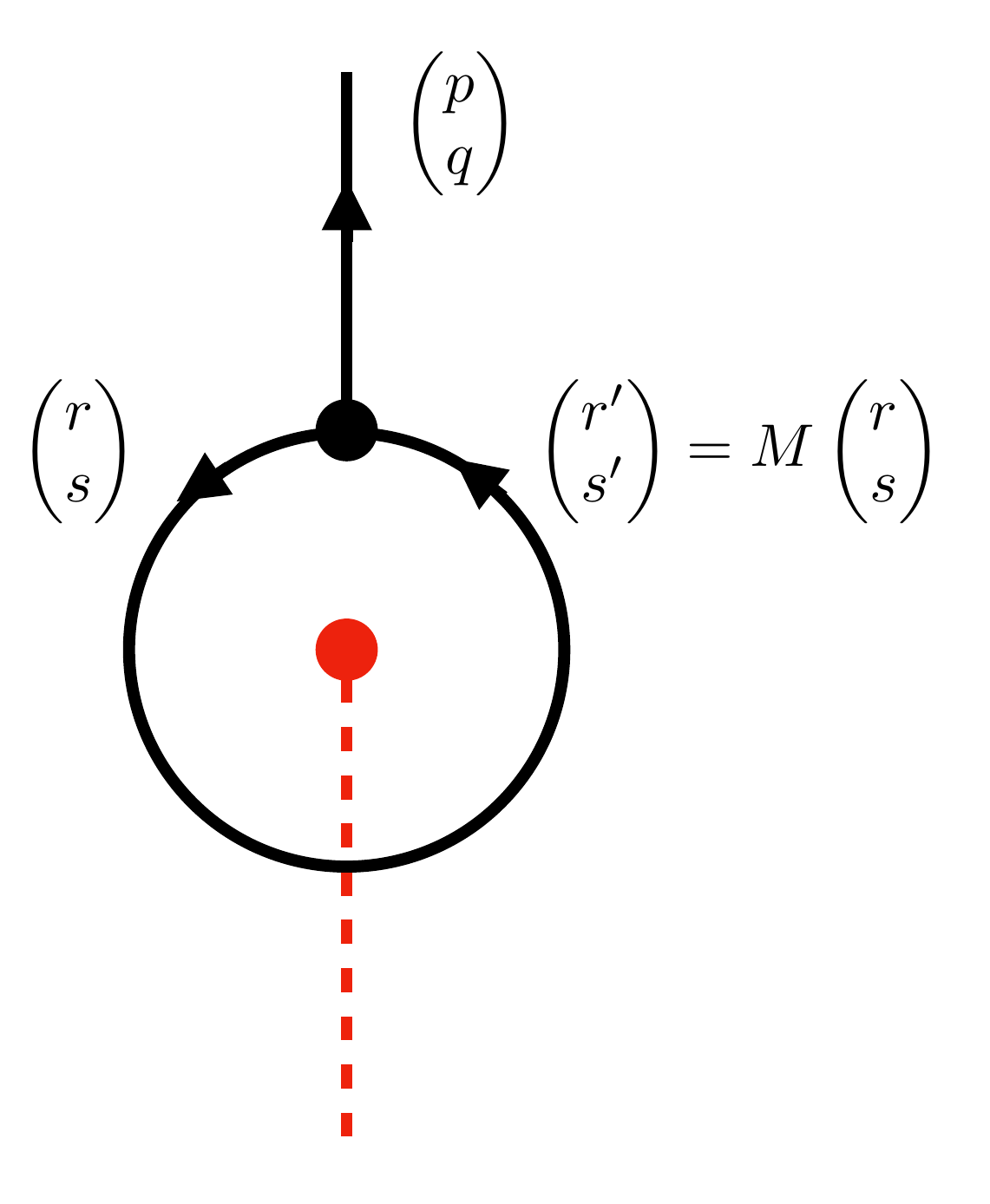}
    \caption[includes a matrix]{A loop junction $\boldsymbol\ell_{(r,s)}$ around a collection of 7-branes with overall monodromy $M$.
    The asymptotic charge $\colpq{p}{q} = \colpq{r'}{s'} - \colpq{r}{s} = (M - \mathbb{1}) \colpq{r}{s}$ is in general non-zero.
    }
    \label{fig:loop_junction}
\end{figure}

A junction type that will be particularly important to our discussions are \emph{loop junctions}.
These are formed by encircling a collection of 7-branes with an $\colpq{r}{s}$-charge, that undergoes $SL(2,\bbZ)$ transformations as it crosses their branch cuts.
As a convention for nomenclature, we use the $\colpq{r}{s}$-charge it starts out with to label the loop junction $\boldsymbol\ell_{(r,s)}$, even if its $(p,q)$-type changes after it comes back, see Figure \ref{fig:loop_junction}.
If the overall monodromy of the encircled stack is $M$, then such a loop has asymptotic charge $\colpq{p}{q} = (M - \mathbb{1}) \colpq{r}{s}$.
For two loops, $\boldsymbol\ell_{(r,s)}$ and $\boldsymbol\ell_{(u,v)}$, encircling the same 7-branes, one clearly has $\boldsymbol\ell_{(r,s)} + \boldsymbol\ell_{(u,v)} = \boldsymbol\ell_{(r+u,s+v)}$.
In principle, any such loop can be turned into the standard basis \ref{eq:Jphys} with prongs on 7-branes by pulling the loop across the encircled 7-branes via a Hanany--Witten transition, which allows to compute pairings involving loop junctions.
However, since the loop does not touch the encircled 7-branes, but only sees their overall monodromy, the self-pairing of a loop should be computable just with this data.

\begin{figure}[ht]
    \centering
    \includegraphics[width = 0.7 \textwidth]{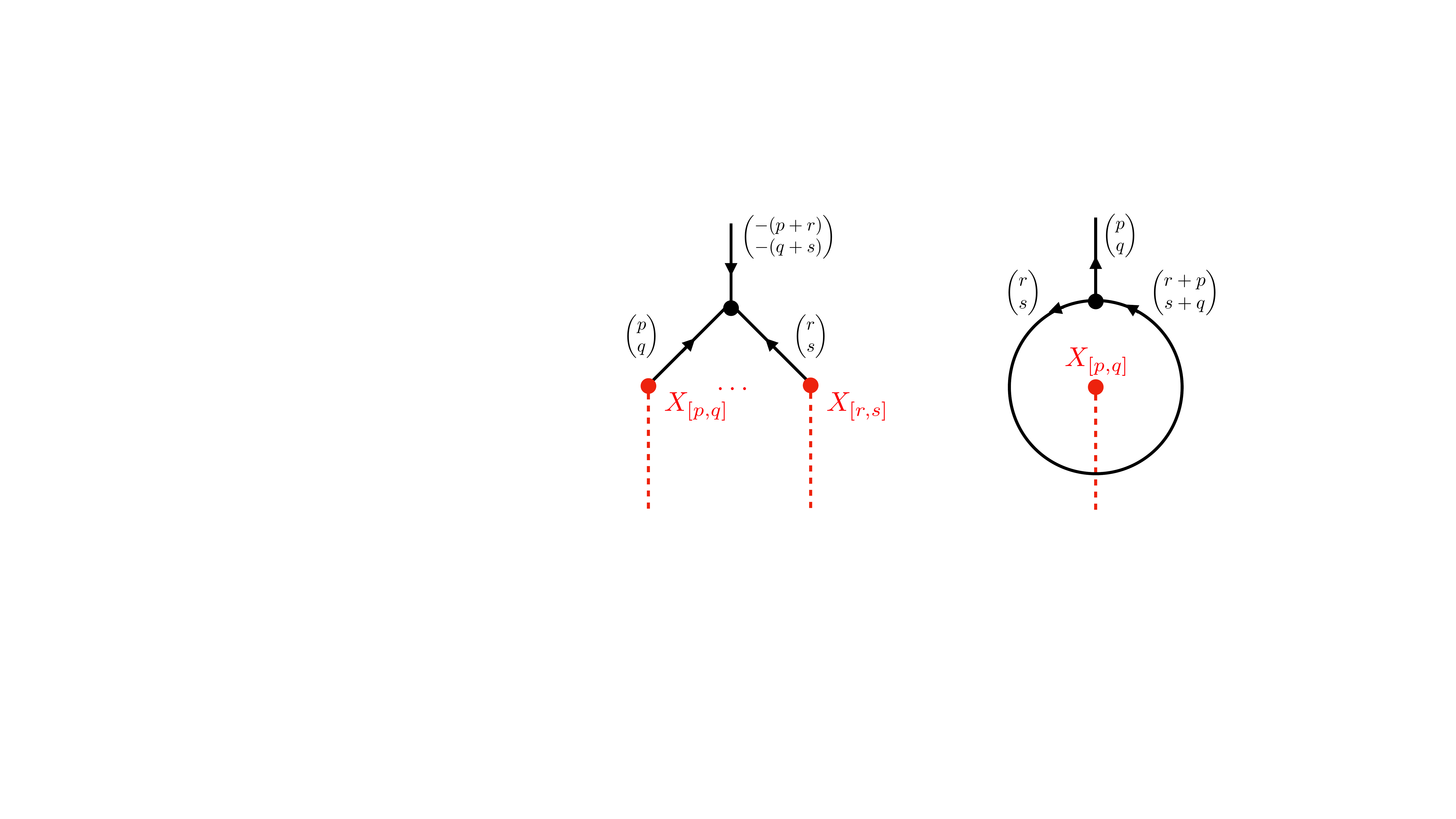}
    \caption{The self-pairing of a 3-pronged junction (left) can be separated into contributions from the ends on 7-branes and the vertex, see \eqref{eq:self_pairing_3-pronged}.
    When there are no prongs ending on 7-branes, such as for loop juncions (right), the only contribution is that of the vertex.
    }
    \label{fig:Juncexa}
\end{figure}

To do so, first consider the junction ${\bf j} = \mathbf{x}_{[p,q]} + \mathbf{x}_{[r,s]}$ as depicted on the left of Figure \ref{fig:Juncexa}.
According to \eqref{eq:selfintersec}, we have
\begin{align}\label{eq:self_pairing_3-pronged}
    ({\bf j}, {\bf j}) = ({\bf x}_{[p,q]}, { \bf x}_{[p,q]}) + ({\bf x}_{[r,s]}, {\bf x}_{[r,s]}) + 2({\bf x}_{[p,q]},{\bf x}_{[r,s]}) = -2 + \det \begin{pmatrix} p & r \\ q& s \end{pmatrix} \, .
\end{align}
As pointed out in \cite{DeWolfe:1998zf}, this result can also be interpreted as the sum of the contributions from the two end points of the 7-branes (each contribution $-1$), and the contribution of the 3-pronged vertex.
The latter must therefore be
\begin{align}
    \det \begin{pmatrix} p & r \\ q& s \end{pmatrix} = ps - rq = \det \begin{pmatrix} r & -(p+r) \\ s& -(q+s) \end{pmatrix} = \det \begin{pmatrix} -(p+r) & p \\ -(q+s) & q \end{pmatrix} \, ,
\end{align}
i.e., the determinant of two of the three $\colpq{p}{q}$-charge vectors, arranged in their counter-clockwise ordering (and all prongs either ingoing or outgoing).

This logic can now be easily applied to compute self-pairings of loop junctions.
Since such a junction has no endpoints on 7-branes, the only contribution to the self-pairing must come from the 3-pronged vertex.
For the junction $\boldsymbol\ell_{(r,s)}$ in Figure \ref{fig:loop_junction}, this contribution evaluates to (after accounting for the signs necessary to have all prongs in- or outgoing)
\begin{align}
    (\boldsymbol\ell_{(r,s)}, \boldsymbol\ell_{(r,s)}) = \det \begin{pmatrix} p & r \\ q & s \end{pmatrix} = - \det \begin{pmatrix} r & r' \\ s & s' \end{pmatrix} \, .
\end{align}
As a consistency check, consider a loop junction $\boldsymbol{\ell}_{(r,s)}$ around a single $[p,q]$-7-brane such that the asymptotic charge is $\colpq{p}{q}$ (see right of Figure \ref{fig:Juncexa}), i.e.,
\begin{align}
    (M_{[p,q]} - \mathbb{1}) \colpq{r}{s} = (qr - ps) \colpq{p}{s} \stackrel{!}{=} \colpq{p}{q} \quad \Leftrightarrow \quad (qr - ps) \stackrel{!}{=} 1 \, ,
\end{align}
which always has a solution for $(r,s)$ since the labels of a single $\bX_{[p,q]}$ must be coprime.
Then, the self-pairing is $(\boldsymbol\ell_{(r,s)}, \boldsymbol\ell_{(r,s)}) = \det \left( \begin{smallmatrix} p & r \\ q & s \end{smallmatrix} \right) = ps - qr = -1 = (\bx_{[p,q]}, \bx_{[p,q]})$.
This was expected, since by construction, this loop is equivalent, by a Hanany--Witten transition, to the unit junction $\bx_{[p,q]}$.

\subsubsection*{(Co-)weight lattices from junctions}

For a single brane stack of ADE type \eqref{eq:ADE_brane_stacks}, the physical junctions \emph{without} asymptotic charges are generated by
\begin{equation}
\begin{split}
\mathfrak{su}_n:& \quad \boldsymbol\alpha_i = \mathbf{a}_i - \mathbf{a}_{i + 1} \,, \enspace i \in \{1, \dots, n-1\} \,, \\
\mathfrak{so}_{2n}:& \quad \boldsymbol\alpha_i = \mathbf{a}_i - \mathbf{a}_{i + 1} \,, \enspace i \in \{1, \dots, n-1 \} \,, \enspace \boldsymbol\alpha_n = \mathbf{a}_{n-1} + \mathbf{a}_n - \mathbf{b} - \mathbf{c} \,, \\
\mathfrak{e}_n:& \quad \boldsymbol\alpha_i = \mathbf{a}_i - \mathbf{a}_{i + 1} \,, \enspace i \in \{1, \dots, n-2 \} \,, \enspace \boldsymbol\alpha_{n-1} = \mathbf{a}_{n-2} + \mathbf{a}_{n-1} - \mathbf{b} - \mathbf{c}_1 \,, \enspace \boldsymbol\alpha_n = \mathbf{c}_1 - \mathbf{c}_2 \,,
\end{split}\label{eq:roots_junctions_ADE}
\end{equation}
where we have indexed 7-branes and their associated unit junctions of the same $[p,q]$-type.
Computing their mutual bi-linear pairing of $\boldsymbol\alpha_i$ one finds
\begin{align}
(\boldsymbol\alpha_i, \boldsymbol\alpha_j) = A_{ij} \,,
\end{align}
with $A_{ij}$ the \textit{negative} Cartan matrix. 
Indeed, strings represented by the junctions above are associated to the W-bosons which lead to the enhanced gauge symmetry on the 7-brane stack.
We will call them \emph{root junctions} for obvious reasons, and they span the root junction lattice of the ADE algebra $\Lambda_\text{r} \subset J_\text{phys}$.

In complete analogy to representation theory (save for a minus sign for the pairing), the bi-linear pairing allows the definition of the \emph{coroot junctions}, whose span is the coroot junction lattice $\Lambda_\text{cr}$, as follows
\begin{align}
\boldsymbol\alpha^{\vee}_i = \frac{2}{-(\boldsymbol\alpha_i, \boldsymbol\alpha_i)} \boldsymbol\alpha_i \, .
\end{align}
Since for ADE algebras all roots have length-square $2$, these coincide with the root junctions.
However, physically, these should be thought of as the magnetically dual states, and hence arise from 5-brane webs represented by the junctions.
We can therefore also identify the pairing between two junctions, where one represents a string and the other a 5-brane, as the Dirac-pairing between electric and magnetic operators of the 8d gauge theory.

One further defines the \emph{weight junctions} ${\bf w}_i$, which are dual to the coroot junctions with respect to $(.,.)$ (or, more precisely, its $\mathbb{Q}$-linear extension),
\begin{align}
    ({\bf w}_i, \boldsymbol\alpha^{\vee}_j) = -\delta_{ij} \,.
\end{align}
They span the weight junction lattice $\Lambda_\text{w}$, and correspond to the electric states of the gauge symmetry if they represent a string. Similarly, one defines the \emph{coweight junctions} ${\bf w}_i^\vee$ and their lattice $\Lambda_\text{cw}$ via
\begin{align}
({\bf w}_i^{\vee}, \boldsymbol\alpha_j) = -\delta_{ij} \,,
\end{align}
which, when representing a 5-brane, is a magnetic state.

Note that the (co-)weights and (co-)roots are in a very real sense localized degrees of freedom.
For any additional 7-brane $\bX_{[r,s]}$ that is added to the system, we can explicitly compute from \eqref{eq:roots_junctions_ADE} that $(\bx_{[r,s]}, \boldsymbol\alpha^{(\vee)}_j) = 0$.
Therefore, any junction that has no prong on the 7-brane stack represents an uncharged state under the gauge symmetry on that stack.

For ADE algebras the coweights and weights again agree, and there the distinction between string and 5-brane junctions is only of formal nature.
However, it will become important once we include O7$^+$-planes.
Before that, we have to introduce the concept of so-called extended (co-)weight junctions \cite{DeWolfe:1998zf}.

\subsection{Extended (co-)weights and higher-form center symmetries}

In general, the (co-)weight junctions,
\begin{align}
    {\bf w}^\vee_i = \sum_j (-A^{-1})_{ij} \boldsymbol\alpha_j \, , \quad {\bf w}_i = \sum_j (-\tilde{A}^{-1})_{ij} \boldsymbol\alpha^\vee_j \, ,
\end{align}
with $\widetilde{A}_{ij} = (\boldsymbol\alpha_i^\vee, \boldsymbol\alpha_j^\vee)$, will have fractional coefficients in front of the unit prongs $\mathbf{x}_{[p_i,q_i]}$.
This implies that they are not physical junctions on their own.
However, they can be made physical by adding certain other fractional junctions with \emph{non-zero} asymptotic $\colpq{p}{q}$-charges, resulting in an integer (i.e., physical) junction with a prong that extends away from the 7-brane stack.
Equivalently, it formalizes the intuition that non-adjoint matter states (carrying weights that are not roots) on a 7-brane stack arise from open strings that have ends on other 7-branes (possibly at infinity).

As a simple example, consider $\fkg = \mathfrak{su}_2$, realized on an $\bA_1 \bA_2$-stack.
While the (co-)weight junction ${\bf w}^{(\vee)} = \tfrac12 (\ba_1 - \ba_2)$ without any asymptotic $(p,q)$-charge is non-physical, we can consider the unit string junctions $\ba_1$ or $\ba_2$, each of which carries an asymptotic $\colpq{p}{q} = \colpq{1}{0}$ charge.
From $(\ba_1 , \boldsymbol\alpha^\vee) = (\ba_1, \ba_1 - \ba_2) = -(\ba_2, \ba_1 - \ba_2) = -1$, we expect these (string) junctions to be fundamental matter of the $\mathfrak{su}_2$.
Note that we can formally write
\begin{align}
    \ba_1 = \tfrac12 (\ba_1 + \ba_2) + {\bf w} \, , \quad \ba_2 = \tfrac12 (\ba_1 + \ba_2) - {\bf w} \, .
\end{align}
Because $\tfrac12 (\ba_1 + \ba_2) \equiv \boldsymbol\omega$ has asymptotic charge $\colpq{1}{0}$, and satisfies $(\boldsymbol\omega, \boldsymbol\alpha) = 0$, we can interpret the above rewriting as separating the $\mathfrak{su}_2$ gauge charges of the unit junctions, captured by the summand proportional to $\bf w$, from the asymptotic $SL(2,\bbZ)$-charges, captured by $\boldsymbol\omega$.
By linearity, this separation can be done for any physical junction ${\bf j} = n_1 \ba_1 + n_2 \ba_2$.
For $\mathfrak{su}_2$, the state corresponding to ${\bf j} = s {\bf w} + k \boldsymbol\omega \in J_\text{phys}$ is a weight of an spin-$s/2$ representation, which has charge $s \mod 2$ under the $\bbZ_2$-center.
It is easy to see in this case, the physicality condition, i.e., for ${\bf j}$ to have integer number of prongs on the 7-branes, relates $s \equiv k \mod 2$.
Therefore, the coefficient of any physical junction in front of $\boldsymbol\omega$ provides an equivalent way to encode the center charge of that corresponding state.
This line of argument can be generalized to any ADE-stack \cite{DeWolfe:1998zf}.

\begin{figure}
    \centering
    \includegraphics[width = 0.5 \textwidth]{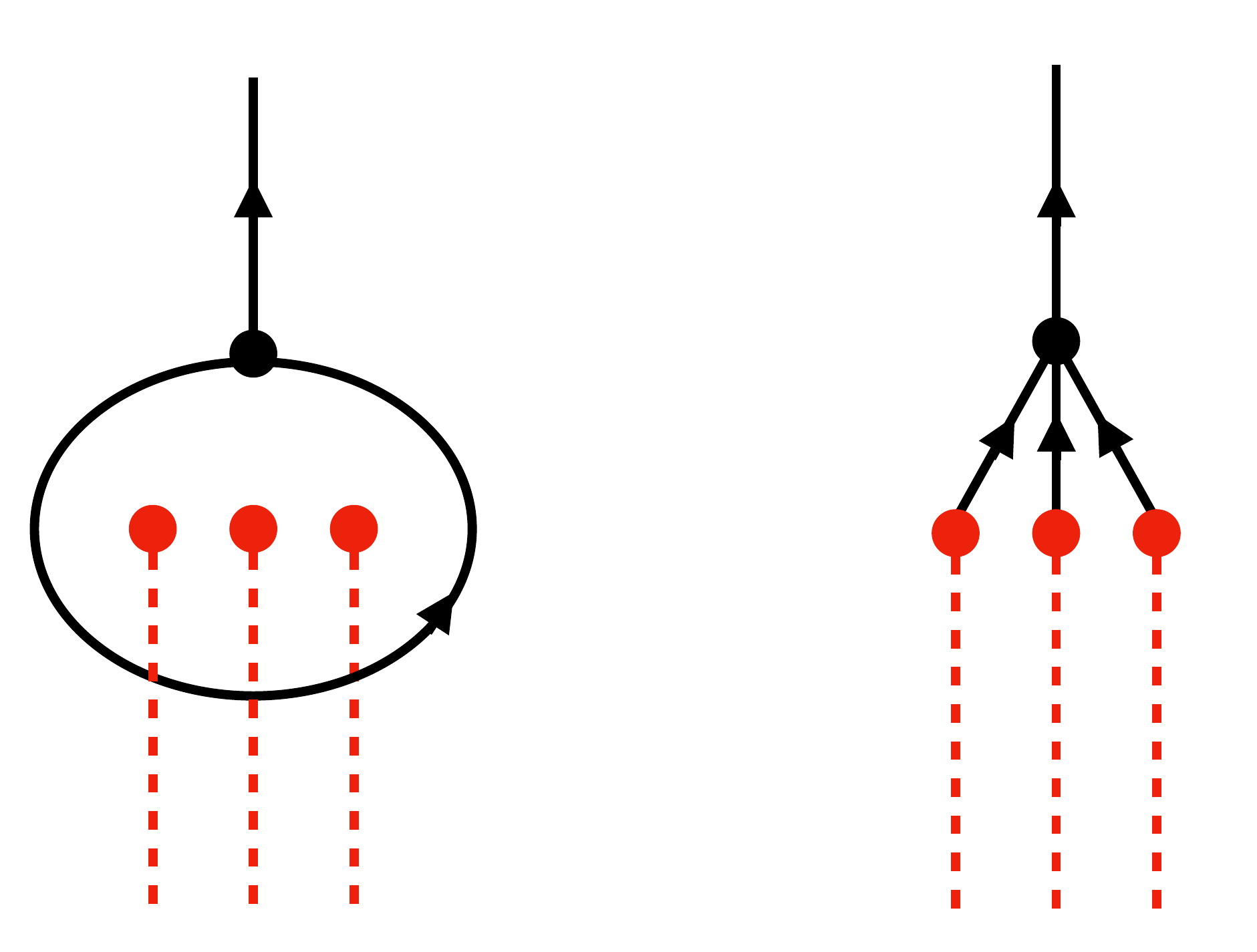}
    \caption[matrix]{Construction of extended weight junctions. Since the $\colpq{r}{s}$-charges that appear in the loop are in general fractional, the prongs ending on the 7-branes after pulling the loop across also have fractional coefficients.}
    \label{fig:ExtWeight}
\end{figure}

First, notice that, by a Hanany--Witten transition, $\boldsymbol\omega = \boldsymbol\ell_{(0,-\nicefrac{1}{2})}$ is a loop junction around the $\bA_1 \bA_2$ stack.
For a general stack with monodromy $M$, one defines $\boldsymbol\omega$, which are called extended weight junctions, as the generators of all loop junctions $\boldsymbol\ell_{(r,s)}$ (with possibly fractional $(r,s)$) encircling the stack that have integer asymptotic $\colpq{p}{q}$-charge, i.e.,
\begin{align}
    (M - \mathbb{1}) \colpq{r}{s} \in \mathbb{Z}^2 \, .
\end{align}
For the ADE algebras realized via the stacks as given in \eqref{eq:ADE_brane_stacks}, a standard basis for these are denoted $\boldsymbol\omega_{p,q}$ with asymptotic charges
\begin{align}
\boldsymbol\omega_p: \quad \colpq{p}{q}_{\text{asymp}} = \colpq{1}{0} \,, \quad \boldsymbol\omega_q: \quad \colpq{p}{q}_{\text{asymp}} = \colpq{0}{1} \,.
\label{eq:basisextweight}
\end{align}
For $\mathfrak{su}_n$ stacks, $(M-\mathbb{1})$ has only rank 1, so there is only one generator, $\boldsymbol\omega_p$ with asymptotic $\colpq{1}{0}$ charge.
Due to the generally fractional $(r,s)$-prong that crosses the branch cuts, the prongs that end on the constituent branes of the stack are also fractional after a Hanany--Witten transition, see Figure \ref{fig:ExtWeight}.
Explicitly, the extended weight junctions and their pairings are given by\footnote{Note that we can infer $\big(\boldsymbol{\omega}_p, \boldsymbol{\omega}_q\big)$ from the self-pairing of $\boldsymbol{\omega}_p + \boldsymbol{\omega}_q = \boldsymbol\ell_{(r_p, s_p)} + \boldsymbol\ell_{(r_q, s_q)} = \boldsymbol\ell_{(r_p+r_q, s_p+s_q)}$, which can be computed from the contribution of the single 3-pronged vertex, as in Figure \ref{fig:Juncexa}.}
\begin{align}
\begin{split}
    \mathfrak{su}_{n} \quad (\bA^n): & \quad \boldsymbol\omega_p = \boldsymbol\ell_{(0, -\nicefrac{1}{n})} = \tfrac{1}{n} \sum_{i = 1}^{n} \mathbf{a}_i \,, \quad (\boldsymbol\omega_p , \boldsymbol\omega_p) = -\tfrac{1}{n} \, ,  \\
    \mathfrak{so}_{2n} \quad (\bA^n \bB \bC): & \quad 
    \begin{cases}
        \boldsymbol\omega_p = \boldsymbol\ell_{(-\nicefrac{1}{2},0)} = \tfrac{1}{2}(\mathbf{b} + \mathbf{c}) \,, \\
        \boldsymbol\omega_q = \boldsymbol\ell_{(1-\nicefrac{n}{4}, -\nicefrac{1}{2})} = \tfrac{1}{2}(\sum_{i} \mathbf{a}_i - \mathbf{b} + \mathbf{c} - n \boldsymbol{\omega}_p) \,,
    \end{cases}  
    \ (\boldsymbol\omega_\alpha, \boldsymbol\omega_\beta) = 
    \begin{pmatrix}    0 & 0 \\ 0 & \tfrac{n}{4}-1    \end{pmatrix}_{\alpha\beta} ,
\\
    \mathfrak{e}_6 \quad (\bA^5 \bB \bC^2): & \quad
    \begin{cases}
        \boldsymbol\omega_{p} = \boldsymbol\ell_{(0,\nicefrac{1}{3})} = -\tfrac{1}{3} \sum_{i=1}^{5} \mathbf{a}_{i}+\tfrac{4}{3} \mathbf{b}+\tfrac{2}{3} \sum_{i=1}^{2} \mathbf{c}_{i} \,, \\ 
        \boldsymbol{\omega}_{q} = \boldsymbol\ell_{(-1,-1)} =\sum_{i=1}^{5} \mathbf{a}_{i}-3 \mathbf{b}-\sum_{i=1}^{2} \mathbf{c}_{i} \,,
    \end{cases}
    \ (\boldsymbol\omega_\alpha, \boldsymbol\omega_\beta) = 
    \begin{pmatrix}    \tfrac13 & -\tfrac12 \\ -\tfrac12 & 1 \end{pmatrix}_{\alpha\beta} ,
 \\
    \mathfrak{e}_7 \quad (\bA^6 \bB \bC^2): & \quad 
    \begin{cases}
        \boldsymbol{\omega}_{p} = \boldsymbol\ell_{(\nicefrac12, \nicefrac12)} =-\tfrac{1}{2} \sum_{i=1}^{6} \mathbf{a}_{i}+2 \mathbf{b}+\sum_{i=1}^{2} \mathbf{c}_{i} \,, \\
        \boldsymbol{\omega}_{q} = \boldsymbol\ell_{(-\nicefrac52, -\nicefrac32)} =\tfrac{3}{2} \sum_{i=1}^{6} \mathbf{a}_{i}-5 \mathbf{b}-2 \sum_{i=1}^{2} \mathbf{c}_{i} \,,
    \end{cases} 
    \ (\boldsymbol\omega_\alpha, \boldsymbol\omega_\beta) = 
    \begin{pmatrix}    \tfrac12 & -1 \\ -1 & \tfrac52 \end{pmatrix}_{\alpha\beta} ,
    \\
    \mathfrak{e}_8 \quad (\bA^7 \bB \bC^2): & \quad 
    \begin{cases}
        \boldsymbol{\omega}_{p}= \boldsymbol\ell_{(2,1)} =-\sum_{i=1}^{7} \mathbf{a}_{i}+4 \mathbf{b}+2 \sum_{i=1}^{2} \mathbf{c}_{i} \,, \\
        \boldsymbol{\omega}_{q}= \boldsymbol\ell_{(-7,-3)} =3 \sum_{i=1}^{7} \mathbf{a}_{i}-11 \mathbf{b}-5 \sum_{i=1}^{2} \mathbf{c}_{i} \,,
    \end{cases} 
    \ (\boldsymbol\omega_\alpha, \boldsymbol\omega_\beta) = 
    \begin{pmatrix}    1 & -\tfrac52 \\ -\tfrac52 & 7 \end{pmatrix}_{\alpha\beta} .
\end{split}
\label{eq:extADE}
\end{align}
All physical junctions associated to a 7-brane stack, i.e., junctions with prongs of only integer $\colpq{p}{q}$-charge, can be written uniquely in terms of a linear combination of weight and extended weight junctions 
\begin{align}
\mathbf{j} = \sum_i a^i {\bf w}_i + a^p \boldsymbol\omega_p + a^q \boldsymbol\omega_q \,, \quad a^i, a^p, a^q \in \mathbb{Z} \,.
\label{eq:nAstates}
\end{align}
Physically, this means that a physical $(p,q)$-string/-5-brane is fully characterized by its asymptotic electric/magnetic $\colpq{p}{q}$-charge under $(B_2, C_2)$, and the weight/coweight charges under the 7-brane gauge algebras.

In turn, it can be verified that every possible weight junction ${\bf w} = \sum_i a^i {\bf w}_i$ ($a^i \in \bbZ$) of the gauge algebra $\fkg$ can be completed into a physical junction by the addition of an integer linear combination ${\bf j}_e = a^p \boldsymbol\omega_p + a^q \boldsymbol\omega_q$ of extended weights \cite{DeWolfe:1998zf}.
Such integer linear combination is not unique and is determined only up to multiples $n^p \boldsymbol\omega_p + n^q \boldsymbol\omega_q$ which have integer charges for each prong.
This non-uniqueness can be understood as the fact that ${\bf j}_e$ is determined by the charge of ${\bf w}$ under the center $Z(\widetilde{G})$ of the simply-connected group $\widetilde{G}$ associated to $\mathfrak{g}$.
Intuitively, this is expected because the charge under the center of a specific state ${\bf w}$ is encoded in its prefactors of the weight basis ${\bf w}_i$, which in turn introduces fractional prongs that can only be cancelled by the extended weights.
Analogously to how weights can be screened by the W-bosons for observers ``at infinity'', there are multiples of specific asymptotic $\colpq{p}{q}$-charges that can be added and subtracted without affecting the local gauge dynamics on the 7-branes.\footnote{Describing the gauge dynamics by F-theory on a non-compact K3, this is reflected by the homology of the asymptotic boundary exhibiting discrete torsion, associated to the fact that $n^p \times \text{(A-cycle)} + n^q \times \text{(B-cycle)}$ on the generic torus fiber shrinks at the singularity \cite{Cvetic:2021sxm}.}

The precise connection between higher-form symmetries and extended weights have been described in \cite{Cvetic:2021sxm}.
Formally, we can define the lattice,
\begin{align}
J_{\text{ext}} = \{ \mathbf{j}_e = a^p \boldsymbol{\omega}_p + a^q \boldsymbol{\omega}_q \, | \, a^p, a^q \in \mathbb{Z} \} \, ,
\end{align}
whose elements are arbitrary integer linear combinations of extended weights that may be fractional.
Then, the screening arguments for the center symmetries, together with the junction characterization of gauge degrees of freedom, translates into:
\begin{align}
Z (\widetilde{G}_{\text{ADE}}) = \frac{\text{weights}}{\text{roots}} = \frac{\text{coweights}}{\text{coroots}} = \frac{J_{\text{ext}}}{J_{\text{phys}} \cap J_{\text{ext}}} \,,
\end{align}
where $(J_{\text{phys}} \cap J_{\text{ext}})$ denotes extended weight junctions that are themselves physical, i.e., do not contain fractional prongs.
Note that, since $\boldsymbol\omega_\circ = \boldsymbol\ell_{(r_\circ, s_\circ)}$ are loop junctions of the form depicted on the left of Figure \ref{fig:ExtWeight}, $(J_{\text{phys}} \cap J_{\text{ext}})$ are precisely the loops $\ell_{(r,s)} = n^p \ell_{(r_p, s_p)} + n^q \ell_{(r_q, s_q)}$ with integer $(r,s)$.
Concretely, in terms of the extended weights summarized in \eqref{eq:extADE}, one finds
\begin{align}\label{eq:center_charge_table_ADE}
    \renewcommand{\arraystretch}{1.8}
    \begin{array}{c|c}
        \mathfrak{g} & J_{\text{ext}} / ( J_\text{phys} \cap J_{\text{ext}} ) \\ \hline
        \mathfrak{su}_n & \displaystyle\frac{\{a^p \boldsymbol{\omega}_p\}}{(n \boldsymbol{\omega}_p)} \cong \big\{ a^p \ \text{mod } n \big\} \cong \bbZ_n \\
        \mathfrak{so}_{4n} & \displaystyle\frac{\{a^p \boldsymbol{\omega}_p + a^q \boldsymbol{\omega}_q\}}{(2 \boldsymbol{\omega}_p, 2\boldsymbol{\omega}_q ) } \cong \{ (a^p \ \text{mod } 2, a^q \ \text{mod } 2)\} \cong  \mathbb{Z}_2 \oplus \mathbb{Z}_2 \\
        \mathfrak{so}_{4n+2} & \displaystyle\frac{\{a^p \boldsymbol{\omega}_p + a^q \boldsymbol{\omega}_q\}}{(2 \boldsymbol{\omega}_p, 4 \boldsymbol{\omega}_q, \boldsymbol{\omega}_p + 2 \boldsymbol{\omega}_q )} \cong \big\{ 2a^p + a^q \mod 4 \big\} \cong \bbZ_4 \\
        \mathfrak{e}_6 & \displaystyle\frac{\{a^p \boldsymbol{\omega}_p + a^q \boldsymbol{\omega}_q\}}{(3 \boldsymbol{\omega}_p, \boldsymbol{\omega}_q)} \cong \big\{ a^p \ \text{mod } 3 \} \cong \bbZ_3 \\
        \mathfrak{e}_7 & \displaystyle\frac{\{a^p \boldsymbol{\omega}_p + a^q \boldsymbol{\omega}_q\}}{(2 \boldsymbol{\omega}_p, 2 \boldsymbol{\omega}_q, \boldsymbol{\omega}_p + \boldsymbol{\omega}_q )} \cong \big\{ a^p + a^q \mod 2 \big\} \cong \bbZ_2 
    \end{array}
\end{align}
In the language of higher-form symmetries (see also \cite{Cvetic:2021sxm}), a physical string/5-brane junction ${\bf j} = \sum_i a^i {\bf w}_i + a^p \boldsymbol\omega_p + a^q \boldsymbol\omega_q$ carries an electric/magnetic $Z(\widetilde{G})$ 1-form/5-form symmetry charge prescribed by \eqref{eq:center_charge_table_ADE}.

\section[Junctions on \texorpdfstring{\textbf{O7}$^+$}{O7+} and center symmetries of \texorpdfstring{$\mathfrak{sp}$}{sp} dynamics]{Junctions on O7\boldmath{$^+$} and center symmetries of \boldmath{$\mathfrak{sp}$} dynamcis}
\label{subsec:junctions_on_O7}

So far, we have reviewed the junction framework for ordinary $[p,q]$-7-branes, which succinctly encode the 8d ${\cal N}=1$ gauge dynamics with simply-laced gauge algebras.
However, field theoretically, one can also have $\mathfrak{sp}_n$ algebras. 
In the type IIB string constructions these are linked to the presence of O7$^+$-planes, which was not considered systematically within the junction framework previously.
Therefore, we need to generalize the above analysis.

First, we note that the O7$^+$-plane, unlike the the O7$^-$-plane, does not split, at finite string coupling, into constituents represented by ordinary $[p,q]$-7-branes.
Therefore, we will represent it by a single, albeit special, 7-brane.
The monodromy generated by one O7$^+$-plane is in the same $SL(2,\bbZ)$ conjugacy class as a 7-brane stack with $\mathfrak{g} = \mathfrak{so}_{16}$.
In the following local analysis, we use the same presentation as in \eqref{eq:ADE_brane_stacks},
\begin{align}\label{eq:O7_monodromy}
M_{\text{O7}^+} = \begin{pmatrix} -1 & 4 \\ 0 & -1 \end{pmatrix} \,.
\end{align}
There are multiple ways of arguing for this physically.
The prevalent interpretation of an O7$^+$ in recent literature \cite{Witten:1997bs, deBoer:2001wca} is as the remnant of ``freezing'' the $\mathfrak{so}_{16}$ gauge dynamics on an ordinary 7-brane stack, see also \cite{Tachikawa:2015wka, Bhardwaj:2018jgp}.

However, even after two decades, the freezing operation remains somewhat mysterious.
In particular, a geometric derivation of its effect on higher-form symmetries in the M-theory frame \cite{Morrison:2020ool,Albertini:2020mdx} appears to be challenging.
However, as we will argue now, one can obtain a complete picture, at least in the IIB duality frame, of the mechanism using junctions.
The key distinction to $[p,q]$-branes is that the physicality condition for prongs that end on an O7$^+$ \emph{differ} between strings and 5-branes.

From the perturbative IIB picture, only pairs of fundamental strings can end on an O7$^+$ \cite{Imamura:1999uf}, which one can see in a perturbative picture via Chan--Paton factors.
Via various dualities it can also be argued that only an even number of D-strings can end on the O7$^+$-plane, see \cite{Imamura:1999uf, Bergman:2001rp}. 
Thus, the physical string junctions emanating from the O7$^+$-plane have $\colpq{p}{q}$-charge restricted by $p, q \in 2 \mathbb{Z}$.
In contrast, $(p,q)$-5-branes can end with any integer number on O7$^+$.
This is relevant, e.g., in the construction of 5d SCFTs via 5-brane webs \cite{Bergman:2015dpa}.
Hence, a prong of a physical 5-brane junction can end with arbitrary integer $\colpq{p}{q}$-charge on an O7$^+$.
As we will see momentarily, these conditions naturally give rise to a consistent description of $\mathfrak{sp}_n$ gauge algebras, including their center symmetries, from the junction lattice.
Moreover, once we have set up the notation in Section \ref{sec:global}, we can derive these conditions independently in global models with one O7$^+$-plane that realize 8d rank $(2,10)$, by appealing to the dual description of these models via the CHL-string (see Appendix \ref{apdx:evenness}).

It is worthwhile to compare the boundary conditions for strings and 5-branes on O7$^+$ with the unfrozen $\mathfrak{so}_{16}$-stack.
First, since both generate the same monodromy, the loop junctions that generate all integer asymptotic $\colpq{p}{q}$-charges in the presence of a single O7$^+$ are the same as for an $\mathfrak{so}_{16}$ stack \eqref{eq:extADE},
\begin{align}
\boldsymbol{\omega}_p^{\text{O7}^+} = \boldsymbol{\ell}_{(- \nicefrac{1}{2},0)} \,, \quad \boldsymbol{\omega}_q^{\text{O7}^+} = \boldsymbol{\ell}_{(-1, - \nicefrac{1}{2})} \,.
\end{align}
We will call these the extended weight junctions of the O7$^+$, even though a single O7$^+$ (unlike its unfrozen cousin) has no gauge dynamics, and hence no root or weight lattice to begin with.
By collapsing the loop or, equivalently, performing a Hanany--Witten transition across the orientifold plane, these extended weight junctions have a $\colpq{1}{0}$ and a $\colpq{0}{1}$ prong, respectively, on the O7$^+$.
The set of junctions ending/emanating from one O7$^+$, which is entirely characterized by its total $\colpq{p}{q} = \colpq{a^p}{a^q}$-charge, can therefore be written as
\begin{align}
    {\bf j} = a^p \, \boldsymbol{\omega}_p^{\text{O7}^+} + a^q \, \boldsymbol\omega_q^{\text{O7}^+} \, , \quad a^p, a^q \in \bbZ \, .
\end{align}
Physical string junctions must then have $(a^p, a^q) \equiv (0 \ \text{mod 2}, 0\ \text{mod 2}) $, which agrees with the physicality condition for junctions on an $\mathfrak{so}_{16}$-stack that has no (unscreenable) gauge charge, see \eqref{eq:center_charge_table_ADE}.
Equivalently, any string loop $\boldsymbol\ell_{(r,s)}$ that encircles the O7$^+$ is physical if and only if $r$ and $s$ are both integer.
In contrast, a physical 5-brane junction with odd $(a^p, a^q)$, which can end on an O7$^+$, would not be admissible on an $\mathfrak{so}_{16}$-stack without picking up (unscreenable) gauge charge.
In particular, this means that \emph{physical} 5-brane loops $\boldsymbol\ell_{(r,s)}$ around the O7$^+$ could have \emph{half-integer} valued $r$ and $s$.

To fully incorporate O7$^+$'s in the junction framework, we also need to define the bi-linear pairing.
It is hard to come up with a rule for prongs ending on the O7$^+$ by appealing to any geometric counterpart in a dual M-/F-theory picture, because of the presence of frozen singularities there.
However, since the extended weights can be viewed as loops which is only sensitive to the induced $SL(2,\bbZ)$ monodromy $M$, but not the ``microscopics'' of a 7-brane stack, one would naturally expect that the pairing of such junctions is insensitive to whether $M$ is sourced by an O7$^+$, or an $\mathfrak{so}_{16}$ stack.
Following the discussion around Figure \ref{fig:Juncexa}, we can therefore directly compute (rather than define, which would require further justifications) from the loop junction representation:
\begin{align}
\big(\boldsymbol{\omega}_p^{\text{O7}^+}, \boldsymbol{\omega}_p^{\text{O7}^+}\big) = 0 \,, \quad \big(\boldsymbol{\omega}_q^{\text{O7}^+}, \boldsymbol{\omega}_q^{\text{O7}^+}\big) = 1 \,, \quad \big(\boldsymbol{\omega}_p^{\text{O7}^+}, \boldsymbol{\omega}_q^{\text{O7}^+}\big) = 0 \,.
\label{eq:bilinext}
\end{align}
Considering O7$^+$-planes together with general $[r,s]$-7-branes (which we assume to be on the left of the O7$^+$), one further finds
\begin{align}
(\boldsymbol{\omega}_p^{\text{O7}^+}, \mathbf{x}_{[r,s]}) = -\tfrac{s}{2} \,, \quad (\boldsymbol{\omega}_q^{\text{O7}^+}, \mathbf{x}_{[r,s]}) = \tfrac{r}{2} \,.
\end{align}

\subsubsection*{$\mathfrak{sp}$ gauge algebras and their higher-form symmetries from junctions}

From the perturbative IIB picture, it is well-known that we can generate 8d $\mathfrak{sp}_n$ gauge dynamics on a 7-brane stack formed by $n$ $\bA$-branes on top of one O7$^+$.
This 7-brane stack, of the form $\mathbf{A}^n {\bf O7}^+$, has the same monodromy as an $\mathfrak{so}_{16+2n}$ stack:
\begin{align}\label{eq:monodromy_sp}
M_{\mathfrak{sp}_n} = M_{\mathbf{A}^n {\bf O7}^+} = \begin{pmatrix} -1 & 4 + n \\ 0 & -1 \end{pmatrix} \,.
\end{align}
This allows us to straightforwardly define the extended weight junctions, as loops $\ell_{(r,s)}$ around the entire stack (including the O7$^+$) such that an asymptotic $\colpq{1}{0}$- (for $\boldsymbol\omega_p$) or $\colpq{0}{1}$-charge (for $\boldsymbol\omega_q$) remains.
Then, after performing the suitable Hanany--Witten transitions, we find
\begin{align}\label{eq:sp_extended_weights}
\begin{split}
    & \boldsymbol\omega_p^{\mathfrak{sp}_n} = \boldsymbol\ell_{(-\nicefrac12,0)} = \boldsymbol\omega_p^{\text{O7}^+} \,, \quad \boldsymbol\omega_q^{\mathfrak{sp}_n} = \boldsymbol\ell_{(-1 - \nicefrac{n}{4}, -\nicefrac12)} =\tfrac{1}{2} \sum_{i = 1}^n \mathbf{a}_i - \tfrac{n}{2} \boldsymbol\omega_p^{\text{O7}^+} + \boldsymbol\omega_q^{\text{O7}^+} \, .
\end{split}
\end{align}
Since $M_{\mathfrak{sp}_n} = M_{\mathfrak{so}_{16+2n}}$, the loop junctions look identical to those of $\mathfrak{so}_{16+2n}$.
Hence, with $J_\text{ext} = \{ a^p \boldsymbol\omega_p^{\mathfrak{sp}_n} + a^q \boldsymbol\omega_q^{\mathfrak{sp}_n} \ | \ a^p, a^q \in \bbZ \}$, the physicality condition on linear combinations of extended junctions is captured by
\begin{align}\label{eq:wrong_electric_center_sp}
    J_\text{ext} / \big( J_\text{ext} \cap J_\text{phys, strings}) = \begin{cases}
        \big\{ (a^p \ \text{mod 2} , a^q \ \text{mod 2}) \} \cong \bbZ_2 \times \bbZ_2 \, , & n \ \text{even} \, , \\
        \big\{ 2a^p + a^q \ \text{mod 4} \} \cong \bbZ_4 \, , & n \ \text{odd} \, .
    \end{cases}
\end{align}
On the other hand, for 5-brane junctions, we have
\begin{align}
    J_\text{ext} / \big( J_\text{ext} \cap J_\text{phys, 5-branes}) = 
        \big\{  a^q \ \text{mod 2} \} \cong \bbZ_2 \, .
\end{align}
Comparing to the discussion around \eqref{eq:center_charge_table_ADE}, one might be tempted to identify $Z(Sp(n))_\text{electric} = \bbZ_2 \times \bbZ_2$ or $\bbZ_4$, and $Z(Sp(n))_\text{magnetic} = \bbZ_2$, which is clearly not correct.

\begin{figure}[ht]
    \centering
    \includegraphics[width = 0.4 \textwidth]{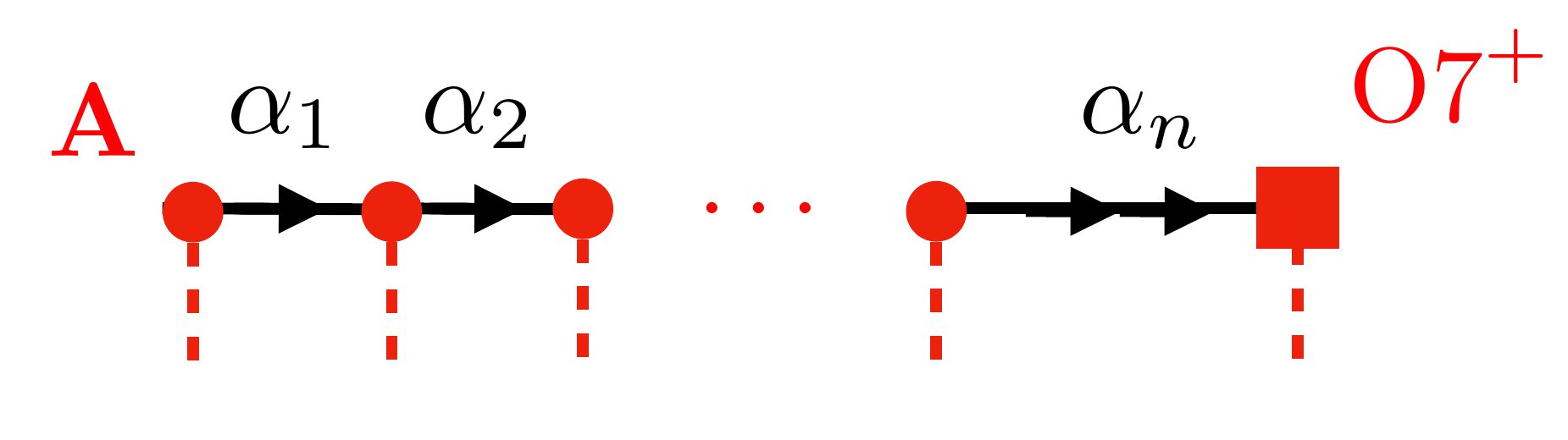}
    \caption{Root junctions of $\mathfrak{sp}$ algebra (the double arrow denotes the factor of 2 required by evenness on O7$^+$).}
    \label{fig:sproots}
\end{figure}

To rectify this, we must instead consider in detail the role of the extended weights as completing weights and coweights into physical strings and 5-branes, respectively.
In analogy to the ADE-stacks, we first construct the $\mathfrak{sp}_n$ roots $\boldsymbol\alpha_i$ as string junctions with no asymptotic $\colpq{p}{q}$-charge, that stretch between the constituents of this stack.
A basis for such junctions are
\begin{align}
\boldsymbol{\alpha}_i = \mathbf{a}_i - \mathbf{a}_{i+1} \,, \enspace i \in \{1, \dots, n-1\} \,, \enspace \boldsymbol{\alpha}_n = 2 \mathbf{a}_n - 2 \boldsymbol{\omega}_p^{\text{O7}^+} \,,
\end{align}
see also Figure \ref{fig:sproots}.
With this and the bi-linear pairings in \eqref{eq:bilinext}, one straightforwardly verifies
\begin{align}
(\boldsymbol{\alpha}_n , \boldsymbol{\alpha}_n) = -4 \, , \quad (\boldsymbol{\alpha}_i , \boldsymbol{\alpha}_n) = 2 \, \delta_{i, n-1} \, , \quad  (\boldsymbol{\alpha_i}, \boldsymbol{\alpha}_j) = 
    \begin{cases} -2 \, , & i = j \, , \\
        1 \, , & | i -j | = 1\, , \\
        0 \, , & \text{else} \, ,
    \end{cases}
     \quad 1 \leq i,j \leq n-1 \,,
\end{align}
which precisely reproduces the negative of the Cartan matrix of an $\mathfrak{sp}_n$ algebra.
In particular, we see that, while the short roots (those with length squared 2) arise from single-pronged strings between the $\bA$-branes, just as for ADE-algebras, the long root $\boldsymbol{\alpha}_n$ is only a generator due to the evenness condition for strings ending on the O7$^+$.

As a non-simply-laced algebra, $\mathfrak{sp}$ has different coroots than roots:
\begin{align}
\boldsymbol{\alpha}_i^\vee = \frac{2 \boldsymbol{\alpha}_i}{- (\boldsymbol{\alpha}_i, \boldsymbol\alpha_i)} = 
\begin{cases} 
    \boldsymbol\alpha_i \, , & 1 \leq i \leq n-1 \, ,\\
    \frac{1}{2} \boldsymbol\alpha_n = {\bf a}_n - \boldsymbol\omega^{O7{^+}}_p \, , & i = n \,.
\end{cases}
\end{align}
Since these are the magnetic objects in a gauge theory context, they arise from 5-brane junctions, which, consistent with the boundary conditions, can have a single prong on the O7$^+$ that is needed to form the short coroot $\boldsymbol\alpha^\vee_n$ with length squared 1.

The distinction between strings and 5-branes are of course also important for the weights and coweights.
The weight junctions are obtained as dual to the coroot junctions. Defining the matrix $\widetilde{A}_{ij} = \big( \boldsymbol\alpha^{\vee}_i, \boldsymbol\alpha_j^{\vee} \big)$, these can be written as
\begin{align}
\mathbf{w}_i = \big( - \widetilde{A}^{-1} \big)_{ij} \, \boldsymbol\alpha^{\vee}_j = \sum_j \text{min}\{i,j\} \, \boldsymbol\alpha_j^\vee = \sum_{j=1}^{n-1} \text{min}\{i,j\} \, \boldsymbol\alpha_j + \tfrac12 \text{min}\{i,n\} \, \boldsymbol\alpha_n  \,.
\label{eq:spweights}
\end{align}
Similarly, one obtains the coweight junctions as duals of the root junctions. With the negative Cartan matrix $A = (\alpha_i, \alpha_j)$ one has
\begin{align}
\mathbf{w}_i^\vee = \big( - A^{-1} \big)_{ij} \, \boldsymbol\alpha_j = 
\begin{cases}
    \sum_{j=1}^{n-1} \text{min}\{i,j\} \, \boldsymbol\alpha^\vee_j + i \, \boldsymbol\alpha^\vee_n \, , & i < n \, , \\
    \sum_{j=1}^{n} \frac{j}{2} \, \boldsymbol\alpha_j^\vee \, , & i = n \,,
\label{eq:spcoweights}
\end{cases}
\end{align}
expressed as fractional linear combinations of coroots.

In terms of the junctions, the electric center symmetry,
\begin{align}
    Z\big(Sp(n) \big) = \text{weights} / \text{roots} \cong \frac{ \{ \tfrac{k}{2} \boldsymbol\alpha_n \} }{(\boldsymbol\alpha_n)} \cong \bbZ_2,
\end{align}
is precisely generated by multiples of $\tfrac12 \boldsymbol\alpha_n = \ba_n - \boldsymbol\omega_p^{\text{O7}^+}$, which is unphysical as a string junction because of the odd prong on the O7$^+$.
To obtain a physical junction with the same $\mathfrak{sp}_n$ gauge charge, we must therefore add linear combinations of the extended weights \eqref{eq:sp_extended_weights} with integer prongs on $\bA$, and odd numbers of prong of $\colpq{p}{q} = \colpq{1}{0}$-charge on the O7$^+$.
For even $n$, this requirement is met by $a^p \boldsymbol\omega_p^{\mathfrak{sp}_n} + a^q \boldsymbol\omega_q^{\mathfrak{sp}_n}$ with $a^p \equiv 1 \ \text{mod } 2$ and $a^q \equiv 0 \ \text{mod } 2$, whereas for odd $n$, we need $2a^p + a^q \equiv 2 \ \text{mod } 4$.
Each of these generate a $\bbZ_2$-subgroup of the putative center \eqref{eq:wrong_electric_center_sp}, which is the correct presentation of $Z\big(Sp(n) \big)_\text{electric} \cong Z\big( Sp(n) \big)$.
The mismatch from \eqref{eq:wrong_electric_center_sp} is because not all integer linear combinations of extended weight junctions can be completed into a physical string junction with an $\mathfrak{sp}_n$ weight.
For the magnetic center symmetry,
\begin{align}
    \text{Hom}( Z\big(Sp(n) \big), \bbZ) \cong Z\big(Sp(n) \big) = \text{coweights} / \text{coroots} \cong \frac{ \{ k{\bf w}_n^\vee \} }{\text{coroots}} \, ,
\end{align}
the unphysical 5-brane coweight junctions come from the half-integer valued prongs in
\begin{align}
    {\bf w}^{\vee}_n = \sum_{i=1}^n \tfrac{i}{2} \boldsymbol\alpha_i^\vee = \tfrac{1}{2} \sum_{i = 1}^n \mathbf{a}_i - \tfrac{n}{2} \boldsymbol\omega_p^{\text{O7}^+} = \boldsymbol\omega_q^{\mathfrak{sp}_n} - \boldsymbol\omega_q^{\text{O7}^+}\, ,
\end{align}
which can be made physical by adding $a^p \boldsymbol\omega_p^{\mathfrak{sp}_n} + a^q \boldsymbol\omega_q^{\mathfrak{sp}_n}$ with $a^q \equiv 1 \ \text{mod 2}$.

In summary, for a physical string junction that ends on an $\mathfrak{sp}_n$ stack, which can be uniquely decomposed as
\begin{align}
    {\bf j} = \sum_i a^i {\bf w}_i + a^p \boldsymbol\omega_p^{\mathfrak{sp}_n} + a^q \boldsymbol\omega_q^{\mathfrak{sp}_n}
\end{align}
in terms of the extended weights \eqref{eq:sp_extended_weights} and weight junctions \eqref{eq:spweights}, the $\bbZ_2$ charge of the corresponding state under the electric center symmetry is
\begin{align}\label{eq:center_charge_string_sp}
\begin{cases}
    a^p \ \text{mod 2} \, , & \text{$n$ even,} \\
    a^p + \tfrac12 a^q \ \text{mod 2} \, , & \text{$n$ odd (then $a^q$ must be even).}
\end{cases}
\end{align}
For a physical 5-brane junction, decomposed into extended weights and coweights \eqref{eq:spcoweights},
\begin{align}
    {\bf j} = \sum_i a^i {\bf w}^\vee_i + a^p \boldsymbol\omega_p^{\mathfrak{sp}_n} + a^q \boldsymbol\omega_q^{\mathfrak{sp}_n} \, ,
\end{align}
its magnetic $\bbZ_2$-center charge is
\begin{align}\label{eq:center_charge_5-brane_sp}
    a^q \ \text{mod 2} \, .
\end{align}

This completes the list of local building blocks of simple gauge algebras that can be combined into a global model describing 8d supergravity.
As we will discuss now, the consistent combination of the individual brane stacks then determines the global structure of the gauge dynamics in these models.

\section{8d string vacua and their global structure from junctions}\label{sec:global}

In this section, we combine the above local descriptions of simple gauge algebras into a compact setting, to classify all 8d $\cN = 1$ string vacua using junctions.
These vacua fall into three moduli branches, which have gauge rank $(2,18)$, $(2,10)$, and $(2,2)$, respectively.
As shown in \cite{Hamada:2021bbz}, the gauge symmetries of the effective supergravity descriptions can be classified by the $SL(2,\bbZ)$ monodromies assoiated with each simple gauge factor, with a few additional consistency conditions.
In theories of rank $(2,18)$, which enjoy a description as F-theory on an elliptically-fibered K3 surface, these restrictions are met by 24 $[p,q]$-7-branes with trivial total monodromy \cite{Morrison:1996na,Morrison:1996pp,Douglas:2014ywa}.
Via the freezing procedure \cite{Witten:1997bs,Tachikawa:2015wka,Bhardwaj:2018jgp}, one can further obtain all rank $(2,10)$ or $(2,2)$ vacua, if one replaces any rank $(2,18)$ 7-brane configuration with one or two $\mathfrak{so}_{16+2n}$ stacks with the corresponding frozen $\mathfrak{sp}_{n}$ algebra that contains an O7$^+$ \cite{Hamada:2021bbz}.
Based on this, junctions provide a unified description of the gauge dynamics, in particular, the gauge group topology, for all these vacua.

Before we dive into the details, let us give a schematic description of this approach.
For a given 7-brane configuration (with or without O7$^+$) in a global model, the set of characters or cocharacters (i.e., a sublattice of the weight or coweight lattice that is occupied by dynamical states) correspond to physical (string or 5-brane) junctions that have zero asymptotic $\colpq{p}{q}$-charge.
Since the $\colpq{p}{q}$-charge of a prong ending on a stack is entirely captured by extended weight junctions, which in turn encodes the center charge of the (co-)characters represented by the junction, enumerating all linear combinations of extended weights from different stacks that add up to zero $\colpq{p}{q}_\text{asymp}$ also enumerates the center charge of all dynamical gauge charges.
In particular, computing the center charges of all string junctions that have no $\mathfrak{u}(1)$-charges determines $Z(G)$, and those of 5-brane junctions determine $\pi_1(G) \cong {\cal Z}$, where $G = \widetilde{G} / {\cal Z}$ is the physically realized non-Abelian gauge group, with simply-connected cover $\widetilde{G}$.

For the rank $(2,18)$ branch, the information about $\mathcal{Z}$ has been shown to be conveniently encoded in so-called fractional null junctions \cite{Guralnik:2001jh}, which are certain fractional multiples of physical loop junctions $\boldsymbol\ell^N_{(r,s)}$ around all 7-branes (with trivial total monodromy).
As we will see, this correspondence continues in realizations of rank $(2,10)$ with one O7$^+$-plane, and rank $(2,2)$ theories with two O7$^+$'s.
While for rank $(2,10)$, we can crosscheck the results with those obtained from a dual CHL string description \cite{Font:2021uyw,Cvetic:2021sjm}, the junction description provides a prediction for the gauge group topology of rank $(2,2)$ vacua which are inaccessible via the heterotic/CHL string.
To illustrate the procedure we will explicitly work out an example for each of the three branches of the 8d moduli space.

\subsection{Gauge group topology from global null junctions}
\label{subsec:global_junciton_lattice_null_junctions}

The construction of supersymmetric 8d theories with a dynamical gravity sector and rank $(2,18 - 8k)$ gauge sector requires the identification of a set of $(24 - 10k)$ $[p,q]$-7-branes and $k$ O7$^+$-planes with vanishing overall monodromy. These configurations can then be placed on a $\mathbb{P}^1$ which compactifies the underlying type IIB theory from ten to eight dimensions\footnote{The number of branes in the setup can also be understood as the demand that their cumulative gravitational backreaction in terms of the induced deficit angle adds up to $4 \pi$ as appropriate for the 2-sphere $S^2 \sim \mathbb{P}^1$.}.

Following the conventions laid out in the previous section, we arrange the 7-branes along a horizontal axis in the perpendicular plane (which is now a compact $\mathbb{P}^1$), and enumerate (from the left to right) the ordinary $[p,q]$-7-branes and the O7$^+$'s separately.
Within the vector space of all possible junctions (which carries a pairing given by simply linearly-extending the rules \eqref{eq:selfintersec} and \eqref{eq:bilinext}),
\begin{align}\label{eq:junction_lattice_compact}
    J = \left\{ \sum_{i=1}^{24-10k} a^i \bx_{[p_i,q_i]} + \sum_{j=1}^k \left( b^j \boldsymbol\omega_p^{\text{O7}^+_j} + \tilde{b}^j \boldsymbol\omega_q^{\text{O7}^+_j} \right) \ \big| \ a^i, b^j, \tilde{b}^j \in \mathbb{Q} \right\} \, ,
\end{align}
string junctions giving rise to the electrically charged states must have integral number of prongs on 7-branes, with ``integrality'' on the O7$^+$ being defined as having even number of prongs. 
Analogously, magnetically charged states are described by physical 5-branes with integral number of prongs on all 7-branes, including the O7$^+$'s.
However, since the 7-branes move on a compact $\mathbb{P}^1$, physical junctions must have vanishing asymptotic $\colpq{p}{q}$-charge, i.e., have no open ends.
This means that the \emph{physical} string / 5-brane junction lattice, corresponding to dynamical electric / magnetic states of the 8d supergravity theory, is
\begin{align}
\begin{aligned}
    J_\text{phys}^\text{el} = & \left\{
    \renewcommand{\arraystretch}{1.5}
    \begin{array}{c}
        \mathbf{j} = \sum_{i = 1}^{24 - 10k} a^i \, \mathbf{x}_{[p_i,q_i]} + \sum_{j = 1}^k \Big( 2 b^j \, \boldsymbol\omega^{\text{O7}^+_j}_p + 2 \tilde{b}^j \, \boldsymbol\omega_q^{\text{O7}^+_j} \Big) \, , \\
    \text{with} \ a^i, b^j, \tilde{b}^j \in \mathbb{Z} \, , \quad \sum_i a^i p_i + \sum_j 2b^j = 0 \, , \quad \sum_i a^i q_i + \sum_j 2\tilde{b}^j = 0 
    \end{array}
    \right\} \, , \\
    J_\text{phys}^\text{mag} = & \left\{
    \renewcommand{\arraystretch}{1.5}
    \begin{array}{c}
        \mathbf{j} = \sum_{i = 1}^{24 - 10k} a^i \, \mathbf{x}_{[p_i,q_i]} + \sum_{j = 1}^k \Big( b^j \, \boldsymbol\omega^{\text{O7}^+_j}_p + \tilde{b}^j \, \boldsymbol\omega_q^{\text{O7}^+_j} \Big) \, , \\
    \text{with} \ a^i, b^j, \tilde{b}^j \in \mathbb{Z} \, , \quad \sum_i a^i p_i + \sum_j b^j = 0 \, , \quad \sum_i a^i q_i + \sum_j \tilde{b}^j = 0 
    \end{array}
    \right\} \, .
\end{aligned}
\end{align}
Obviously, these lattices are of rank $24 - 10k -2 = 22-10k$.

Furthermore, by Hanany--Witten transitions, different elements in these lattices can represent the same physical junction.
Equivalently, we can add arbitrary multiples of so-called \emph{(global) integer null junctions}, 
\begin{align}
    J^{N,\text{el/mag}}_\text{int} = \left\{ \boldsymbol\delta_{(r,s)}^N \in J^\text{el/mag}_\text{phys} \ \big| \  \boldsymbol\delta_{(r,s)}^N = \boldsymbol\ell_{(r,s)} \ \text{loops around \emph{all} 7-branes} \right\} \, ,
\end{align}
where it is understood that, \textit{a priori}, there are different integral null junctions for strings and 5-branes.
Because of the compactness, such a loop can be shrunk to a point ``on the other side'' of the $\mathbb{P}^1$ without crossing any 7-branes, and thus are physically trivial.
However, they would appear as a non-trivial element in $J_\text{phys}$ after pulling them through the 7-branes, which must therefore be modded out before we can identify the junction lattice with the physical charge lattice.
Note that, by construction, $\boldsymbol\delta_{(r,s)}^N \in J^N_\text{int}$ has trivial pairing with any other $\boldsymbol\delta^N \in J^N_\text{int}$, as well as no asymptotic $\colpq{p}{q}$-charge (since they encircle a configuration with trivial overall monodromy).
Moreover, $(\boldsymbol\delta_{(r,s)}^N, {\bf j}) = 0$ for all $\boldsymbol\delta_{(r,s)}^N \in J_\text{int}^N$ if and only if ${\bf j}$ has zero asymptotic charge \cite{DeWolfe:1998zf}.
Hence, $\boldsymbol\delta_{(r,s)}^N \in J_\text{int}^N$ has trivial pairing with all physical junctions, explaining the prefix ``null''.
As a notational convention, we shall denote any loop junctions $\boldsymbol\ell_{(r,s)}$ with no asymptotic charge by $\boldsymbol\delta_{(r,s)}$.

This allows us now to identify the {\it (co-)character lattice} $\Lambda_\text{c}$ ($\Lambda_\text{cc}$), the lattice of all electrically (magnetically) charged states present in the supergravity theory, as
\begin{align}
    \Lambda_\text{c} \cong J^\text{el}_\text{phys} / J_\text{int}^{N,\text{el}} \, , \quad \Lambda_\text{cc} \cong J^\text{mag}_\text{phys} / J_\text{int}^{N,\text{mag}} \, ,
\end{align}
which are rank $20 - 8k$ lattices.
Since we mod out a sublattice which is null, the junction pairing on \eqref{eq:junction_lattice_compact} induces a non-degenerate pairing on these lattices, whose signature can be shown to be $(2, 18-8k)$.
For $[{\bf j}_e] \in \Lambda_\text{c}$ and $[{\bf j}_m] \in \Lambda_\text{cc}$ with representatives ${\bf j}_{e}, {\bf j}_m \in J^{\text{el/mag}}_{\text{phys}}$, integrality of the Dirac pairing requires $({\bf j}_e , {\bf j}_m) \in \bbZ$.
Moreover, the Completeness Hypothesis for quantum gravity implies that the two lattices are dual to each other, $\Lambda_\text{c} = (\Lambda_\text{cc})^*$, i.e., for any $[{\bf j}_e]$ there is a $[{\bf j}_m]$ such that $({\bf j}_e, {\bf j}_m) = 1$ and vice versa.
This can be explicitly checked, as we will discuss later.

Now suppose that the 7-branes give rise to the full 8d gauge algebra
\begin{align}
    \mathfrak{g} = \bigoplus_\sigma \mathfrak{g}_\sigma \oplus \mathfrak{u}(1)^{\oplus r_A} \,, \quad \text{with} \enspace r_A = 20 - 8k - \sum_{a} \text{rank} (\mathfrak{g}_a) \,.
\end{align}
Since for each 7-brane stack with gauge factor $\fkg_\sigma$, we have (co-)weight junctions ${\bf w}^{(\vee)}_{\sigma;i_\sigma}$, $i_\sigma=1,...,\text{rank}(\fkg_\sigma)$, we can uniquely (up to global null junctions) decompose\footnote{Note that each stack $\sigma$ can appear with a monodromy that is conjugated by $g_\sigma \in SL(2,\bbZ)$ compared to the ``standard frame'' \eqref{eq:ADE_brane_stacks} or \eqref{eq:monodromy_sp} chosen in the previous section, so that the extended weights have $(p,q)$-charges $g_\sigma\colpq{1}{0}$ for $\boldsymbol\omega_p^\sigma$ and $g_\sigma \colpq{0}{1}$ for $\boldsymbol\omega_q^\sigma$, respectively.}
\begin{align}\label{eq:decomp_j_em}
    {\bf j}_{e \, (m)} = \sum_{\sigma} \left( \sum_{i_\sigma} a^{i_\sigma}_\sigma {\bf w}^{(\vee)}_{\sigma;i_\sigma} + a^p_\sigma \boldsymbol\omega_{p}^{\sigma} + a^q_\sigma \boldsymbol\omega_{q}^{\sigma}  \right) + \sum_s b_s \bx_s \in J_\text{phys}^{\text{el (mag)}}\, ,
\end{align}
where $s$ labels the remaining 7-branes (including potential O7$^+$'s) that are not part of the non-Abelian stacks, on which the prongs must be integral, $b_s \in \bbZ$ (or satisfy the corresponding integrality condition on O7$^+$'s).
Since gauge charges under $\fkg_\sigma$ are carried by the (co-)weights ${\bf w}^{(\vee)}_{\sigma;i_\sigma}$, the states with \emph{only} Abelian charges live in the subspace orthogonal to the (co-)weights,
\begin{align}\label{eq:abelian_junctions}
    J^{\text{el/mag}}_A := \left\{ P_A({\bf j}_{e (m)}) \ | \ {\bf j}_{e (m)} \in J_\text{phys}^\text{el (mag)} \right\}_\text{phys} \equiv \left\{ \sum_{\sigma} \left( a^p_\sigma \boldsymbol\omega_{p}^{\sigma} + a^q_\sigma \boldsymbol\omega_{q}^{\sigma}  \right) + \sum_s b_s \bx_s \right\} \cap J_\text{phys}^{\text{el/mag}} \, ,
\end{align}
where $P_A$ is the projection onto the orthogonal complement of the non-Abelian (co-)weights.
In particular, since global null junctions have zero pairing with all physical junctions, we have $J_\text{int}^{N, \text{el/mag}} \subset J^{\text{el/mag}}_A$.

Because the overall $\colpq{p}{q}$ charge of any physical junction must be zero, only specific linear combinations of extended weights, and therefore, only (co-)weights of $\fkg_\sigma$ with specific center charges, can be completed into a physical junction in \eqref{eq:decomp_j_em} with the available singlet branes.
If the resulting string junctions give rise to representations that are all invariant under a subgroup ${\cal Z}$ of the center, then the gauge group has some non-trivial global structure.

More precisely, the most general global gauge group structure is 
\begin{align}\label{eq:general_gauge_group_topology}
G = \frac{[\prod_\sigma \widetilde{G}_\sigma / \mathcal{Z} ] \times U(1)^{r_A}}{\mathcal{Z}'} \,,
\end{align}
where $\widetilde{G}_\sigma$ is the simply-connected realization of the gauge algebra $\mathfrak{g}_\sigma$. The finite group $\mathcal{Z}$ embeds into the overall center $\prod_\sigma Z(\widetilde{G}_\sigma)$ of the non-Abelian factors with trivial map to the Abelian groups, whereas $\mathcal{Z}'$ does have a non-trivial map into the Abelian sector.
We will now explain how to extract these discrete groups from junctions.

We first focus on the factor $\mathcal{Z}$, which demands that electric states can appear only in certain irreducible representations under $\prod_\sigma \widetilde{G}_\sigma$ that are invariant under ${\cal Z}$.
Equivalently, this can be understood as the existence of magnetic states $[{\bf j}^\text{nA}_m] \in \Lambda_\text{cc}$ charged only under the non-Abelian gauge factors, which via the Dirac pairing condition $({\bf j}_e, {\bf j}^\text{nA}_m) \in \bbZ$ enforces the absence of electric states that are not invariant under ${\cal Z} \subset \prod_\sigma Z(\widetilde{G}_\sigma)$.
Decomposing such a junction,
\begin{align}
    {\bf j}^\text{nA}_{m} = \sum_{\sigma} \left( \sum_{i_\sigma} a^{i_\sigma}_\sigma {\bf w}^{\vee}_{\sigma;i_\sigma} + a^p_\sigma \boldsymbol\omega_{p}^{\sigma} + a^q_\sigma \boldsymbol\omega_{q}^{\sigma}  \right) + \sum_s b_s \bx_s \in J_\text{phys}^\text{mag} \, ,
\end{align}
the assumption that this junction is only charged under the non-Abelian factors implies that the Abelian part,
\begin{align}
    \sum_{\sigma} \left( a^p_\sigma \boldsymbol\omega_{p}^{\sigma} + a^q_\sigma \boldsymbol\omega_{q}^{\sigma}  \right) + \sum_s b_s \bx_s \in J_A^\text{mag} \otimes \mathbb{Q} \, ,
\end{align}
has also zero pairing with every junction in $J_A^\text{mag}$.
However, this is only possible if it is proportional to a linear combination of global integer null junction with rational coefficients, i.e.,
\begin{align}\label{eq:fracnull}
    P_A({\bf j}_m^\text{nA}) = \sum_{\sigma} \left( a^p_\sigma \boldsymbol\omega_{p}^{\sigma} + a^q_\sigma \boldsymbol\omega_{q}^{\sigma}  \right) + \sum_s b_s \bx_s = b_m \boldsymbol\delta_{(r_m,s_m)}^N \, , \quad \boldsymbol\delta_{(r_m,s_m)}^N \in J_\text{int}^{N, \text{mag}} \, , \ b_m \in \mathbb{Q} \, .
\end{align}

Considering such decompositions for all 5-brane junctions ${\bf j}_m^\text{nA} \in J_\text{phys}^\text{mag}$ with no Abelian charge, one obtains the lattice
\begin{align}
    J^{N, \text{mag}}_\text{frac} = \left\{ b_m \boldsymbol\delta_{(r_m,s_m)}^N \ \bigg| \ {\bf j}_m^\text{nA} = \sum_{\sigma} \sum_{i_\sigma} a^{i_\sigma}_\sigma {\bf w}^{\vee}_{\sigma;i_\sigma} + b_m \boldsymbol\delta_{(r_m,s_m)}^N \in J_\text{phys}^\text{mag} \right\} \supset J^{N,\text{mag}}_\text{int} \, ,
\end{align}
of what is called (global) fractional null junctions \cite{Fukae:1999zs,Guralnik:2001jh}.
Let us further denote the smallest positive integer $n_m$ such that $n_m b_m \boldsymbol\delta_{(r_m,s_m)}^N \in J_\text{int}^{N, \text{mag}}$.
Since the prongs on $\bX_s$ are already integral, due to the physicality of ${\bf j}_m^\text{nA}$, $n_m$ is also the smallest positive integer such that $n_m (a^p_\sigma \boldsymbol\omega_{p}^{\sigma} + a^q_\sigma \boldsymbol\omega_{q}^{\sigma}) \in J^{\sigma}_\text{phys, 5-branes}$ is physical on every non-Abelian stack $\sigma$.
At the same time, according to the discussions around \eqref{eq:center_charge_table_ADE} and \eqref{eq:center_charge_5-brane_sp}, the coefficients $(a^p_\sigma, a^q_\sigma)$ specify an element 
\begin{align}
    z_m = (z_\sigma) \in \prod_\sigma \frac{J^{\sigma}_\text{ext}}{J_\text{phys}^\sigma \cap J^\sigma_\text{ext}} \cong \prod_\sigma Z(\widetilde{G}_\sigma) \, .
\end{align}
Since the set of all such $z_m$ generate the discrete factor ${\cal Z}$ in \eqref{eq:general_gauge_group_topology}, we find
\begin{align}
    {\cal Z} \cong \frac{J^{N, \text{mag}}_{\text{frac}}}{J^{N, \text{mag}}_{\text{int}}} \,.
\label{eq:juncZ}
\end{align}

The main advantage of this formula is that we can conveniently compute $J_\text{frac}^{N, \text{mag}}$ from pulling the two generators $\boldsymbol\delta^N_a$ of $J_\text{int}^{N, \text{mag}}$ across all 7-brane stacks, which yields
\begin{align}
    \boldsymbol\delta^N_a = \sum_\sigma \big( c_{a;\sigma}^p \boldsymbol\omega_p^\sigma + c_{a;\sigma}^q \boldsymbol\omega_q^\sigma) + \sum_s c_{a;s} \bx_s \in J_\text{phys}^\text{mag} \, .
\end{align}
Then $J_\text{frac}^{N, \text{mag}}$ is generated by $\mathbb{Q}$-linear combinations,
\begin{align}
    \lambda_1 \boldsymbol\delta^N_1 + \lambda_2 \boldsymbol\delta^N_2 = \sum_\sigma \left( \big( \lambda_1 c_{1;\sigma}^p + \lambda_2 c_{2;\sigma}^p \big) \boldsymbol\omega_p^\sigma + \big( \lambda_1 c_{1;\sigma}^q + \lambda_2 c_{2;\sigma}^q \big) \boldsymbol\omega_q^\sigma \right) + \sum_s \big( \lambda_1 c_{1;s} + \lambda_2 c_{2;s} \big) \bx_s \, ,
\end{align}
such that $\big( \lambda_1 c_{1;\sigma}^p + \lambda_2 c_{2;\sigma}^p \big)$, $\big( \lambda_1 c_{1;\sigma}^q + \lambda_2 c_{2;\sigma}^q \big)$ are integer, and $\big( \lambda_1 c_{1;s} + \lambda_2 c_{2;s} \big)$ satisfies the physicality condition on $\bX_s$.
As advertised, this procedure applies indiscriminately to configurations with or without O7$^+$-planes, as long as the integrality conditions on O7$^+$'s and $\mathfrak{sp}_n$ stacks follow the prescription in Section \ref{subsec:junctions_on_O7}.

The second discrete factor ${\cal Z}' \subset \prod_\sigma Z(\widetilde{G}_\sigma) \times U(1)^{r_A}$ in \eqref{eq:general_gauge_group_topology} correlates the representations under the non-Abelian factors $\prod_\sigma \widetilde{G}_\sigma$ of electric states to their $\mathfrak{u}(1)$ charges, such that their transformation under $\prod_\sigma Z(\widetilde{G}_\sigma)$ is compensated by a ${\cal Z}'$ subgroup in $U(1)^{r_A}$.
Analogously as above, this subgroup can be viewed as being enforced by the presence of magnetic states, now with non-trivial $U(1)$-charges, and hence have a junction representation $[{\bf j}_m] \in \Lambda_\text{cc}$ with
\begin{align}
    J_\text{phys}^\text{mag} \ni {\bf j}_{m} = \sum_{\sigma} \left( \sum_{i_\sigma} a^{i_\sigma}_\sigma {\bf w}^{\vee}_{\sigma;i_\sigma} + a^p_\sigma \boldsymbol\omega_{p}^{\sigma} + a^q_\sigma \boldsymbol\omega_{q}^{\sigma}  \right) + \sum_s b_s \bx_s = \sum_{\sigma} \sum_{i_\sigma} a^{i_\sigma}_\sigma {\bf w}^{\vee}_{\sigma;i_\sigma} + {\bf j}_{A,m}\,,
    \label{eq:magdecompA}
\end{align}
where ${\bf j}_{A,m} = P_A({\bf j}_m) \in J_A \otimes \mathbb{Q}$ is no longer a null junction.
Nevertheless, as the mismatch of ${\bf j}_{A,m}$ from being a physical junction is still determined by the coefficients $(a_\sigma^p, a_\sigma^q)$, we have
\begin{align}\label{eq:abelian_quotient_from_junctions}
\mathcal{Z}' = \frac{P_A\left( J_\text{phys}^\text{mag} \right)}{J_A^\text{mag}}  \,.
\end{align}
Intuitively, this measures the ``fractionality'' of the $\mathfrak{u}(1)$-charges of all magnetic objects (living in $J_\text{phys}^\text{mag}$) with respect to the charges of those that are uncharged under any non-Abelian symmetry (and hence live in $J_A^\text{mag})$.
Note that, since $P_A(J_\text{int}^N) = J_\text{int}^N$, the null junctions do not affect these quotients.

\subsection{Duality to heterotic and CHL descriptions}

Theories of rank $(2,18)$ and $(2,10)$ have a dual construction in terms of the heterotic and CHL string, respectively, which provides a crosscheck for the junction description. 
In both cases, electrically charged states are identified as elements from a momentum lattice for perturbative string excitations.
For the heterotic string the momentum lattice is the so-called Narain lattice \cite{Narain:1985jj}:
\begin{align}
\Lambda_{\text{Narain}} = (- \text{E}_8 ) \oplus (- \text{E}_8 ) \oplus U \oplus U \,,
\label{eq:Narainlattice}
\end{align}
with $(- \text{E}_8)$ the negative of the E$_8$ root lattice and $U$ the hyperpolic lattice defined by the bi-linear form
\begin{align}
U: \quad \begin{pmatrix} 0 & 1 \\ 1 & 0 \end{pmatrix} \,.
\label{eq:bilinU}
\end{align}
The CHL string is determined by the Mikhailov lattice \cite{Mikhailov:1998si}
\begin{align}
\Lambda_{\text{Mikhailov}} = (- \text{D}_8) \oplus U \oplus U = (- \text{E}_8) \oplus U \oplus U(2) \,,
\label{eq:Mikh}
\end{align}
with the negative Spin$(16)$ root lattice $(- \text{D}_8)$. Here, $U(x)$ denotes a lattice of rank two with bi-linear form \eqref{eq:bilinU} multiplied by $x$.

Following the notation of \cite{Cvetic:2021sxm}, we will denote them collectively as $\Lambda_S$. 
Then, if the duality holds, we expect $\Lambda_S = \Lambda_\text{c}$.
Equivalently, points in the dual momentum lattice\footnote{Note that $\Lambda_{\text{Narain}}^* = \Lambda_{\text{Narain}}$ is self-dual while $\Lambda_{\text{Mikhailov}}^* \neq \Lambda_{\text{Mikhailov}}$ is not.} correspond to physical magnetically charged states and therefore must be associated to 5-brane junctions (modulo null junctions) in $\Lambda_\text{cc}$.
The non-Abelian gauge factors are then specified by an embedding of the corresponding (negative) root lattice into $\Lambda_S$; the coroot lattice then naturally embeds in the dual lattice $\Lambda^*_S$.
It is worth noting that the computation of the gauge group topologies from this data \cite{Font:2020rsk,Font:2021uyw,Cvetic:2021sxm} is in a sense complimentary to the junctions approach outlined above.
While in both scenarios, the setting is fully characterized by the non-Abelian gauge algebras (by specifying either the 7-brane stacks or the embedding of the (co-)root lattices), the gauge group is concisely encoded in the projection of the full physical lattice onto the Abelian junctions (see \eqref{eq:juncZ} and \eqref{eq:abelian_quotient_from_junctions}), the methods in \cite{Font:2020rsk,Font:2021uyw,Cvetic:2021sxm} extract the gauge group from the projection onto the (co-)root lattice.

To corroborate the equivalence of the two approaches, we describe in the following the precise identification of the momentum lattices with string junctions on 7-brane configurations with zero or one O7$^+$-plane.

\subsubsection*{Narain lattice from junctions}

To construct the Narain lattice \eqref{eq:Narainlattice} from junctions, it is easiest to find a 7-brane configuration in which the two $(-\text{E}_8)$ factors are manifest via the root junctions on two $\mathfrak{e}_8$ 7-brane stacks, and make use of the fact that the lattice structure does not change as we move 7-branes.
Each of these $\mathfrak{e}_8$ stacks contains ten 7-branes, leaving a remaining four branes to specify the compact type IIB background.
A convenient configuration of this sort has been presented in Section 7 of \cite{DeWolfe:1998pr}, and takes the form
\begin{align}
\bA (\bA^7 \bB \bC \bC) \bX_{[3,1]}\bA' (\bA^7 \bB \bC \bC)' \bX'_{[3,1]} \,,
\label{eq:parent_brane_config}
\end{align}
up to possible $SL(2,\bbZ)$ conjugations. 
In fact, the above configuration has two identical parts, consisting of $\bA (\bA^7 \bB \bC \bC) \bX_{[3,1]}$, each having a trivial SL$(2,\mathbb{Z})$ monodromy.\footnote{Note however that the Spin cover of the monodromy is non-trivial and given by $(-1)^F \in \text{Mp}(2,\mathbb{Z})$, see e.g. \cite{Pantev:2016nze, GarciaEtxebarria:2020xsr, Dierigl:2020lai}.}
In addition, by pushing the twelve branes onto a single stack, one enhances the symmetry to the so-called double loop algebra $\hat{\mathfrak{e}}_9$, whose significance we will explain further in Section \ref{sec:9d}.
In an F-theory description, one may interpret this configuration as a stable degeneration limit of the elliptically-fibered K3 into two $dP_9$ surfaces.

\begin{figure}[ht]
    \centering
    \includegraphics[width = 0.6 \textwidth]{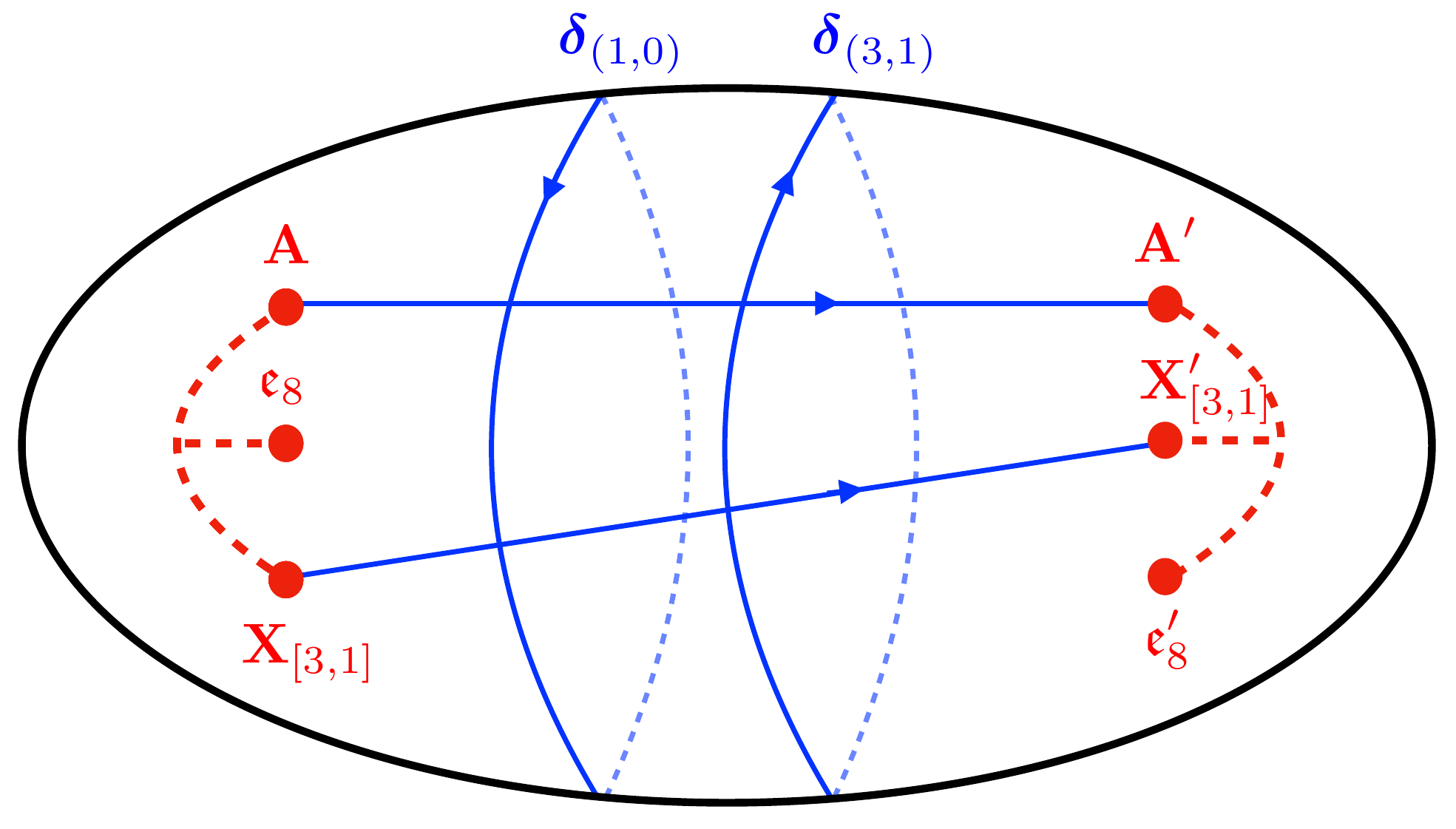}
    \caption{String junction lattice for rank $(2,18)$ theories.}
    \label{fig:rank18}
\end{figure}

Note that for $\mathfrak{e}_8$, whose extended weights \eqref{eq:extADE} are physical, roots and weights junctions agree, so the span of all physical string prongs on each $\mathfrak{e}_8$ stack (with decomposition as in \eqref{eq:nAstates}) contains two copies of the $(-\text{E}_8)$-lattice.
Next we need to find the factor $U \oplus U$ in the orthogonal complement of the $\mathfrak{e}_8$ root lattices.
A convenient set of generators for these two hyperbolic lattices can be expressed as\footnote{This is the same result as in \cite{DeWolfe:1998pr} (see their Figure 8).}
\begin{align}\label{eq:U_lattices_gens_Narain}
\begin{split}
U:& \quad \big( \boldsymbol{\delta}_{(1,0)} \,, \boldsymbol\delta_{(1,0)} + \bx_{[3, 1]} - \bx'_{[3, 1]} \big) \,, \\
U:& \quad \big( \boldsymbol{\delta}_{(3,1)} \,, \boldsymbol\delta_{(3, 1)} - \ba + \ba' \big) \,.
\end{split}
\end{align}
Here $\boldsymbol{\delta}_{(r,s)} = \boldsymbol\ell_{(r,s)}$ denotes a $(r,s)$-string loop around one $\bA (\bA^7 \bB \bC \bC) \bX_{[3,1]}$ configuration, which has no asymptotic charge since this stack has no overall monodromy.
Its orthogonality to the $\mathfrak{e}_8$ root lattice is evident from the fact that this junction has no prongs on any of the two stacks.
Note that, as is evident from Figure \ref{fig:rank18}, such a loop automatically encircles the other $\bA (\bA^7 \bB \bC \bC) \bX_{[3,1]}$ configuration.
Using Hanany--Witten transition one can rewrite them in terms of integer strings ending on the brane constituents in the interior (say, on the left in Figure \ref{fig:rank18}), e.g., 
\begin{align}
\begin{split}
\boldsymbol{\delta}_{(1,0)} &= 3 \boldsymbol{\omega}^{\mathfrak{e}_8}_p + \boldsymbol{\omega}^{\mathfrak{e}_8}_q - \bx_{[3,1]} \,, \\
\boldsymbol{\delta}_{(3,1)} &= - \ba + \boldsymbol{\omega}_p^{\mathfrak{e}_8} \,,
\end{split}
\end{align}
or equivalently for the primed stack.

This accounts for a rank 20 sublattice $(-\text{E}_8)^{\oplus 2} \oplus U^{\oplus 2}$ of the full string junction lattice $J_\text{phys}^\text{el}$.
The remaining two directions are spanned by the global null junctions $J_\text{int}^N$, for which the canonical basis is
\begin{align}
\begin{split}
    &\boldsymbol\delta_{(1,0)}^N = -(3 \boldsymbol\omega_p^{\mathfrak{e}_8} + \boldsymbol\omega_q^{\mathfrak{e}_8}) + \bx_{[3,1]} - (3 \boldsymbol\omega_p^{\mathfrak{e}'_8} + \boldsymbol\omega_q^{\mathfrak{e}'_8}) + \bx_{[3,1]}' \, , \\
    &\boldsymbol\delta_{(0,1)}^N = -\ba  + 10 \boldsymbol\omega_p^{\mathfrak{e}_8} + 3 \boldsymbol\omega_q^{\mathfrak{e}_8} - 3\bx_{[3,1]} -\ba'  + 10 \boldsymbol\omega_p^{\mathfrak{e}_8'} + 3 \boldsymbol\omega_q^{\mathfrak{e}_8'} - 3\bx_{[3,1]}' \, .
\end{split}
\end{align}
Since these have trivial pairing with all physical junctions, the lattice pairing of the above generators (including the $\mathfrak{e}_8$ roots and those of the $U$-lattices) desecend, without modification, to the quotient
\begin{align}
\Lambda_\text{c} = \frac{J_\text{phys}^\text{el}}{J_\text{int}^{N, \text{el}}} = \frac{(-\text{E}_8)^{\oplus 2} \oplus U^{\oplus 2} \oplus J_\text{int}^{N, \text{el}}}{J_\text{int}^{N, \text{el}}} \cong \Lambda_{\text{Narain}} \, .
\end{align}
Since there are no O7$^+$-planes, the 5-brane junction lattices are the same as their stringy counterparts, so we immediately find
\begin{align}
    \Lambda_\text{cc} = J_\text{phys}^\text{mag} / J_\text{int}^{N, \text{mag}} \cong J_\text{phys}^\text{el} / J_\text{int}^{N, \text{el}} \cong \Lambda_\text{Narain} = \Lambda_\text{Narain}^* \, .
\end{align}

\subsubsection*{Mikhailov lattice from junctions}

\begin{figure}[ht]
    \centering
    \includegraphics[width = 0.6 \textwidth]{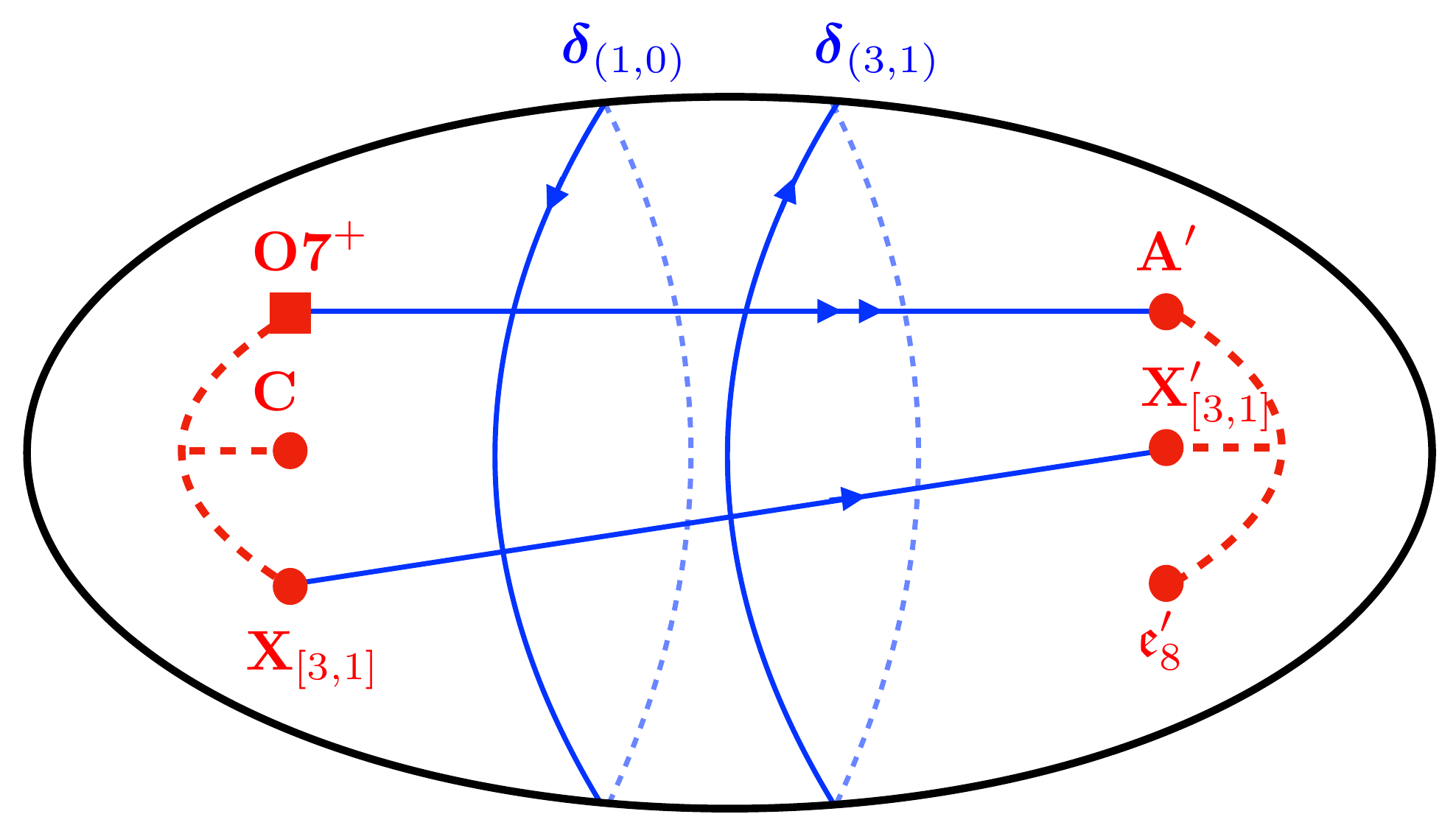}
    \caption{String junction lattice for rank $(2,10)$ theories.}
    \label{fig:rank10}
\end{figure}

The Mikhailov lattice describing the momentum lattice for the 8d CHL string is obtained as follows. 
We keep one of the $\hat{\mathfrak{e}}_9$ configurations unchanged, which still leads to an $(- \text{E}_8)$ factor in the string junction lattice. 
On the other side, we remove a $\mathbf{C}$ brane from the $\mathfrak{e}_8$ stack, but add to it the singlet $\mathbf{A}$-brane, which leads to an $\mathfrak{so}_{16}$ brane stack, which we then ``freeze'' into an O7$^+$-plane:
\begin{align}
\bA {\bf E}_8 \bX_{[3,1]} = \bA (\bA^7 \bB \bC \bC) \bX_{[3,1]} \rightarrow (\bA^8 \bB \bC) \bC \bX_{[3,1]} \rightarrow {\bf O7}^+ \bC \bX_{[3,1]} \,.
\end{align}
The resulting complete 7-brane configuration,
\begin{align}
{\bf O7}^+ \bC \bX_{[3, 1]} \bA' (\bA^7 \bB \bC^2) \bX'_{[3, 1]} \,,
\end{align}
and is depicted in Figure \ref{fig:rank10}.
The total rank of the junction lattice is now 14.

Inside the string junction lattice $J_\text{phys}^\text{el}$ of this configuration, we now need to identify $U \oplus U(2)$ orthogonal to the $\mathfrak{e}_8$ roots.
Given the similarities to the rank $(2,18)$ configuration, a natural choice for the generators would be a variation of \eqref{eq:U_lattices_gens_Narain}.
While the first set exists also for the frozen configuration, the second $U$-factor has a generator with a single $\ba$-prong, which would become part of the O7$^+$, and not be physical.
Therefore, the set of generators orthogonal to the E$_8$ root lattce are
\begin{align}
\begin{split}
U:& \quad \big( \boldsymbol{\delta}_{(1,0)} \,, \boldsymbol{\delta}_{(1,0)} + \bx_{[3,1]} - \bx_{[3,1]}' \big) \,, \\
U(2):& \quad \big( \boldsymbol{\delta}_{(3,1)} \,, 2 \boldsymbol{\delta}_{(3,1)} + 2 \boldsymbol{\omega}^{\text{O7}^+}_p - 2 \ba' \big) \,,
\end{split}
\label{eq:Mlattices}
\end{align}
where the second generator of $U(2)$ is primitive because of the evenness condition for strings ending on O7$^+$.
These have an equivalent representation with
\begin{align}
\begin{split}
\boldsymbol{\delta}_{(1,0)} &= 2 \boldsymbol{\omega}_p^{\text{O7}^+} + \bc - \bx_{[3,1]} \,, \\
\boldsymbol{\delta}_{(3,1)} &= - 2 \boldsymbol{\omega}^{\text{O7}^+}_p - 2 \boldsymbol{\omega}^{\text{O7}^+}_q + 2 \bc \,.
\end{split}
\label{eq:Mloops}
\end{align}
After quotienting out the global null junctions $J_\text{phys}^{N,\text{el}}$, with generators
\begin{align}\label{eq:null_junctions_rank_10}
    \begin{split}
    &\boldsymbol\delta_{(1,0)}^N = -2\boldsymbol\omega_p^{\text{O7}^+} - \bc + \bx_{[3,1]} - (3 \boldsymbol\omega_p^{\mathfrak{e}'_8} + \boldsymbol\omega_q^{\mathfrak{e}'_8}) + \bx_{[3,1]}' \, , \\
    &\boldsymbol\delta_{(0,1)}^N = 4\boldsymbol\omega_p^{\text{O7}^+} -2\boldsymbol\omega_q^{\text{O7}^+} + 5\bc - 3\bx_{[3,1]} -\ba'  + 10 \boldsymbol\omega_p^{\mathfrak{e}_8'} + 3 \boldsymbol\omega_q^{\mathfrak{e}_8'} - 3\bx_{[3,1]}' \, ,
\end{split}
\end{align}
one finds
\begin{align}
\Lambda_\text{c} \cong (-\text{E}_8) \oplus U \oplus U(2) = \Lambda_{\text{Mikhailov}} \,.
\end{align}

In this case it is interesting to also analyze the 5-brane junctions that correspond to the dual lattice. Here, it is important to remember that 5-branes have different physicality conditions, which allow for an arbitrary integer number of them to end on the O7$^+$-plane.
This does not affect the $\mathfrak{e}_8$ root junctions, and the $U$-factor in \eqref{eq:Mlattices}, but does imply that $\boldsymbol\delta_{(3,1)}$ in \eqref{eq:Mloops} is no longer a primitive 5-brane junction.
Instead, it is a multiple of
\begin{align}
\tfrac{1}{2} \boldsymbol{\delta}_{(3,1)} = \boldsymbol{\delta}_{(\nicefrac{3}{2},\nicefrac{1}{2})} = - \boldsymbol{\omega}^{\text{O7}^+}_p - \boldsymbol{\omega}^{\text{O7}^+}_q + \bc \,.
\end{align}
This implies that the there is an overlattice of $U(2)$ in \eqref{eq:Mlattices} inside the physical 5-brane junction lattice, given by
\begin{align}
U(\tfrac{1}{2}): \quad \big( \tfrac{1}{2} \boldsymbol{\delta}_{(3,1)} \,, \boldsymbol{\delta}_{(3,1)} + \boldsymbol{\omega}^{\text{O7}^+}_p - \ba' \big) \,.
\end{align}
Note that the null junction lattice spanned by \eqref{eq:null_junctions_rank_10} remain primitive as a sublattice of $J_\text{phys}^\text{mag}$.
Therefore one has for the magnetic 5-brane junction lattice
\begin{align}
\Lambda_{cc} = (- \text{E}_8) \oplus U \oplus U(\tfrac{1}{2}) = \Lambda_{\text{Mikhailov}}^* \,,
\end{align}
which precisely coincides with the dual of the Mikhailov lattice \eqref{eq:Mikh}.

\subsection[A rank \texorpdfstring{$(2,2)$}{(2,2)} momentum lattice]{A rank \boldmath{$(2,2)$} momentum lattice}

\begin{figure}[ht]
    \centering
    \includegraphics[width = 0.6 \textwidth]{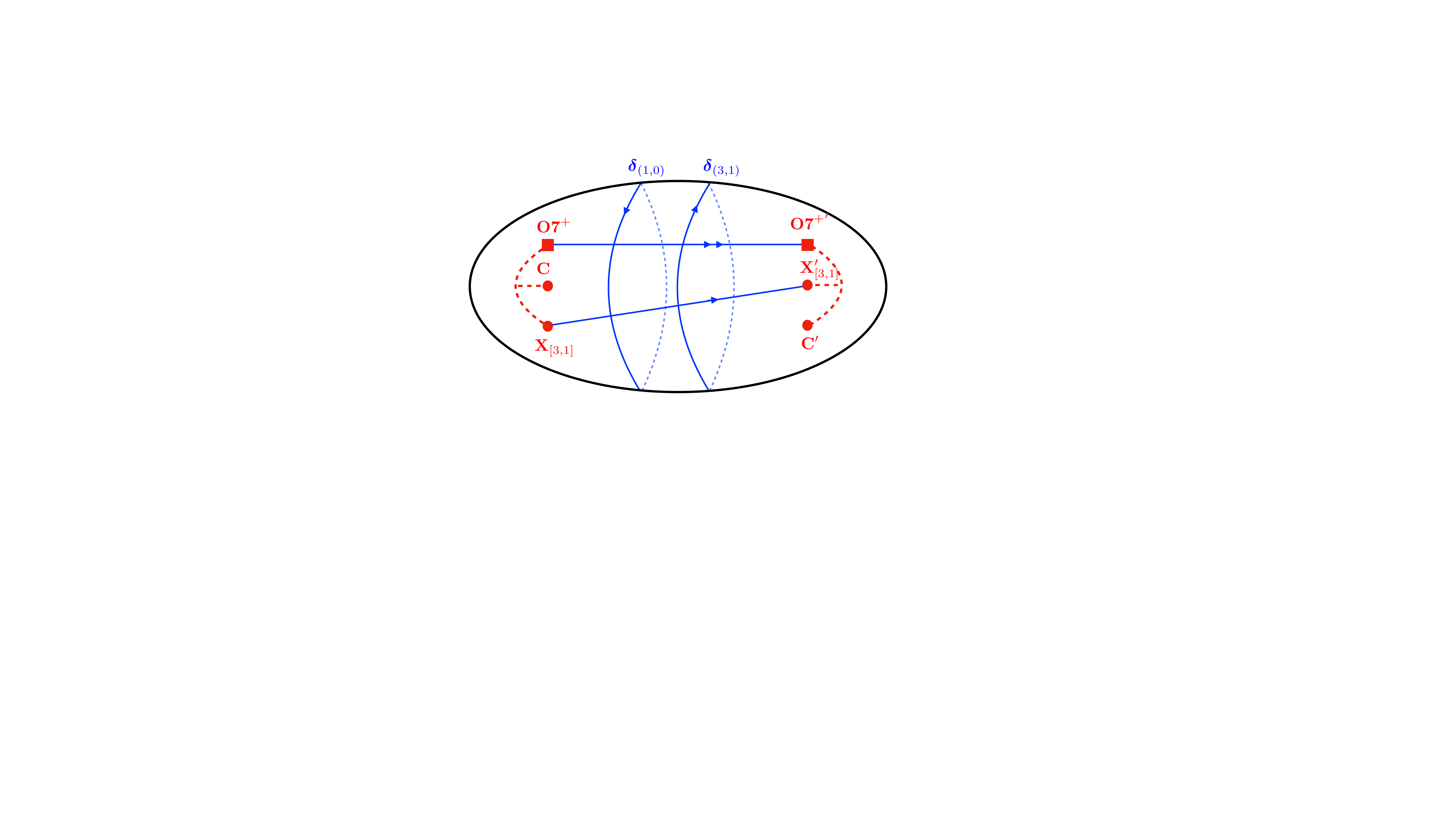}
    \caption{String junction lattice for rank $(2,2)$ theories.}
    \label{fig:rank2}
\end{figure}

8d rank $(2,2)$ string vacua have no known constructions as $T^2$- or $S^1$-reductions of the heterotic or CHL string.
However, using the junctions, we propose an analogue of a momentum lattice description, which can be applied, in particular, to determine the gauge group topologies of these theories.
To this end, we start with a 7-brane configuration with two O7$^+$'s, that we obtain from further freezing an $\mathfrak{so}_{16}$ stack on the primed side of \eqref{eq:parent_brane_config}.
The overall brane configuration is then given by
\begin{align}
{\bf O7}^+ \bC \bX_{[3,1]} {{\bf O7}^+}' \bC' \bX'_{[3,1]} \,,
\end{align}
and is depicted in Figure \ref{fig:rank2}.

The string junction lattice in this case has rank $6$, whose non-null directions are isomorphic to 
\begin{align}
    U \oplus U(2) \, .
\end{align}
The explicit generators in terms of physical string junctions are given by
\begin{align}
\begin{split}
U:& \quad \big( \boldsymbol{\delta}_{(1,0)} \,, \boldsymbol{\delta}_{(1,0)} + \bx_{[3,1]} - \bx_{[3,1]}' \big) \,, \\
U(2):& \quad \big( \boldsymbol{\delta}_{(3,1)} \,, 2 \boldsymbol{\delta}_{(3,1)} + 2 \boldsymbol{\omega}^{\text{O7}^+}_p - 2 \boldsymbol{\omega}^{{\text{O7}^+}'}_p \big) \,,
\end{split}
\end{align}
where the loop junctions satisfy the same relation as in \eqref{eq:Mloops}.
The string null junctions $J_\text{phys}^{N, \text{el}}$ have generators
\begin{align}
\begin{split}
    & \boldsymbol\delta_{(1,0)}^N = -2 \boldsymbol\omega_p^{\text{O7}^+} - \bc + \bx_{[3,1]} -2 \boldsymbol\omega_p^{{\text{O7}^+}'} - \bc' + \bx_{[3,1]}' \, , \\
    & \boldsymbol\delta_{(0,1)}^N = 4 \boldsymbol\omega_p^{\text{O7}^+} - 2 \boldsymbol\omega_q^{\text{O7}^+} + 5 \bc - 3 \bx_{[3,1]} + 4 \boldsymbol\omega_p^{{\text{O7}^+}'} - 2 \boldsymbol\omega_q^{{\text{O7}^+}'} + 5 \bc' - 3 \bx_{[3,1]}' \, .
\end{split}
\end{align}
The full physical string junction lattice is therefore $J_\text{phys}^\text{el} = U \oplus U(2) \oplus J_\text{phys}^{N, \text{el}}$.

As for the rank $(10,2)$ case above, we find that for 5-branes the $U(2)$ turns into a $U(\tfrac{1}{2})$, i.e., $J_\text{phys}^\text{mag} = U \oplus U(\tfrac12) \oplus J_\text{phys}^{N, \text{mag}}$.
A novel modification that will affect the gauge group computation is that also the null junctions are refined.
Namely, the generators of $J_\text{phys}^{N,\text{mag}}$ are
\begin{align}\label{eq:null_junction_gens_rank_2}
\begin{split}
    \tfrac12 (\boldsymbol\delta^N_{(1,0)} + \boldsymbol\delta^N_{(0,1)}) = \boldsymbol\delta^N_{(\nicefrac12,\nicefrac12)} & = \boldsymbol\omega_p^{\text{O7}^+} - \boldsymbol\omega_q^{\text{O7}^+} + 2\bc - \bx_{[3,1]} + \boldsymbol\omega_p^{{\text{O7}^+}'} - \boldsymbol\omega_q^{{\text{O7}^+}'} + 2\bc' - \bx_{[3,1]}' \, , \\
    \tfrac12 (\boldsymbol\delta^N_{(1,0)} - \boldsymbol\delta^N_{(0,1)}) = \boldsymbol\delta^N_{(\nicefrac12,-\nicefrac12)} & = -3 \boldsymbol\omega_p^{\text{O7}^+} + \boldsymbol\omega_q^{\text{O7}^+} - 3\bc - 2\bx_{[3,1]} - 3 \boldsymbol\omega_p^{{\text{O7}^+}'} + \boldsymbol\omega_q^{{\text{O7}^+}'} - 3\bc' + 2\bx_{[3,1]}' \, .
\end{split}
\end{align}
In summary, after modding out the null junctions, we have
\begin{align}
    \Lambda_\text{c} = U \oplus U(2) \, , \quad \Lambda_\text{cc} = U \oplus U(\tfrac{1}{2}) = \Lambda_\text{c}^*\,.
\end{align}

\subsection{Examples}

Using the techniques outlined in Section \ref{subsec:global_junciton_lattice_null_junctions}, we can determine the brane configurations and the resulting gauge group topologies for all 8d ${\cal N}=1$ string vacua.
This is done explicitly for all maximally-enhanced cases on each branch of the moduli space, as summarized in Appendix \ref{app:results}.
In the following we demonstrate the general procedure in specific examples.
For convenience, we focus 8d theories that were discussed in \cite{Font:2021uyw,Cvetic:2021sjm} from the perspective of the heterotic or CHL momentum lattice.
The generalization to rank $(2,2)$ theories is, to our knowledge, the first time in the literature the global gauge group topology has been computed for these string vacua.

\subsubsection{A rank \boldmath{$(2, 18)$} example}

The non-Abelian gauge algebra of the model is given by
\begin{align}
\mathfrak{g} = \mathfrak{so}_{20} \oplus \mathfrak{su}_4 \oplus \mathfrak{su}_4 \oplus \mathfrak{su}_2 \oplus \mathfrak{su}_2 \,,
\end{align}
which can be generated by the following brane configuration:
\begin{align}
(\bA^{10} \bB \bC) \bN^4 \bX_{[1,3]}^4 \bX_{[2,5]}^2 \bC^2 \,.
\end{align}
Note that for a consistent overall monodromy, the $\mathfrak{su}$ algebras are not associated to a stack of $\bA$-branes, but rather in some $SL(2,\mathbb{Z})$ conjugated frame. 
Accordingly, the associated extended weight junctions summarized in \eqref{eq:extADE} need to be conjugated, and are given by
\begin{align}
\begin{split}
\mathfrak{so}_{20} \quad (\bA^{10} \bB \bC):& \quad \boldsymbol\omega_{p} = \tfrac{1}{2} (\bb + \bc) \,, \quad \boldsymbol\omega_q = \tfrac{1}{2} \sum_{i = 1}^{10} \ba_i - 3\bb - 2\bc \,, \\
\mathfrak{su}_4 \quad (\bN^4):& \quad \boldsymbol\omega_{(0,1)} = \tfrac{1}{4}\sum_{i = 1}^4\bn_i \,, \\
\mathfrak{su}_4 \quad (\bX_{[1,3]}^4):& \quad \boldsymbol\omega_{(1, 3)} = \tfrac{1}{4}\sum_{i = 1}^4 \bx_{[1,3],i} \,, \\
\mathfrak{su}_2 \quad (\bX^2_{[2,5]}):& \quad \boldsymbol\omega_{(2, 5)} = \tfrac{1}{2}\sum_{i = 1}^2 \bx_{[2,5],i} \,, \\
\mathfrak{su}_2 \quad (\bC^2):& \quad \boldsymbol\omega_{(1,1)} = \tfrac{1}{2}\sum_{i = 1}^2 \bc_{i} \,,
\end{split}
\label{eq:exaext20}
\end{align}
where $\boldsymbol\omega_{(p,q)}$ is the extended weight of the corresponding $\mathfrak{su}$-stack with asymptotic $\colpq{p}{q}$-charge. 
In terms of these extended weights the two linearly independent integer null junctions are given by
\begin{align}
\begin{split}
\boldsymbol{\delta}^N_{(1,0)} &= - 2 \boldsymbol\omega_p - 4 \boldsymbol\omega_{(0,1)} + 4 \boldsymbol\omega_{(1, 3)} - 2 \boldsymbol\omega_{(2, 5)} + 2 \boldsymbol\omega_{(1,1)} \,, \\
\boldsymbol{\delta}^N_{(0,1)} &= 6 \boldsymbol\omega_p - 2 \boldsymbol\omega_q + 24 \boldsymbol\omega_{(0,1)} - 20 \boldsymbol\omega_{(1, 3)} + 8 \boldsymbol\omega_{(2, 5)} - 2 \boldsymbol\omega_{(1,1)} \,.
\end{split}
\label{eq:exa20null}
\end{align}
Notice how, in both junctions, the greatest common divisor of the coefficients is 2.
Therefore, the fractional null junctions are generated by
\begin{align}
J^{N, \text{mag}}_{\text{frac}} = \Big\{ \lambda^N_{(1,0)} \, \boldsymbol\delta^N_{(1,0)} + \lambda^N_{(0,1)} \, \boldsymbol\delta^N_{(0,1)} \ \Big| \ \lambda^N_{(1,0)}, \lambda^N_{(0,1)} \in \tfrac{1}{2} \mathbb{Z} \Big\} \,, 
\end{align}
and the global realization of the non-Abelian gauge group is determined by
\begin{align}
\mathcal{Z} = \frac{J^N_{\text{frac}}}{J^N_{\text{int}}} = \frac{ \{\tfrac{x}{2} \boldsymbol\delta_{(1,0)}^N \ | \ x \in \bbZ \} }{(\boldsymbol\delta_{(1,0)}^N)} \times \frac{ \{\tfrac{y}{2} \boldsymbol\delta_{(1,0)}^N \ | \ y \in \bbZ\} }{(\boldsymbol\delta_{(0,1)}^N)} \cong \mathbb{Z}_2 \times \mathbb{Z}_2 \,.
\end{align}
Moreover, the coefficients in front of the extended weights in $\tfrac12 \boldsymbol\delta^N$ determine, according to \eqref{eq:center_charge_table_ADE}, the generators of ${\cal Z} \subset Z(Spin(20)) \times Z(SU(4))^2 \times Z(SU(2))^2$ to be
\begin{align}
\begin{split}
    & \tfrac12 \boldsymbol\delta_{(1,0)}^N \simeq (1,0; 2; 2; 1; 1) \in (\bbZ_2 \times \bbZ_2) \times \bbZ_4 \times \bbZ_4 \times \bbZ_2 \times \bbZ_2 \, , \\
    & \tfrac12 \boldsymbol\delta_{(0,1)}^N \simeq (1,1; 0; 2 ; 0; 1) \in (\bbZ_2 \times \bbZ_2) \times \bbZ_4 \times \bbZ_4 \times \bbZ_2 \times \bbZ_2 \, .
\end{split}
\end{align}

Beyond the non-Abelian gauge factors, the theory contains two gravi-photons generating two $\mathfrak{u}(1)$ gauge factors.
These arise from Abelian junctions \eqref{eq:abelian_junctions} that are not null junctions, which for the present model can be easily determined, from \eqref{eq:exaext20}, to be linear combinations of
\begin{align}\label{eq:exaAbelian}
    \begin{split}
        & {\bf u}_1 = 4 (\boldsymbol\omega_{(0,1)} + \boldsymbol\omega_{(1,3)} - \boldsymbol\omega_{(2,5)} + \boldsymbol\omega_{(1,1)} ) \equiv 4 {\bf v}_1 \, , \\
        & {\bf u}_2 = 4 (\boldsymbol\omega_{(0,1)} - \boldsymbol\omega_q) \equiv 4 {\bf v}_2\, .
    \end{split}
\end{align}
${\bf v}_a$, while themselves non-physical, can be made physical by adding coweight junctions, so lie in $P_A(J_\text{phys}^\text{mag})$.
At the same time, there are no ``finer'' coweights to make fractions of ${\bf v}_a$ physical, so ${\bf v}_a$ generate $P_A(J_\text{phys}^\text{mag})$.
Then, the Abelian quotient ${\cal Z'}$, according to \eqref{eq:abelian_quotient_from_junctions}, is
\begin{align}
\mathcal{Z}' = \frac{P_A(J_\text{phys}^\text{mag})}{J_A^\text{mag}} = \frac{\{x{\bf v}_1 \ | \ x \in \bbZ\} }{({\bf u}_1)} \times \frac{\{ y{\bf v}_2 \ | \ y \in \bbZ\}}{({\bf u}_2)} = \mathbb{Z}_4 \times \mathbb{Z}_4 \,.
\end{align}
The generators of ${\cal Z}'$ are
\begin{align}\label{eq:abelian_quotient_generators_rk-20_example}
\begin{split}
    & {\bf v}_1 \simeq (0,0; 1; 1; 1; 1 \ | \ e^{i \pi /2}; 1) \in (\bbZ_2 \times \bbZ_2) \times \bbZ_4 \times \bbZ_4 \times \bbZ_2 \times \bbZ_2 \times U(1)_1 \times U(1)_2\, , \\
    & {\bf v}_2 \simeq (0,1; 1; 0; 0; 0 \ | \ 1; e^{i \pi /2}) \in (\bbZ_2 \times \bbZ_2) \times \bbZ_4 \times \bbZ_4 \times \bbZ_2 \times \bbZ_2 \times U(1)_1 \times U(1)_2 \, ,
\end{split}
\end{align}
where we have used a vertical line to separate the finite groups from the $U(1)$'s, whose trivial element is $1$.

Consequently, the global form of the gauge group is given by
\begin{equation}
    \frac{\left[\left(Spin(20) \times SU(2)^{2} \times SU(4)^{2}\right) /\left(\mathbb{Z}_{2} \times \mathbb{Z}_{2}\right)\right] \times U(1)^{2}}{\mathbb{Z}_{4} \times \mathbb{Z}_{4}} \,,
\end{equation}
in perfect agreement with the heterotic analysis in \cite{Cvetic:2021sjm}.

\subsubsection{A rank \boldmath{$(2,10)$} example}

Since the brane configuration above already containes an $\mathfrak{so}_{16}$ stack we can simply reinterpret this as an O7$^+$-plane, leading to the brane configuration
\begin{align}
(\bA^2 {\bf O7}^+) \bN^4 \bX_{[1,3]}^4 \bX_{[2,5]}^2 \bC^2 \,.
\end{align}
This setup has a non-Abelian gauge algebra given by
\begin{align}
\mathfrak{g} = \mathfrak{sp}_2 \oplus \mathfrak{su}_4 \oplus \mathfrak{su}_4 \oplus \mathfrak{su}_2 \oplus \mathfrak{su}_2 \,,
\end{align}
leading to a model with rank $(2,10)$ dual to a specific CHL background. 
Except for the $\mathfrak{sp}$ factor, the extended weights are the same as in \eqref{eq:exaext20}. 
For the $\mathfrak{sp}$ algebra one has
\begin{align}
    \mathfrak{sp}_2 \quad (\bA^2 {\bf O7}^+): \quad \boldsymbol\omega_p = \boldsymbol\omega_p^{\text{O7}^+} \, , \quad \boldsymbol\omega_q = \tfrac12 (\ba_1 + \ba_2) - \boldsymbol\omega_p^{\text{O7}^+} + \boldsymbol\omega_q^{\text{O7}^+} \, .
\end{align}
Formally, the global null junctions $\boldsymbol\delta^N_{(p,q)}$ are the same as in \eqref{eq:exa20null}, except that the $\boldsymbol\omega_{p,q}$ appearing there are now the extended weight junctions of $\mathfrak{sp}$.
Again, we can divide both by 2, obtaining the fractional 5-brane junctions
\begin{align}
    J^{N, \text{mag}}_{\text{frac}} = \Big\{ \lambda^N_{(1,0)} \, \boldsymbol\delta^N_{(1,0)} + \lambda^N_{(0,1)} \, \boldsymbol\delta^N_{(0,1)} \ \Big| \ \lambda^N_{(1,0)}, \lambda^N_{(0,1)} \in \tfrac{1}{2} \mathbb{Z} \Big\} \, ,
\end{align}
implying ${\cal Z} \cong \bbZ_2 \times \bbZ_2$, with generators
\begin{align}
\begin{split}
    & \tfrac12 \boldsymbol\delta_{(1,0)}^N \simeq (0; 2; 2; 1; 1) \in \bbZ_2 \times \bbZ_4 \times \bbZ_4 \times \bbZ_2 \times \bbZ_2 = Z(Sp(2) \times SU(4)^2 \times SU(2)^2) \, , \\
    & \tfrac12 \boldsymbol\delta_{(0,1)}^N \simeq (1; 0; 2; 0; 1) \in \bbZ_2 \times \bbZ_4 \times \bbZ_4 \times \bbZ_2 \times \bbZ_2 = Z(Sp(2) \times SU(4)^2 \times SU(2)^2) \, ,
\end{split}
\end{align}
where the entry for $\bbZ_2 = Z(Sp(2))$ is determined only by the coefficient in front of $\boldsymbol\omega_q$, see \eqref{eq:center_charge_5-brane_sp}.

In a similar way, there are two $\mathfrak{u}(1)$ generators that formally are the same as in \eqref{eq:exaAbelian}.
Since $({\bf u}_1, {\bf v}_1)$ have no prongs on the $\mathfrak{sp}$ stack, we get a $\bbZ_4$ factor in ${\cal Z}'$, with generator
\begin{align}
    {\bf v}_1 \simeq (0;1;1;1;1 \ | \ e^{i\pi /2} ; 1) \in \bbZ_2 \times \bbZ_4^2 \times \bbZ_2^2 \times U(1)_1 \times U(1)_2 \, .
\end{align}
Moreover, since ${\bf v}_2$ has an order-4 prong that is not on the O7$^+$, physicality conditions do not change the fact that we obtain another $\bbZ_4$ factor in ${\cal Z}'$, now with generator
\begin{align}
    {\bf v}_2 \simeq (1;1;0;0;0 \ | \ 1 ; e^{i\pi/2} ) \in \bbZ_2 \times \bbZ_4^2 \times \bbZ_2^2 \times U(1)_1 \times U(1)_2 \, .
\end{align}
To summarize, the global gauge group of this rank $(2,10)$ model is
\begin{equation}
    \frac{\left[\left(Sp(2) \times SU(2)^{2} \times SU(4)^{2}\right) /\left(\mathbb{Z}_{2} \times \mathbb{Z}_{2}\right)\right] \times U(1)^{2}}{\mathbb{Z}_{4} \times \mathbb{Z}_{4}} \, ,
\end{equation}
agreeing with the CHL result computed in \cite{Cvetic:2021sjm}.

\subsubsection{A rank \boldmath{$(2,2)$} example}\label{subsubsec:rank4}

The rank $(2,2)$ moduli branch has six special points with non-Abelian symmetry enhancements \cite{Hamada:2021bbz}.
We have enumerated the 7-brane configurations for all of these, as well as the resulting gauge group topologies in Appendix \ref{subapdx:rank4_catalog}. 

It turns out that there is only one whose non-Abelian gauge group is non-simply-connected.
For illustration, we consider this example in more detail.
The brane configuration is
\begin{equation}\label{eq:rank_2_example_config}
     {\bf O7}^+ {{\bf O7}^+}' \bB^2 \bC^2 \, ,
\end{equation}
where the monodromy of the ${{\bf O7}^+}'$ is $SL(2,\bbZ)$ conjugated to be $\left( \begin{smallmatrix} 7 & 16 \\ -4 & -9 \end{smallmatrix} \right)$.\footnote{This can be viewed as freezing both $\mathfrak{so}_{16}$ algebras of the rank $(2,18)$ configuration $(\bA^8 \bB \bC) (\bX_{[2, -1]}^8 \bB \bX_{[3, -1]}) \bB^2 \bC^2$.}
Notice that, since the extended weights of O7$^+$ generate all $\colpq{p}{q}$-charges, whose parity is invariant under $SL(2,\bbZ)$ conjugation, we use the canonical basis $\boldsymbol\omega_{p,q}^{{\text{O7}^+}'}$ for the ${{\bf O7}^+}'$, which have $\colpq{1}{0}$ and $\colpq{0}{1}$ prongs, respectively.

The non-Abelian gauge algebra of \eqref{eq:rank_2_example_config} is $\mathfrak{su}_2 \oplus \mathfrak{su}_2$, with extended weights
\begin{align}
    \begin{split}
        \mathfrak{su}_2 \quad (\bB^2) : & \quad \boldsymbol\omega_{(1,-1)} \equiv \boldsymbol\omega_b = \tfrac12 (\bb_1 + \bb_2) \, , \\
        \mathfrak{su}_2 \quad (\bC^2) : & \quad \boldsymbol\omega_{(1,1)} \equiv \boldsymbol\omega_c = \tfrac12 (\bc_1 + \bc_2) \, .
    \end{split}
\end{align}
The physical 5-brane null junctions $J_\text{phys}^{N,\text{mag}}$ (see \eqref{eq:null_junction_gens_rank_2}) are then
\begin{align}
    \begin{split}
        \boldsymbol\delta^N_{(\nicefrac12, \nicefrac12)} &= \boldsymbol\omega_p^{\text{O7}^+} - \boldsymbol\omega_q^{\text{O7}^+} + \boldsymbol\omega_p^{{\text{O7}^+}'} - \boldsymbol\omega_q^{{\text{O7}^+}'} - 2 \boldsymbol\omega_b \, , \\
        \boldsymbol\delta^N_{(\nicefrac12, -\nicefrac12)} &= -3\boldsymbol\omega_p^{\text{O7}^+} + \boldsymbol\omega_q^{\text{O7}^+} - 7\boldsymbol\omega_p^{{\text{O7}^+}'} + 5\boldsymbol\omega_q^{{\text{O7}^+}'} + 8\boldsymbol\omega_b + 2 \boldsymbol\omega_c\, ,
    \end{split}
\end{align}
from which we find that the fractional null junctions $J_\text{frac}^{N,\text{mag}}$ is generated by
\begin{align}
    \tfrac12 \big( \boldsymbol\delta^N_{(\nicefrac12, \nicefrac12)} + \boldsymbol\delta^N_{(\nicefrac12, -\nicefrac12)} \big) = \boldsymbol\delta^N_{(\nicefrac12, 0)} = -\boldsymbol\omega^{\text{O7}^+}_p - 3\boldsymbol\omega^{{\text{O7}^+}'}_p + 2\boldsymbol\omega^{{\text{O7}^+}'}_q + 3\boldsymbol\omega_b + \boldsymbol\omega_c \, .
\end{align}
It corresponds to the generator
\begin{align}
    (3 \ \text{mod 2}, 1 \ \text{mod 2}) = (1,1) \in \bbZ_2 \times \bbZ_2 = Z(SU(2) \times SU(2))
\end{align}
of ${\cal Z} = J_\text{frac}^{N,\text{mag}} / J_\text{phys}^{N,\text{mag}} = \bbZ_2$.

Together with the null junctions, the Abelian junction lattice $J_A^\text{mag}$ is generated by
\begin{align}
    {\bf u}_1 = 2{\bf v}_1 = 2 \big( -\boldsymbol\omega_p^{{\text{O7}^+}} + \boldsymbol\omega_q^{{\text{O7}^+}} + \boldsymbol\omega_b \big) \, , \quad {\bf u}_2 = 2{\bf v}_2 = 2 \big( -\boldsymbol\omega_p^{{\text{O7}^+}} - \boldsymbol\omega_q^{{\text{O7}^+}} + \boldsymbol\omega_c \big) \, .
\end{align}
Clearly, this leads to ${\cal Z}' = \bbZ_2 \times \bbZ_2$, with generators
\begin{align}
    \left. \begin{array}{c}
         {\bf v}_1 \simeq (1;0 \ | \ -1 ; 1) \\
         {\bf v}_2 \simeq (0;1 \ | \ 1 ; -1)
    \end{array} \right\}
    \in \bbZ_2 \times \bbZ_2 \times U(1)_1 \times U(1)_2 = Z\big(SU(2)_{\bB} \times SU(2)_{\bC} \times U(1)_1 \times U(1)_2  \big) \, .
\end{align}
The full gauge group is thus
\begin{align}
    \frac{ [SU(2) \times SU(2)]/\bbZ_2 \times U(1)^2}{\bbZ_2 \times \bbZ_2} \, .
\end{align}

\section{9d vacua via affine 7-brane stacks}\label{sec:9d}

Recently, it was argued that one can recover any 9d ${\cal N}=1$ string vacuum with gauge rank $(1,17)$ from F-theory on a suitably degenerated K3 geometry that lies at infinite distance in the complex structure moduli space \cite{Lee:2021qkx,Lee:2021usk}.
As shown in these works, such decompactification limits have a particularly convenient description in terms of $[p,q]$-7-branes and junctions that realize affine algebras.
In the following, we demonstrate how the methods from the previous sections naturally apply also to these limiting configurations, and compute the 9d gauge group topologies for rank $(1,17)$ vacua.

Since the affinization is characterized entirely by the $SL(2,\bbZ)$ monodromy, a natural proposition is that these configurations also describe 9d uplifts when we include O7$^+$-planes.
Indeed, (after resolving an ambiguity by string dualities) this straightforwardly reproduces the landscape of 9d rank $(1,9)$ vacua \cite{Font:2021uyw}, including their global gauge group structures.
Moreover, applying the same reasoning to configurations with two O7$^+$'s, we consistently find two branches of 9d rank $(1,1)$ vacua \cite{Aharony:2007du}, which are only connected through circle-reductions to 8d.

The key ingredient that enters the description for all ranks are 7-brane stacks realizing an \emph{affine} Lie algebra $\widehat{\mathfrak{e}}_k$, which we will now briefly recall.

As found in \cite{DeWolfe:1998yf,DeWolfe:1998eu,DeWolfe:1998pr}, the $\mathfrak{e}_n$ and $\tilde{\mathfrak{e}}_n$ algebras can be enhanced to their affine versions, by including a specific 7-brane on top:
\begin{align}\label{eq:affine_stacks}
\begin{split}
    \widehat{{\bf E}}_{n \geq 1} = & \underbrace{ {\bf A}^{n-1} {\bf B} {\bf C}^2}_{{\bf E}_n} {\bf X}_{[3,1]} = {\bf A}^{n-1} {\bf B} {\bf C} {\bf B} {\bf C} \, , \\
    \widehat{\tilde{{\bf E}}}_{n \geq 0} = & \underbrace{ {\bf A}^{n} {\bf X}_{[2,-1]} {\bf C}}_{\tilde{\bf E}_n} {\bf X}_{[4,1]} \, .
\end{split}
\end{align}
Note that for $n\geq 2$, these are equivalent up to 7-brane moves and $SL(2,\bbZ)$ conjugations \cite{DeWolfe:1998eu}.
It is straightforward to check that, in this $SL(2,\bbZ)$-frame, they have monodromy
\begin{align}
    M(\widehat{\bf E}_n) = M(\widehat{\tilde{\bf E}}_n) = \begin{pmatrix}
    1 & 9 - n \\ 0 & 1
    \end{pmatrix} \, .
\end{align}
The hallmark of these stacks is the existence of a special loop junction $\boldsymbol{\delta}_{(1,0)} \equiv \boldsymbol\delta$ around them (with no asymptotic charge), satisfying $(\boldsymbol{\delta}, \boldsymbol{\delta}) = (\boldsymbol{\delta}, \boldsymbol{\alpha}_i) = 0$, with $\boldsymbol\alpha_i$ the root junctions of ${\bf E}_n$ or $\tilde{\bf E}_n$.
By a Hanany--Witten transition, one finds the equivalent presentation \cite{DeWolfe:1998pr}
\begin{align}
    \boldsymbol\delta_{\widehat{\text{E}}} =  {\bf x}_{[3,1]} - {\bf b} - {\bf c}_1 - {\bf c}_2= {\bf b}_2 + {\bf c}_2 - {\bf b}_1 - {\bf c}_1 \, , \quad \boldsymbol\delta_{\widehat{\tilde{\text{E}}}} = {\bf x}_{[4,1]} - 2{\bf c} - {\bf x}_{[2,-1]} \, .
\end{align}

Representation theoretically, $\boldsymbol\delta$ plays the role of the imaginary root required for the affinization of $\mathfrak{e}_n$ or $\tilde{\mathfrak{e}}_n$.
They generate an infinite dimensional Kac--Moody algebra with roots $\{\boldsymbol\alpha + k \boldsymbol\delta \, | \, k \in \bbZ\}$, where $\boldsymbol\alpha$ is any root of $\mathfrak{e}_n$ or $\tilde{\mathfrak{e}}_n$.
When we seperate the affinizing $\bX$-branes in \eqref{eq:affine_stacks} from the $\mathfrak{e}_n$ or $\tilde{\mathfrak{e}}_n$ stacks, these junctions, as strings, give rise to BPS states with masses proportional to $k$.
In the affine limit, we thus obtain an infinite tower of massless BPS states.
Physically, string junctions of these type give rise to an infinite tower of massless BPS states.

A special extension exists for $n=8$.
Here, by adding an ${\bf A}$-brane from the left to the $\widehat{\bf E}_8$ or $\widehat{\tilde{\bf E}}_8$, the monodromy becomes $\mathbb{1} \equiv M(\widehat{\bf E}_9) = M(\widehat{\tilde{\bf E}}_9)$.
This would give rise to two independent towers of massless BPS states from loops of $\colpq{1}{0}$ and $\colpq{0}{1}$ string junctions, which lead to the double loop enhancement of $\mathfrak{e}_8$.
These special enhancements reflect a decompactification to 10d \cite{Lee:2021qkx,Lee:2021usk}.
For discussions of 9d vacua, we will not consider such double loop brane-stacks, but the necessary constituent branes form one half of the rank $(2,18)$ configuration \eqref{eq:parent_brane_config} that correspond geometrically to the singular fibers of a $dP_9$ surface.

Among the various types of infinite distance limits of F-theory compactified on K3 surfaces, those describing decompactification from 8d to 9d are captured by so-called Kulikov models of type III.a \cite{Lee:2021qkx,Lee:2021usk}.
In these geometries, the complex structure moduli have been tuned such that the K3 degenerates into a collection of intersecting elliptic and/or rational surfaces.
While we refer to those references for details, the relevant fact about these deformations is that they correspond to brane motions which generate one or two 7-brane stacks carrying an $\widehat{\mathfrak{e}}_n$ or $\widehat{\tilde{\mathfrak{e}}}_n$ algebra (with $n \leq 8$).
The tower of massless states from the imaginary root may then be identified with the momentum states of a Kaluza--Klein (KK) tower on a circle whose size becomes infinite at the infinite distance limit.
In the case with two affine stacks, the individual imaginary root junctions turn out to be identical in the global setting, consistent with having just one KK-tower \cite{Lee:2021usk}.

\subsection{Global structure of 9d vacua of rank 17}

As for the classification of 8d vacua, one can also categorize all brane configurations with such affine stacks.
Then, if the non-Abelian brane stacks correspond to the algebra $\mathfrak{h} \oplus \widehat{{\cal E}}_n$ or $\mathfrak{h} \oplus \widehat{{\cal E}}_n \oplus \widehat{{\cal E}}_m$ (where ${\cal E} = \mathfrak{e}$ or $\tilde{\mathfrak{e}}$) for some finite semi-simple, simply-laced algebra $\mathfrak{h}$, the associated non-Abelian gauge algebra in 9d is $\mathfrak{h} \oplus {\cal E}_n$ or $\mathfrak{h} \oplus {\cal E}_n \oplus {\cal E}_m$, respectively \cite{Lee:2021qkx,Lee:2021usk}.
This reproduces, e.g., all the maximally enhanced non-Abelian algebras (i.e., with rank 17) determined in the dual heterotic frame \cite{Font:2020rsk}.

To also analyze the gauge group topologies in this description, we need to examine the full junction lattice, including the branes away from the affine stack.
An important detail here is that the overall gauge rank reduces by 2 as we decompactify from 8d to 9d, corresponding to the re-interpretation of the KK-states (which become massless) and the decoupling of the winding states (which become infinitely heavy) as we increase the size of the compactification circle.
In the momentum lattice description of the 8d and 9d theories of rank $(2,18)$ and $(1,17)$, respective, we have
\begin{align}\label{eq:affinization_lattice}
    \Lambda^\text{het}_{\text{8d}} = \Lambda^\text{het}_{\text{9d}} \oplus U \quad \Rightarrow \quad \Lambda^\text{het}_{\text{9d}} \cong \Lambda^\text{het}_{\text{8d}} / U \, ,
\end{align}
with $U$ the rank 2 hyperbolic lattice that is spanned by the KK and winding states.
Since the momentum lattice is equivalently described by junctions $J_\text{phys}^{\text{el}} = J_\text{phys}^{\text{mag}} \cong \Lambda_\text{8d}^\text{het} \oplus J_\text{phys}^{N}$, with the KK-tower being generated by the junction $\boldsymbol\delta$, there must exist another non-null junction $\boldsymbol\epsilon$ that generates this $U$ factor with $\boldsymbol\delta$, i.e., satisfying
\begin{align}
    (\boldsymbol\delta, \boldsymbol\epsilon) = 1 \, , \quad (\boldsymbol\delta, \boldsymbol\delta) = (\boldsymbol\epsilon, \boldsymbol\epsilon) = (\boldsymbol\delta, {\bf j}) = (\boldsymbol\epsilon, {\bf j}) = 0 \, ,
\end{align}
for ${\bf j}$ any (co-)weight or (co-)root junction, or a non-null generator of the Abelian junctions $J_A$.
Such a $\boldsymbol\epsilon$-junction always exists, but the details depend on the specific configuration.

Since the junction lattice reproduces the 9d momentum lattice, it must also encode the global structure of the gauge group.
In particular, it allows us to use the intuition in terms of fractional null junctions to re-derive the results of \cite{Font:2020rsk}.
Let us demonstrate this for 9d models with maximally enhanced non-Abelian symmetries, for which there are two classes of 8d brane configurations \cite{Lee:2021usk}.

In the first class, the non-Abelian algebra (with the place holder ${\cal E} = \mathfrak{e}$ or $\tilde{\mathfrak{e}}$) is
\begin{equation}
\fkg_{\text{8d}, \infty} = \mathfrak{su}_{18 - m - n} \oplus \widehat{{\cal E}}_m \oplus \widehat{{\cal E}}_{n} \quad \Rightarrow \quad \fkg_\text{9d} = \mathfrak{su}_{18 - m - n} \oplus \mathcal{E}_m \oplus \mathcal{E}_n, \quad m, n \in \{0, 1, 3, \dots, 8\} \, ,
\end{equation}
whose brane configurations (together with the $U$-lattice generators) are
\begin{subequations}\label{eq:decomp_series_A}
\begin{align}
\begin{split}
    m \geq n \geq 1:& \quad \bA^{18 - m - n} (\overbrace{\bA^{m-1} \bX_{[n-10, 1]} \bX_{[n-8, 1]}^2 \bX_{[n-6, 1]}}^{\widehat{{\bf E}}_m}) (\overbrace{\bA^{n-1} \bB \bC^2 \bX_{[3, 1]}}^{\widehat{{\bf E}}_n}) \, , \\
    & \quad (\boldsymbol\delta, \boldsymbol\epsilon) = (\boldsymbol\delta^R_{(1,0)},\ \ (n-5)\boldsymbol\delta^R_{(1,0)} + {\boldsymbol\ell}^R_{(0, 1)} + \bx_{[n-6, 1]} - \bx_{[3, 1]} ) \, ,
\end{split} \label{eqn:2dP9_a}\\
\begin{split}
    m = 1, n = 0:& \quad \bA^{17}(\overbrace{\bX_{[10, -1]} \bX_{[8, -1]}^2 \bX_{[6, -1]}}^{\widehat{{\bf E}}_1}) (\overbrace{\bX_{[2, -1]} \bC \bX_{[4, 1]}}^{\widehat{\tilde{{\bf E}}}_0}) \, , \\
    & \quad (\boldsymbol\delta, \boldsymbol\epsilon) = (\boldsymbol\delta^R_{(1,0)},\ \ -17 \boldsymbol\omega_A + \bx_{[10, -1]} + 2\bx_{[4, 1]} - \bc) \, ,
\end{split} \label{eqn:2dP9_b} \\
\begin{split}
    m > n = 0;\ m \neq 1:& \quad \bA^{18-m} (\overbrace{ \bA^m \bX_{[11, -1]} \bX_{[8, -1]} \bX_{[5, -1]}}^{\widehat{\tilde{{\bf E}}}_{m \neq 1}}) (\overbrace{\bX_{[2, -1]} \bC \bX_{[4, 1]}}^{\widehat{\tilde{\bf E}}_0}) \, ,  \\
    & \quad (\boldsymbol\delta, \boldsymbol\epsilon) = (\boldsymbol\delta^R_{(1,0)},\ \ -\boldsymbol\delta^R_{(1,0)} + \bx_{[5, -1]} - 2\bx_{[2,-1]} - \bc) \, .
\end{split}\label{eqn:2dP9_c}
\end{align}
\end{subequations}
where $\boldsymbol\delta^R$ and $\boldsymbol\ell^R$ are loop junction that encircle counterclockwise around the second affine stack.
Note that, because the $\colpq{1}{0}$ loop is mutually-local with respect to the ${\bf A}$-branes, it is evident that, by pulling $\boldsymbol\delta_{(1,0)}$ across these, one obtains a loop junction around the other affine stack, showing explicitly that imaginary roots of each affine stack are identical.

The second class has non-Abelian gauge algebras
\begin{equation}
    \fkg_{\text{8d}, \infty} = \mathfrak{so}_{34-2k} \oplus \widehat{{\cal E}}_{k} \quad \Rightarrow \quad \fkg_\text{9d} = \mathfrak{so}_{34-2k} \oplus {\cal E}_k \, , \quad 0 \leq k \leq 8 , k\neq 2 \, , 
\end{equation}
whose brane configurations (and $U$-lattice generators) are
\begin{subequations}\label{eq:decomp_series_D}
\begin{align}
\begin{split}
    k = 1, 3, \dots, 8: & \quad ( \overbrace{\bA^{17 - k} \bX_{[k - 10, 1]} \bX_{[k - 8, 1]}}^{\textbf{D}_{17-k}}) \bX_{[k-8, 1]} (\overbrace{\bA^{k - 1} \bB \bC^2 \bX^{(1)}_{[3, 1]}}^{\widehat{\textbf{E}}_k}) \bX^{(2)}_{[3, 1]} \, , \\
    & \quad (\boldsymbol\delta, \boldsymbol\epsilon) = ( \boldsymbol\delta^R_{(1,0)},\ \ \boldsymbol\delta^R_{(1,0)} + \bx^{(1)}_{[3, 1]} - \bx^{(2)}_{[3, 1]}) \label{eqn:1dP9_a} 
\end{split}\\
\begin{split}
    k = 0 : & \quad (\overbrace{\bA^{17} \bX_{[10, -1]} \bX^{(1)}_{[8, -1]}}^{\textbf{D}_{17}}) \bX^{(2)}_{[8, -1]} (\overbrace{\bX_{[2, -1]} \bC \bX_{[4, 1]}}^{\widehat{\tilde{\textbf{E}}}_0}) \bX_{[3, 1]}  \label{eqn:1dP9_b} \, , \\
    & \quad (\boldsymbol\delta, \boldsymbol\epsilon) = (\boldsymbol\delta^R_{(1,0)},\ \ -2\boldsymbol\delta^R_{(1,0)} + \bx^{(2)}_{[8, -1]}  - 3\bx_{[2,-1]} - 2\bx_{[1,1]}) \, ,
\end{split}
\end{align}
\end{subequations}
again, with the imaginary root junction $\boldsymbol\delta^R_{(1,0)}$ being the $\colpq{1}{0}$-loop around the affine stack to the right.

By separating the ${\bf X}$-brane responsible for the affinization from each affine stack, we obtain a genuine 8d configuration.
For these configurations, we can apply the same procedure as in the previous section, and construct the global fractional null junctions that encode to the cocharacters of the 8d gauge symmetry.
Since the affinization \eqref{eq:affinization_lattice} mods out by physical junctions that are orthogonal to the root lattices of the 9d gauge factors, it does not affect the coefficients of global null junctions in front of the extended weights.
Therefore, the fractional null junctions are the same for the 8d configuration as for its affinized version.

As a concrete example, consider \eqref{eqn:1dP9_a} with $k=7$, which in 9d gives rise to $\fkg_{9d} = \mathfrak{so}_{20} \oplus \mathfrak{e}_7$.
Separating the singlet brane responsible for the affinization,
\begin{align}
    (\underbrace{\bA^{10} \bX_{[-3,1]} \bX_{[-1,1]}}_{\textbf{D}_{10}'}) \, \bX_{[-1,1]} \, (\underbrace{\bA^6 \bB \bC^2}_{\textbf{E}_7}) \, \bX^{(1)}_{[3,1]} \, \bX^{(2)}_{[3,1]} \ ,
\end{align}
we find the $\mathfrak{so}_{20}$-stack to have monodromy $\left( \begin{smallmatrix} -1 & 6 \\ 0 & -1 \end{smallmatrix} \right)$ in this $SL(2,\bbZ)$ frame, with extended weight junctions $\boldsymbol\omega'_{p,q}$ carrying asymptotic $\colpq{p}{q}$-charges as follows:
\begin{align}
    \boldsymbol\omega_p': \, \, \colpq{1}{0} \, , \qquad \boldsymbol\omega_q' : \, \, \colpq{-2}{1} \, .
\end{align}
The two global null junctions $\boldsymbol\delta^N_{(1,0)}$ and $\boldsymbol\delta^N_{(0,1)}$ can be then expressed in terms of the extended weights $\boldsymbol\omega'_{p,q}$ of $\mathfrak{so}_{20}$ and $\boldsymbol\omega_{p,q}$ of $\mathfrak{e}_7$ as
\begin{align}
    \begin{split}
        & \boldsymbol\delta^{N}_{(1,0)} = -2 \boldsymbol\omega_p' - \bx_{[-1,1]} - 5\boldsymbol\omega_p - \boldsymbol\omega_q + \bx^{(1)}_{[3,1]} + \bx^{(2)}_{[3,1]} \, , \\
        & \boldsymbol\delta^{N}_{(0,1)} = 2 \boldsymbol\omega_p' - 2 \boldsymbol\omega_q' + 5 \bx_{[-1,1]} + 17 \boldsymbol\omega_p + 3 \boldsymbol\omega_q + \bx^{(1)}_{[3,1]} + \bx^{(2)}_{[3,1]} \, .
    \end{split}
\end{align}
It is easy to see that the fractional null junctions are then multiples of
\begin{align}
    \tfrac12 (\boldsymbol\delta^N_{(1,0)} + \boldsymbol\delta^N_{(0,1)}) = - \boldsymbol\omega_q' + 2 \bx_{[-1,1]} + 6 \boldsymbol\omega_p + \boldsymbol\omega_q + \bx_{[3,1]}^{(1)} + \bx_{[3,1]}^{(2)} \, .
\end{align}
By \eqref{eq:center_charge_table_ADE}, this corresponds to the central element
\begin{align}
    (0,1;1) \in \bbZ_2^2 \times \bbZ_2 = Z(Spin(20) \times E_7),
\end{align}
which leads to the 9d non-Abelian gauge group $[Spin(20) \times E_7]/\bbZ_2$.

To determine the full gauge group, including the gravi-photon $U(1)$, we must first find the non-null generator of the Abelian junction lattice that is orthogonal to the $U$-lattice spanned by $(\boldsymbol\delta, \boldsymbol\epsilon)$.
In this example, it can be easily determined (by avoiding prongs on the $\mathfrak{e}_7$ stack or the $\bX_{[3,1]}$ branes),
\begin{align}
    {\bf u} = 2{\bf v} = 2( \boldsymbol\omega_p' + \boldsymbol\omega_q' - \bx_{[-1,1]})\, ,
\end{align}
which immediately gives ${\cal Z}' = \bbZ_2$, with generator
\begin{align}
    {\bf v} \simeq (1,1;0 \ | \ e^{i \pi} ) \in Z(Spin(20) \times E_7 \times U(1)) \, .
\end{align}
Therefore, the 9d gauge group is
\begin{align}
    \frac{[Spin(20) \times E_7]/\bbZ_2 \times U(1)}{\bbZ_2} \, .
\end{align}

By analogous computations, we compute the non-Abelian gauge groups of all models with maximally enhanced non-Abelian symmetry (summarized in Table \ref{table:9Drank17}), which agree with results from the heterotic picture \cite{Font:2020rsk}.

\begin{table}[ht!]
\renewcommand{\arraystretch}{1.3}
\centering
\begin{tabular}{|c|c||c|c|}
    \hline
    $(\fkg_\text{9d}, \pi_1(G_\text{9d}))$  & $\text{FNJ}$ & $(\fkg_\text{9d}, \pi_1(G_\text{9d}))$  & $\text{FNJ}$  \\ \hline \hline
       \multicolumn{4}{|c|}{$\fkg_{\text{8d}, \infty} = \mathfrak{su}_{18 - m - n} \oplus \widehat{\mathfrak{e}}_m \oplus \widehat{\mathfrak{e}}_n$}  \\ \hline
       $(\mathfrak{e}_8 \oplus \mathfrak{e}_8 \oplus \mathfrak{su}_2, -)$    & -  &        $(\mathfrak{e}_6 \oplus \mathfrak{su}_3 \oplus \mathfrak{su}_2 \oplus \mathfrak{su}_9 , \bbZ_3)$ & $\boldsymbol\delta^N_{(0,1)}/3$ \\
       $(\mathfrak{e}_8 \oplus \mathfrak{e}_7 \oplus \mathfrak{su}_3, -)$  & - &        $(\mathfrak{e}_6 \oplus \mathfrak{su}_2 \oplus \mathfrak{su}_{11}, -)$ & - \\
       $(\mathfrak{e}_8 \oplus \mathfrak{e}_6 \oplus \mathfrak{su}_4, -)$  & - &        $(\mathfrak{e}_6 \oplus \mathfrak{su}_{12}, \bbZ_{3})$ & $\boldsymbol\delta^N_{(0,1)}/3$ \\
       $(\mathfrak{e}_8 \oplus \mathfrak{so}_{10} \oplus \mathfrak{su}_5, -)$  & - &        $(\mathfrak{so}_{10} \oplus \mathfrak{so}_{10} \oplus \mathfrak{su}_8, \bbZ_4)$ & $\boldsymbol\delta^N_{(1,1)}/4$ \\
       $(\mathfrak{e}_8 \oplus \mathfrak{su}_5 \oplus \mathfrak{su}_6, -)$  & -  &        $(\mathfrak{so}_{10} \oplus \mathfrak{su}_5 \oplus \mathfrak{su}_9, -)$ & - \\
       $(\mathfrak{e}_8 \oplus \mathfrak{su}_3 \oplus \mathfrak{su}_2 \oplus \mathfrak{su}_7, -)$  & - &        $(\mathfrak{so}_{10} \oplus \mathfrak{su}_3 \oplus \mathfrak{su}_2 \oplus \mathfrak{su}_{10} , \bbZ_2)$ & $\boldsymbol\delta^N_{(1,-1)}/2$ \\
       $(\mathfrak{e}_8 \oplus \mathfrak{su}_9 \oplus \mathfrak{su}_2, -)$  & - &        $(\mathfrak{so}_{10} \oplus \mathfrak{su}_{2} \oplus \mathfrak{su}_{12}, \bbZ_4)$ & $\boldsymbol\delta^N_{(1,1)}/4$ \\
       $(\mathfrak{e}_8 \oplus \mathfrak{su}_{10}, -)$  & - &        $(\mathfrak{so}_{10} \oplus \mathfrak{su}_{13}, -)$ & - \\
       $(\mathfrak{e}_7 \oplus \mathfrak{e}_7 \oplus \mathfrak{su}_4, \bbZ_2)$   & $\boldsymbol\delta^N_{(1,1)}/2$ &        $(\mathfrak{su}_5 \oplus \mathfrak{su}_5 \oplus \mathfrak{su}_{10}, \bbZ_5)$ & $\boldsymbol\delta^N_{(1,2)}/5$ \\       
       $(\mathfrak{e}_7 \oplus \mathfrak{e}_6 \oplus \mathfrak{su}_5, -)$   & -  &        $(\mathfrak{su}_5 \oplus \mathfrak{su}_3 \oplus \mathfrak{su}_2 \oplus \mathfrak{su}_{11}, -)$ & - \\ 
       $(\mathfrak{e}_7 \oplus \mathfrak{so}_{10} \oplus \mathfrak{su}_6, \bbZ_2)$  & $\boldsymbol\delta^N_{(1,-1)}/2$ &        $(\mathfrak{su}_5 \oplus \mathfrak{su}_{2} \oplus \mathfrak{su}_{13}, -)$ & - \\      
       $(\mathfrak{e}_7 \oplus \mathfrak{su}_5 \oplus \mathfrak{su}_7, -)$ & - &        $(\mathfrak{su}_5 \oplus \mathfrak{su}_{14}, -)$ & - \\
       $(\mathfrak{e}_7 \oplus \mathfrak{su}_3 \oplus \mathfrak{su}_2 \oplus \mathfrak{su}_8, \bbZ_2)$  & $\boldsymbol\delta^N_{(1,1)}/2$ &        $((\mathfrak{su}_3 \oplus \mathfrak{su}_2) \oplus (\mathfrak{su}_3 \oplus \mathfrak{su}_2) \oplus \mathfrak{su}_{12}, \bbZ_6)$ & $\boldsymbol\delta^N_{(3,1)}/6$ \\
       $(\mathfrak{e}_7 \oplus \mathfrak{su}_2 \oplus \mathfrak{su}_{10}, \bbZ_2)$  & $\boldsymbol\delta^N_{(1,1)}/2$ &        $((\mathfrak{su}_3 \oplus \mathfrak{su}_2) \oplus \mathfrak{su}_{2} \oplus \mathfrak{su}_{14}, \bbZ_2)$ & $\boldsymbol\delta^N_{(1,-1)}/2$ \\
       $(\mathfrak{e}_7 \oplus \mathfrak{su}_{11}, -)$  & - &        $((\mathfrak{su}_3 \oplus \mathfrak{su}_2) \oplus \mathfrak{su}_{15}, \bbZ_3$ & $\boldsymbol\delta^N_{(0,1)}/3$ \\
       $(\mathfrak{e}_6 \oplus \mathfrak{e}_6 \oplus \mathfrak{su}_6, \bbZ_3)$ & $\boldsymbol\delta^N_{(0,1)}/3$ &        $(\mathfrak{su}_2 \oplus \mathfrak{su}_{2} \oplus \mathfrak{su}_{16}, \bbZ_4)$ & $\boldsymbol\delta^N_{(1,-1)}/4$  \\        
       $(\mathfrak{e}_6 \oplus \mathfrak{so}_{10} \oplus \mathfrak{su}_7, -)$ & - &        $(\mathfrak{su}_2 \oplus \mathfrak{su}_{17}, -)$ & - \\
       $(\mathfrak{e}_6 \oplus \mathfrak{su}_5 \oplus \mathfrak{su}_8, -)$ & -  &        $(\mathfrak{su}_{18}, \bbZ_3)$ & $\boldsymbol\delta^N_{(1,1)}/3$ \\
     \hline\hline
       \multicolumn{4}{|c|}{$\fkg_{\text{8d}, \infty} = \mathfrak{so}_{34 - 2k} \oplus \widehat{\mathfrak{e}}_k$}  \\ \hline
       $(\mathfrak{e}_8 \oplus \mathfrak{so}_{18}, -)$   & - & $(\mathfrak{su}_5 \oplus \mathfrak{so}_{26}, -)$ & - \\
       $(\mathfrak{e}_7 \oplus \mathfrak{so}_{20}, \bbZ_2)$  & $\boldsymbol\delta^N_{(1,1)}/2$ & $((\mathfrak{su}_3 \oplus \mathfrak{su}_2) \oplus \mathfrak{so}_{28}, \bbZ_2)$ & $\boldsymbol\delta^N_{(1,1)}/2$ \\
       $(\mathfrak{e}_6 \oplus \mathfrak{so}_{22}, -)$ & - & $(\mathfrak{su}_2 \oplus \mathfrak{so}_{32}, \bbZ_2)$ & $\boldsymbol\delta^N_{(1,1)}/2$ \\ 
       $(\mathfrak{so}_{10} \oplus \mathfrak{so}_{24}, \bbZ_2)$ & $\boldsymbol\delta^N_{(1,1)}/2$ & $(\mathfrak{so}_{34}, -)$ & - \\ \hline
\end{tabular}
\caption{Non-Abelian gauge group $G_\text{9d}$ of all maximally-enhanced 9d rank $(1,17)$ string vacua, seen as dimensional uplifts of 8d string junction vacua.
The generator of $\pi_1(G_\text{9d}) \cong \bbZ_{\ell}$ is represented as a fractional null junction (FNJ) $\boldsymbol\delta^N_{(p,q)} / \ell = \boldsymbol\delta^N_{(\nicefrac{p}{\ell}, \nicefrac{q}{\ell})}$.
}
\label{table:9Drank17}
\end{table}

\subsection{9d CHL vacua via junctions}

Having reproduced the maximal rank branch of the 9d moduli space, we would like to extend the junction method also to rank-reduced theories.
We start by matching the known circle compactification of the 9d CHL string in terms of junctions in the presence of a single O7$^+$-plane, focusing again on the cases with maximal non-Abelian gauge rank.

A key assumption here is that the decompactification limit of 8d vacua, even in the presence of O7$^+$-planes, is characterized by the appearance of singularities in the axio-dilaton profile that induce $SL(2,\bbZ)$ monodromy of affine type.
Though we do not have a proof for this, we expect the identification of the resulting loop junctions as the only BPS-tower compatible with decompactification to be valid also with O7$^+$-planes, given that the loop can be thought of as a $(p,q)$-string that is only sensitive to the monodromy, but not the details of the 7-branes.
Moreover, as we will see below, the results following this assumption agree with the momentum lattice description for the 8d and 9d CHL string \cite{Mikhailov:1998si,Font:2021uyw}.

Analogous to the procedure in previous sections of describing the O7$^+$ as freezing a $\mathfrak{so}_{16}$ stack in an ``ordinary'' rank $(2,18)$ setting, we therefore focus on those brane configurations in \eqref{eq:decomp_series_A} and \eqref{eq:decomp_series_D}, whose non-Abelian stack can host a $\mathfrak{so}_{16}$.
This is only possible if the configuration includes a ${\bf D}_{n \geq 8}$ or $\widehat{\textbf{E}}_8$ brane stack.
While for \eqref{eq:decomp_series_A}, there is only one rank $(2,18)$ configuration, with $\fkg_{\text{8d},\infty} = \mathfrak{su}_2 \oplus \widehat{\mathfrak{e}}_8 \oplus \widehat{\mathfrak{e}}_8$, there is an ambiguity for the class \eqref{eq:decomp_series_D}, in that we can naively embed the O7$^+$ inside the $\widehat{\mathfrak{e}}_8$ or the $\mathfrak{so}$ stack.
However, inspecting the set of allowed string and 5-brane junctions reveals a striking difference between the two options.

If we embed the O7$^+$ inside the $\mathfrak{so}$-stack, the freezing of the $\mathfrak{so}_{16}$ subalgebra and the modified boundary conditions for the junctions do not affect the $U$ lattice.
This is made explicit in \eqref{eq:decomp_series_D}, since $\boldsymbol\delta$ and $\boldsymbol\epsilon$ junctions only have prongs on the affine stack, which remains unmodified.
On the other hand, if we would embed the O7$^+$ inside an $\widehat{\textbf{E}}_8$, then the freezing procedure restricts the set of allowed string junctions to be orthogonal to the $\mathfrak{so}_{16}$ roots, and have even prongs on the orientifold plane.
As we will explain in detail in Appendix \ref{app:O7_into_E8}, the result is that we can no longer consistently define a $U$-lattice from the allowed junctions.
Instead, the evenness condition can at most accommodate a stretched hyperbolic lattice $U(2)$.

Based on the dual CHL string description, we propose that only the embeddings with a \textit{modified} $U$ lattice gives a consistent 9d uplift.
Namely, unlike the maximal rank case, the momentum lattice $\Lambda^\text{CHL}_{\text{8d}} \cong (-\text{E}_8) \oplus U \oplus U(2)$ of 8d CHL vacua is no longer self-dual, whereas the corresponding 9d lattice $\Lambda^\text{CHL}_{\text{9d}} \cong (-\text{E}_8) \oplus U$ is \cite{Mikhailov:1998si}.
The additional $U(2)$ in 8d arises from the winding and KK-states of the CHL string, and must therefore be represented in terms of the imaginary root junction around the affine stack, and another string junction that emanates from it.
If we embed the O7$^+$ inside the $\textbf{D}_{17-k}$ stack of \eqref{eq:decomp_series_D} instead, we would have an unstretched $U$-lattice for winding and KK-states.
Moreover, if we would naively identify the would-be 9d gauge algebra with that of 8d (replacing the affine symmetry with its non-affine version), this kind of embedding would lead to an $\mathfrak{sp}$ algebra in 9d, which again is not compatible with the CHL string.
While these arguments provide strong evidence in favor of the proposal, we leave a rigorous proof for future works, and discuss the resulting characterization of 9d CHL vacua in terms of string junctions.

Let us start from the 8d rank $(2,18)$ configuration \eqref{eqn:2dP9_a} with $n=8$, which, if we moved $\bX_{[2,1]}$ from $\widehat{\bf E}_m$ across the branch cut of $\widehat{\bf E}_{n=8}$, becomes \eqref{eqn:1dP9_a} with $k=8$.
Using the brane moves described in Appendix \ref{app:O7_into_E8}, we can turn the $\widehat{\textbf{E}}_8$ into an $SL(2,\bbZ)$-conjugated $\textbf{E}_9$ stack:
\begin{align}
    {\bf A}^{10-m} (\overbrace{\bA^{m-1} \bX_{[-2, 1]} \bX_{[0, 1]}^2 \bX_{[2, 1]}}^{\widehat{\textbf{E}}_m}) \underbrace{\overbrace{ {\bf X}^8_{[0,1]} {\bf X}_{[1,4]} {\bf X}_{[1,2]} }^{\textbf{D}_8'} {\bf X}_{[1,2]}}_{\textbf{E}_9' \cong \widehat{\textbf{E}}_8} \, .
\end{align}
The $\widehat{\bf E}_m$ stack can be conjugated by $g = \left( \begin{smallmatrix} 1 & 1 \\ 0 & 1 \end{smallmatrix} \right)$ to obtain the standard form from Section \ref{sec:local_analysis}.
In particular, this means that the standard extended weight junctions now carry asymptotic $\colpq{p}{q}$ charge given as
\begin{align}
    \boldsymbol\omega^{\mathfrak{e}_m}_p : \ g^{-1}\colpq{1}{0} = \colpq{1}{0} \, , \quad \boldsymbol\omega^{\mathfrak{e}_m}_q : \ g^{-1}\colpq{0}{1} = \colpq{-1}{1} \, .
\end{align}
The $\textbf{D}_8'$ stack can be conjugated by $g' = \left( \begin{smallmatrix} 3 & -1 \\ 1 & 0 \end{smallmatrix} \right)$ to the standard representation, $g' M(\textbf{D}_8') {g'}^{-1} = M({\bf A}^8 {\bf B} {\bf C})$.
Introducing the O7$^+$, i.e., ${\bf X}^8_{[0,1]} {\bf X}_{[1,4]} {\bf X}_{[1,2]} \rightarrow {{\bf O7}^+}'$ (where we use the prime to denote the non-standard $SL(2,\bbZ)$-frame), we obtain
\begin{align}\label{eq:9d_CHL_brane_example}
    {\bf A}^{10-m} (\overbrace{ \underbrace{\bA^{m-1} \bX_{[-2, 1]} \bX_{[0, 1]}^2}_{{\bf E}_m} \bX_{[2, 1]}}^{\widehat{\textbf{E}}_m}) \underbrace{ {\bf O7}^{+'} {\bf X}_{[1,2]}}_{\sim \textbf{E}_9' \cong \widehat{\textbf{E}}_8} \, .
\end{align}
Compared to the standard presentations discussed in Section \ref{sec:local_analysis} (i.e., where the monodromy of O7$^+$ is $M({\bf O7}^+) = \left( \begin{smallmatrix} -1 & 4 \\ 0 & -1 \end{smallmatrix} \right)$), the O7$^+$ monodromy in this $SL(2,\bbZ)$ frame is
\begin{align}
    M({{\bf O7}^+}') = {g'}^{-1} M({\bf O7}^+) g' = \left( \begin{smallmatrix} -1 & 0 \\ -4 & -1 \end{smallmatrix} \right) ,
\end{align}
and the standard extended weights $\boldsymbol\omega_{p,q}^{\text{O7}^+}$ have asymptotic $\colpq{p}{q}$-charges
\begin{align}
    \boldsymbol\omega_p' : {g'}^{-1}\colpq{1}{0} = \colpq{0}{-1} \, , \quad \boldsymbol\omega_q' : {g'}^{-1}\colpq{0}{1} = \colpq{1}{3} \, ,
\end{align}
for which the pairing relations \eqref{eq:bilinext} hold.
We can pull the imaginary root junction $\boldsymbol\delta \equiv \boldsymbol\delta^R_{(1,0)}$ across the branch-cuts, and obtain the equivalence 
\begin{align}
    \boldsymbol\delta = 2(-\boldsymbol\omega'_p - \boldsymbol\omega_q' + \bx_{[1,2]}) \, ,
\end{align}
which consistently has only even number of prongs on ${{\bf O7}^+}'$.
Moreover, the pairings are $(\boldsymbol\omega_p', \bx_{[1,2]}) = -(\boldsymbol\omega_q^{\text{O7}^+}, \bx_{[1,2]}) = \tfrac12$, and assert, together with \eqref{eq:bilinext}, that $(\boldsymbol\delta, \boldsymbol\delta) = 0$.
The $\boldsymbol\epsilon$-junction from \eqref{eqn:2dP9_a} cannot be realized as a string junction in the presence of the O7$^+$, because it requires a net $\colpq{p}{q} = \colpq{3}{1}$ charge to end on ${{\bf O7}^+}' {\bf X}_{[1,2]}$ (see Appendix \ref{app:O7_into_E8} for details).
Instead, the prongs of any physical string junction on the ${{\bf O7}^+}' {\bf X}_{[1,2]}$ stack must be
\begin{align}
    2 \lambda_p \boldsymbol\omega_p' + 2 \lambda_q \boldsymbol\omega_q' + \lambda {\bf x}_{[1,2]} \, , \quad \lambda_{p,q}, \lambda \in \bbZ \, ,
\end{align}
which necessarily has even $q$-charge, as well as even pairing with $\boldsymbol\delta$.
This means that, orthogonal to the $\mathfrak{su}_{10-m} \oplus \mathfrak{e}_m$ weight junctions in \eqref{eq:9d_CHL_brane_example}, we must have a $U(2)$ lattice, spanned by string junctions $\boldsymbol\delta$ and $\boldsymbol\epsilon' = - \boldsymbol\delta + 6 \boldsymbol\omega_p' + 4 \bx_{[1,2]} - 2\bx_{[2,1]}$.

In the magnetically dual picture, any integer number of 5-brane prongs can end on ${{\bf O7}^+}'$.
In particular, 5-brane junctions corresponding to $\tfrac12 \boldsymbol\delta$ and $\tfrac12 \boldsymbol\epsilon'$ are then physical, and would span a squeezed hyperbolic lattice $U(\tfrac12)$.
This is consistent with the fact that in the 9d uplift of CHL vacua, the momentum lattice ``loses'' a $U(\tfrac12)$ factor \cite{Mikhailov:1998si}:
\begin{align}
    \left(\Lambda^\text{CHL}_{\text{8d}}\right)^* \cong (-\text{E}_8) \oplus U \oplus U(\tfrac12) \, , \quad \left(\Lambda^\text{CHL}_{\text{9d}}\right)^* \cong \Lambda^\text{CHL}_{\text{9d}} \cong (-\text{E}_8) \oplus U \, .
\end{align}
The remaining moduli available in 9d are then the deformations that move the 7-branes outside the ${{\bf O7}^+}' \bX_{[1,2]}$ stack.
The resulting maximal non-Abelian enhancements can be equally characterized by an 8d configuration of type \eqref{eqn:2dP9_a} (with $n=8$) or \eqref{eqn:1dP9_a} (with $k=8$), but with $\widehat{\textbf{E}}_8$ frozen via the embedding of an O7$^+$ described above (as summarized in Table \ref{tab:9d_CHL_maximal}).

The null junctions for \eqref{eq:9d_CHL_brane_example} are
\begin{align}
    \begin{split}
        & \boldsymbol\delta^N_{(1,0)} = -3 \boldsymbol\omega_p^{\mathfrak{e}_m} - \boldsymbol\omega_q^{\mathfrak{e}_m} + \bx_{[2,1]} -2\boldsymbol\omega'_p - 2\boldsymbol\omega_q' + 2\bx_{[1,2]} \, , \\
        & \boldsymbol\delta^N_{(0,1)} = (m-10) \boldsymbol\omega_{\mathfrak{su}} + (18-m) \boldsymbol\omega^{\mathfrak{e}_m}_p + 3 \boldsymbol\omega^{\mathfrak{e}_m}_q - 3\bx_{[2,1]} + 4\boldsymbol\omega'_p + 2\boldsymbol\omega'_q - x_{[1,2]} \, ,
    \end{split}
\end{align}
from which one can straightforwardly determine the non-Abelian gauge group structure for specific $m$.
It so happens that they are all trivial in the maximally enhanced cases, which agrees with the CHL-string computations \cite{Font:2021uyw}.

\begin{table}[ht]
\renewcommand{\arraystretch}{1.3}
\begin{equation*}
\begin{array}{|c|c|c|c|}
\hline \fkg^\text{CHL}_\text{9d} & \pi_1(G_\text{9d}) & \fkg_{\text{8d}, \infty} & \text{8d brane config.}  \\
\hline 
\hline \mathfrak{su}_{10} & 0 & \mathfrak{su}_{10} + \widehat{\mathfrak{e}}_8 & (\ref{eqn:2dP9_c}),\ n=8 \\
\hline \mathfrak{su}_{9}\oplus \mathfrak{su}_{2} & 0 & \mathfrak{su}_9 \oplus \widehat{\mathfrak{e}}_1 \oplus \widehat{\mathfrak{e}}_8  &  (\ref{eqn:2dP9_c}),\ m=1,\ n=8  \\
\hline \mathfrak{su}_7\oplus \mathfrak{su}_{2} \oplus \mathfrak{su}_{3} & 0 &  \mathfrak{su}_7 \oplus \widehat{\mathfrak{e}}_3 \oplus \widehat{\mathfrak{e}}_8 &  (\ref{eqn:2dP9_a}),\ m=3,\ n=8  \\
\hline  \mathfrak{su}_{6} \oplus \mathfrak{su}_{5} & 0 & \mathfrak{su}_6 \oplus \widehat{\mathfrak{e}}_4 \oplus \widehat{\mathfrak{e}}_8  &  (\ref{eqn:2dP9_a}),\ m=4,\ n=8  \\
\hline \mathfrak{su}_{5} \oplus \mathfrak{so}_{10} & 0 & \mathfrak{su}_5 \oplus \widehat{\mathfrak{e}}_5 \oplus \widehat{\mathfrak{e}}_8  &  (\ref{eqn:2dP9_a}),\ m=5,\ n=8  \\
\hline  \mathfrak{su}_{4} \oplus \mathfrak{e}_{6} & 0 & \mathfrak{su}_4 \oplus \widehat{\mathfrak{e}}_6 \oplus \widehat{\mathfrak{e}}_8  &  (\ref{eqn:2dP9_a}),\ m=6,\ n=8  \\
\hline \mathfrak{su}_{3} \oplus \mathfrak{e}_{7} & 0 & \mathfrak{su}_3 \oplus \widehat{\mathfrak{e}}_7 \oplus \widehat{\mathfrak{e}}_8  &  (\ref{eqn:2dP9_a}),\ m=7,\ n=8  \\
\hline \mathfrak{su}_2 \oplus \mathfrak{e}_8  & 0 & \mathfrak{su}_2 \oplus \widehat{\mathfrak{e}}_8 \oplus \widehat{\mathfrak{e}}_8  & (\ref{eqn:2dP9_a}),\ m=8,\ n=8  \\
\hline \mathfrak{so}_{18} & 0 & \mathfrak{so}_{18} \oplus \widehat{\mathfrak{e}}_8 &  (\ref{eqn:1dP9_a}),\ k=8  \\	

\hline
\end{array}
\end{equation*}
\caption{Maximal non-Abelian enhancements on the 9d rank $(1,9)$ moduli space that has a dual description in terms of the CHL string, obtained from an affine 8d realization in which an $\widehat{\mathfrak{e}}_8$ is frozen.
Note that all cases have trivial non-Abelian gauge group topology $\pi_1(G_\text{9d})$.
\label{tab:9d_CHL_maximal}}
\end{table}

\subsection[Disconnected moduli branches for 9d rank \texorpdfstring{$(1,1)$}{(1,1)} vacua]{Disconnected moduli branches for 9d rank \boldmath{$(1,1)$} vacua}

The description of 9d rank $(1,9)$ theories presented above has a clear interpretation in terms of ``freezing'', i.e., introducing an O7$^+$-plane into the 7-brane system that describes a rank $(1,17)$ theory.
In parallel to the construction of 8d vacua discussed in Section \ref{sec:global}, it then is natural to propose that 9d rank $(1,1)$ theories arise by a further freezing.
Moreover, the duality to the CHL string strongly suggests that, in 9d, the freezing process requires an $\widehat{\mathfrak{e}}_8$ affine algebra, in which the $\mathfrak{e}_8$ root junctions, as well as odd multiples of the winding-state-junction (i.e., $\boldsymbol\epsilon$) are projected out.
Therefore, from the maximally-enhanced cases in Table \ref{tab:9d_CHL_maximal}, only the second to last (with brane configuration \eqref{eq:9d_CHL_brane_example}), but not the last entry, can undergo a further freezing.

After repeating the brane motions discussed in Appendix \ref{app:O7_into_E8}, now for the first affine stack in \eqref{eq:9d_CHL_brane_example}, the corresponding (doubly) frozen configuration looks like
\begin{align}\label{eq:9d_rank_1_example_config}
    \bA \bA \, (\widetilde{{\bf O7}^+} \bX_{[1,-2]}) \, ({{\bf O7}^+}' \bX_{[1,2]}) \, ,
\end{align}
where the $M(\widetilde{{\bf O7}^+}) = \left( \begin{smallmatrix} 3 & 4 \\ -4 & -5 \end{smallmatrix} \right)$ is the monodromy of the left O7$^+$-plane in this $SL(2,\bbZ)$-frame.
We obtain an enhanced $\fkg = \mathfrak{su}_2$ non-Abelian symmetry when the two $\bA$-branes are moved on top of each other, which is the maximal enhancement we can have in 9d.
In fact, if one could separate the two $\bX$-branes from their corresponding O7$^+$ (making the KK modes massive), and move them next to each other, they would be locally-mutual, thus allowing for another $\mathfrak{su}_2$ enhancement --- this would be nothing but the 8d rank $(2,2)$ example \eqref{eq:rank_2_example_config} studied in the previous section, which had an $SU(2)^2 /\bbZ_2$ non-Abelian gauge group.
However, since for the 9d uplift, one of them must be broken, the fractional null junction that generated this $\bbZ_2$ quotient no longer exists for the configuration \eqref{eq:9d_rank_1_example_config}.
Hence, the 9d non-Abelian gauge group must be $SU(2)$.

It is suggestive that this doubly frozen, rank $(1,1)$ moduli branch corresponds to M-theory on a Klein-bottle \cite{Aharony:2007du}.
Namely, starting from the rank $(1,17)$ theories with heterotic description, which is dual to M-theory on a cylinder, the first freezing led to CHL vacua in 9d, which are equivalent to M-theory on a M{\"o}bius strip, or a cylinder with one cross-cap.
Freezing once more, i.e., adding another cross-cap on the other side, then produces a Klein-bottle.

However, as pointed out in \cite{Aharony:2007du}, there is a second branch of 9d rank $(1,1)$ moduli space that is disconnected from M-theory on a Klein-bottle.
That is, it cannot be realized as freezing 9d rank $(1,9)$ models.
However, since after an $S^1$-reduction, the 8d rank $(2,2)$ moduli space \emph{is} connected, there should exist a junction description for this 9d branch, as a suitable infinite distance limit in which KK-states become light.

In fact, starting from the general 8d configuration with two O7$^+$'s, depicted in Figure \ref{fig:rank2}, it is not hard to identify such potential limits.
Starting from ${\bf O7}^+ \bC \bX_{[3,1]} \, {{\bf O7}^+} \bC \bX_{[3,1]}$, where both O7's now have the standard monodromy, we can either push the $\bC$-branes from the left on top of the orientifolds,
\begin{align}\label{eqn:rank2_a}
    (\underbrace{{\bf O7}^+ \bC}_{\text{affine}}) \bX_{[3,1]} \, (\underbrace{{{\bf O7}^+} \bC}_{\text{affine}}) \bX_{[3,1]} \, ,
\end{align}
which is just a slightly rearranged version of \eqref{eq:9d_rank_1_example_config}, or we can generate a $\widehat{\bf E}_1 = \bB \bC \bC \bX_{[3,1]}$ stack, by moving 7-branes as
\begin{equation}
\begin{split}
    & {\bf O7}^+ \, \bC \, \underrightarrow{\bX_{[3, 1]}} \, {\bf O7}^+ \bC \bX_{[3, 1]} \ \longrightarrow \ {\bf O7}^+ \, \underrightarrow{\bC} \, {\bf O7}^+\,  \bB  \bC \bX_{[3, 1]} \\
    \longrightarrow \ & {\bf O7}^+ \, {\bf O7}^+ \, \underrightarrow{\bX_{[3,-1]}} \bB \bC \bX_{[3,1]} \ \longrightarrow \ {\bf O7}^+ \, {\bf O7}^+  \, (\underbrace{\bB \bC \bC \bX_{[3,1]}}_{=\widehat{\bf E}_1}) \, .
\end{split}\label{eqn:rank2_b}
\end{equation}

First, notice that one cannot transition between \eqref{eqn:rank2_a} and \eqref{eqn:rank2_b} without separating branes making up the affine stack.
In other words, these configurations are connected only via the 8d moduli space.
Second, by the brane move $\bC \underleftarrow{\bX_{[3,1]}} \rightarrow \bB \bC$ inside the affine stack, we find that $\widehat{\bf E}_1 = \bB \bC \bC \bX_{[3,1]} \simeq (\bB \bC) (\bB \bC)$ is the strong-coupling version of two O7$^-$-planes on top of each other.
Therefore, \eqref{eqn:rank2_b} is T-dual to IIA on an interval with O8$^\pm$'s at each end, which further dualizes to the 9d Dabholkar--Park background in type IIB \cite{Dabholkar:1996zi,Witten:1997bs}.
This is indeed the branch of 9d rank $(1,1)$ moduli space that is disconnected from M-theory on a Klein-bottle \cite{Aharony:2007du}.

\section{Conclusions and outlook}

In this work, we have extended the framework of string junctions on $[p,q]$-7-branes \cite{Gaberdiel:1997ud,Gaberdiel:1998mv,DeWolfe:1998zf} to include O7$^+$-planes.
The key difference is the distinction between physical $(p,q)$-strings and 5-branes that can end on the O7$^+$: while the latter can end with arbitrary integer $\colpq{p}{q}$-charges on the O7$^+$, only even numbers of integer $(p,q)$-strings may do so.
When applied to the construction of 8d ${\cal N}=1$ gauge theories on stacks including both ordinary $[p,q]$-7-branes and O7$^+$'s, this modification consistently reproduces the root and coroot lattices of non-simply-laced $\mathfrak{sp}$-algebras, as well as their electric 1-form- and magnetic 5-form center symmetries.
Furthermore, this provides a junction description for all 8d rank $(2,10)$ string compactifications with a dual CHL-string description \cite{Font:2021uyw,Cvetic:2021sxm}, including their gauge group topologies that are succinctly characterized by loop junctions encircling all 7-branes.
In addition, using junctions, we find a previously unknown lattice description for 8d string vacua of rank $(2,2)$, that is analogous to the Narain lattice characterization of 8d and 9d heterotic/CHL vacua.
This establishes junctions as a unifying framework to describe gauge enhancements (including the global gauge group structure) of \emph{all} 8d string vacua.

Moreover, in synergy with Swampland ideas \cite{Lee:2021qkx,Lee:2021usk}, we have discovered a full classification of 9d ${\cal N}=1$ string vacua, including their global gauge group structures, by 7-brane configurations with affine stacks characterized by loop junctions for their imaginary roots.
Again, the consistent inclusion of O7$^+$-planes in the analysis of potential infinite distance limits on the 8d moduli space turns out to be vital to capture subtleties, such as the two components of the 9d rank $(1,1)$ moduli space that are connected only through an $S^1$-reduction to 8d \cite{Aharony:2007du}.

The 9d results motivate a string-independent classification of the 9d ${\cal N}=1$ supergravity landscape in a similar fashion to \cite{Hamada:2021bbz}, where the 8d landscape was classified based on a Swampland ``translation'' of the $SL(2,\bbZ)$ characterization of 7-branes and O7$^+$-planes.
While perhaps unexpected from their direct constructions, our work shows that 9d string compactifications also admit a completely analogous characterization.
Hence, it is suggestive that there should also be a parallel story for the moduli space of 9d instantons that can be studied by $SL(2,\bbZ)$ monodromies.
In particular, such a bottom-up analysis could provide an explanation independent of the CHL-string, for why the 9d analog of the freezing mechanism can only be performed with an $\widehat{\bf E}_8$, but not an ${\bf D}_{n \geq 8}$ stack.

Another useful insight from the junction perspective is on the stringy origin of center symmetries in 8d gauge theories with non-simply-laced algebra.
Via dualities, it would be interesting if one can use this insight to generalize the geometric engineering framework for higher-form symmetries in M- and F-theory \cite{Morrison:2020ool,Albertini:2020mdx,Cvetic:2021sjm} to include frozen singularities.
This may have promising applications to the study of 6d SCFTs constructed on such singularities \cite{Bhardwaj:2018jgp} as well as lower dimensional SCFTs, obtained either from dimensionally reducing 6d theories, or directly engineering them with junction techniques \cite{Garcia-Etxebarria:2013tba,Agarwal:2016rvx,Hassler:2019eso,Heckman:2020svr}.

\section*{Acknowledgements}
We are grateful to Jonathan Heckman, Max H\"ubner, Miguel Montero, Ethan Torres, and Timo Weigand for valuable discussions.
This work is supported by the DOE (HEP) Award DE-SC0013528 (MC), the Simons Foundation Collaboration grant \#724069 on ``Special Holonomy in Geometry, Analysis and Physics'' (MC and HYZ), the Fay R.~and Eugene L.~Langberg Endowed Chair (MC), the Slovenian Research Agency (ARRS No. P1-0306) (MC), and the German-Israeli Project Cooperation (DIP) on ``Holography and the Swampland'' (MD).

\appendix

\section[Deriving the evenness condition on O7\texorpdfstring{\boldmath{$^+$}}{+} via 8d CHL strings]{Deriving the evenness condition on O7\boldmath{$^+$} via 8d CHL strings} \label{apdx:evenness}

In this appendix, we present a derivation of the evenness condition for string junctions the O7$^+$-plane in 8d rank $(2,10)$ models.
This proof utilizes the known equivalence between the heterotic Narain lattice \eqref{eq:Narainlattice} and the junction lattice (modulo null junctions) on 24 ordinary 7-branes, and the construction $\Lambda_\text{Mikhailov} \hookrightarrow \Lambda_\text{Narain}$ of the Mikhailov lattice, describing states of rank $(2,10)$ vacua, as a sublattice \cite{Mikhailov:1998si}, also known as ``freezing'' \cite{Font:2021uyw,Fraiman:2021hma}.

Assuming that the freezing mechanism in the IIB / 7-brane picture is a local operation, ${\bf D}_8 = \bA^8 \bB \bC \rightarrow {\bf O7}^+$, we show that the evenness condition discussed in Section \ref{subsec:junctions_on_O7} is the necessary and sufficient condition for the string junction lattice on the O7$^+$ and the unaffected 7-branes to agree with the Mikhailov lattice.

To this end, first recall that there is a particular $\mathfrak{so}_{16}$ root lattice $(- \text{D}_8) \subset \Lambda_\text{Narain}$ along which one defines an orthogonal projection $P$ \cite{Mikhailov:1998si} (see also \cite{Cvetic:2021sjm}).
Since $P(\text{D}_8) = 0$, this $\mathfrak{so}_{16}$ is interpreted as projected out from, or ``frozen'' inside the heterotic model.
Then, $\Lambda_\text{Mikhailov} \subset \Lambda_\text{Narain}$ is the image of $P$ \emph{inside} $\Lambda_\text{Narain}$, i.e.,
\begin{align}
    [{\bf j}] \in \Lambda_\text{Mikhailov} \Leftrightarrow \exists [\hat{\bf j}] \in \Lambda_\text{Narain} : [{\bf j}] = P([\hat{\bf j}]) \in \Lambda_\text{Narain} \, .
\end{align}
This is a non-trivial condition on the choice of $[\hat{\bf j}]$, since not all elements of $\Lambda_\text{Narain}$ map to integer lattice points under $P$.

To make contact with the junction description, it is important to remember that any elemnt of the Narain lattice corresponds to an equivalence class of physical junctions modulo null junctions.
Therefore, we first identify $\hat{\bf j} \in J_\text{phys}^\text{el}$ as a physical string junction in a rank $(2,18)$ 7-brane configuration with a ${\bf D}_8$ stack.
Now, as explained in Section \ref{subsec:global_junciton_lattice_null_junctions}, the junction $\hat{\bf j}$, prior to freezing, enjoys a decomposition into an integer linear combination,
\begin{align}\label{eq:appB1}
    \hat{\bf j} = \sum_{i=1}^8 a^i {\bf w}_i + a^p \boldsymbol\omega_p^{\mathfrak{so}_{16}} + a^q \boldsymbol\omega_q^{\mathfrak{so}_{16}} + \hat{\bf j}'' \, ,
\end{align}
where $\hat{\bf j''}$ has no prongs on ${\bf D}_8$.
This decomposition is unique only up to the addition of physical null junctions $\boldsymbol\delta^N_{(p,q)}$; however, because such junctions carry no physical charge, their prongs on the ${\bf D}_8$ stack must induce no $\mathfrak{so}_{16}$ center charges, which, according to \eqref{eq:center_charge_table_ADE}, requires even multiples of $\boldsymbol\omega_p^{\mathfrak{so}_{16}}$ and $\boldsymbol\omega_q^{\mathfrak{so}_{16}}$.
Hence, any representative $\hat{\bf j}$ of the equivalence class $[\hat{\bf j}]$ modulo null junctions takes the form
\begin{align}
    \hat{\bf j} + \boldsymbol\delta^N = \sum_{i=1}^8 a^i {\bf w}_i + (a^p + 2n^p) \boldsymbol\omega_p^{\mathfrak{so}_{16}} + (a^q + 2n^q) \boldsymbol\omega_q^{\mathfrak{so}_{16}} + \hat{\bf j}' \, ,
\end{align}
for some junction $\hat{\bf j}'$ that has no prongs on the ${\bf D}_8$.

As this stack will be replaced with the O7$^+$, the D$_8$ root lattice defining the projection in the momentum lattice description is identified with the root junction lattice of this stack.
The representative for $P([\hat{\bf j}])$ is then
\begin{align}
    P(\hat{\bf j}) := (a^p + 2n^p) \boldsymbol\omega_p^{\mathfrak{so}_{16}} + (a^q + 2n^q) \boldsymbol\omega_q^{\mathfrak{so}_{16}} + \hat{\bf j}' \, , \quad a^p, n^p, a^q, n^q \in \bbZ \, .
\end{align}
Therefore, the condition $P([\hat{\bf j}]) \in \Lambda_\text{Narain}$ translates into $P(\hat{\bf j}) \in J_\text{phys}^\text{el}$, i.e., its prongs on the ${\bf D}_8$ stack must satisfy the physicality conditions.
Since, by construction, $\hat{\bf j}'$ has no prongs on the ${\bf D}_8$ stack, this means that $a^p, a^q \in 2\bbZ$.
As we identify the extended weights of $\mathfrak{so}_{16}$ with those of the O7$^+$ after freezing (see Section \ref{subsec:junctions_on_O7}), we conclude that the junction $P(\hat{\bf j})$ representing an element of $\Lambda_\text{Mikhailov}$ must have even $\colpq{p}{q}$-charge.

Finally, it is straightforward to verify the condition for the magnetically dual 5-brane junctions from the lattice description of the dual Mikhailov lattice, which in terms of the above projection map is given by $\Lambda_\text{Mikhailov}^* = P(\Lambda_\text{Narain})$ \cite{Mikhailov:1998si,Cvetic:2021sxm}.
Since the projection simply removes the terms proportional to the $\mathfrak{so}_{16}$ weights in \eqref{eq:appB1} from any physical junction $\hat{\bf j}$ in the rank $(2,18)$ configuration, it is obvious that one ends up with any integer-valued $a^p,a^q$.

\section{Embedding O7\texorpdfstring{\boldmath{$^+$}}{+} into \texorpdfstring{\boldmath{$\widehat{\text{E}}_8$}}{affine E8}}
\label{app:O7_into_E8}

In this appendix, we discuss the junctions resulting from embedding an O7$^+$ into an $\widehat{\textbf{E}}_8$ stack.
First describe the embedding $\mathfrak{so}_{16} \hookrightarrow \widehat{\mathfrak{e}}_8$ in terms of 7-branes.
To this end, we use the equivalence $\widehat{\textbf{E}}_8 \cong \textbf{E}_9$ of 7-brane stacks \cite{DeWolfe:1998yf,DeWolfe:1998pr}, where $\textbf{E}_9$ is conjugate to ${\bf A}^8 {\bf B} {\bf C}^2$, see \eqref{eq:ADE_brane_stacks}.
In this presentation, it is straightforward to identify the $\mathfrak{so}_{16}$ subalgebra as the $\textbf{D}_8 = {\bf A}^8 {\bf B} {\bf C}$ part.
By the ``freezing'' procedure, the $\textbf{E}_9$ stack becomes an ${\bf O7}^+ {\bf C}$ stack.
Now, due to the evenness condition discussed in Section \ref{sec:local_analysis}, only a subset of string junctions that were allowed to end on $\textbf{E}_9$ prior to freezing are allowed in the presence of the O7$^+$.

One such junction that we will focus on in the following is the $\boldsymbol\epsilon$-junction given in \eqref{eqn:1dP9_a}.
This junction has a unit $\colpq{3}{1}$-prong on the ${\bf X}_{[3,1]}$-brane that affinizes the stack --- however, this is in the realization $\widehat{\textbf{E}}_8$!
To connect the two descriptions, we repeatedly use brane moves \eqref{eq:brane_move_left-to-right} and \eqref{eq:brane_move_right-to-left}, to obtain
\begin{align}
\begin{split}
    \widehat{\textbf{E}}_8 = {\bf A}^7 {\bf B} {\bf C}^2 {\bf X}_{[3,1]} \, & \rightarrow  {\bf A}^7 {\bf B} {\bf C} {\bf B} {\bf C} \\
    & \rightarrow {\bf B} {\bf X}_{[0,1]}^7 {\bf C}{\bf B}{\bf C} \rightarrow {\bf B} {\bf C} {\bf A}^7 {\bf B} {\bf C}\\
    & \rightarrow {\bf C} {\bf X}_{[3,1]} {\bf A}^7 {\bf B} {\bf C} \rightarrow {\bf C} {\bf A}^7 {\bf X}_{[-4,1]} {\bf B} {\bf C} \rightarrow {\bf C} {\bf A}^7  {\bf B} {\bf X}_{[-1,-2]} {\bf C}\\
    & \rightarrow {\bf C} {\bf A}^7 {\bf B}{\bf C} {\bf X}_{[0,-1]} \rightarrow {\bf C} {\bf A}^7 {\bf B} {\bf X}_{[0,1]} {\bf X}_{[1,2]} \rightarrow {\bf C} {\bf A}^7 {\bf X}_{[0,1]} {\bf A} {\bf X}_{[1,2]}\\
    & \rightarrow {\bf X}_{[0,1]}^7 {\bf C} {\bf X}_{[0,1]} {\bf A} {\bf X}_{[1,2]} \rightarrow {\bf X}_{[0,1]}^8 {\bf X}_{[1,2]} {\bf A} {\bf X}_{[1,2]}\\
    & \rightarrow \underbrace{ {\bf X}^8_{[0,1]} {\bf X}_{[1,4]} {\bf X}^{(1)}_{[1,2]} }_{= \textbf{D}_8'} {\bf X}^{(2)}_{[1,2]} = \textbf{E}_9'\, .
\end{split}
\end{align}
In each step, it is easy to track the changes of the prongs of the $\boldsymbol\epsilon$-junction that starts out with a unit $\bx_{[3,1]}$-prong, simply by requiring that the prongs on the two moving branes change in such a way that the net $\colpq{p}{q}$-charge remains invariant.
For example, after the first step, we have $\bx_{[3,1]} \rightarrow \bb_2 + 2\bc_2$.
After the whole process, we end up with
\begin{align}
    \bx_{[3,1]} \rightarrow -\bx_{[0,1]}^{(8)} - 2 \bx_{[1,4]} + 4 \bx_{[1,2]}^{(1)} + \bx_{[1,2]}^{(2)} \, .
\end{align}
Since the first three summands end on the ${\bf D}_8'$ stack, one can decompose their sum using the extended weights of $\mathfrak{so}_{16}$ and the weight junctions; the important thing to track here is that the net $\colpq{p}{q}$-charge of this part is $\colpq{2}{-1}$.
However, after introducing the O7$^+$, i.e., replace ${\bf D}_8' \rightarrow {{\bf O7}^+}'$, which removes the $\mathfrak{so}_{16}$ weights, there is an odd $q$-charge emanating via this junction from the orientifold, which is not allowed for a string junction.
Indeed, it is easy to check that any physical string junction leaving the $\widehat{\textbf{E}}_8 \rightarrow {{\bf O7}^+}' \bX_{[1,2]}$ stack must have even $q$-charge.
Therefore, only even multiples of $\boldsymbol\epsilon$ are physical string junctions after freezing.
On the other hand, this prong, and therefore also $\boldsymbol\epsilon$ would be acceptable as a 5-brane junction.

\section{All 8d supergravity vacua via \texorpdfstring{\boldmath{$[p,q]$}}{[p,q]}-7-branes}
\label{app:results}

In this appendix, we give the full catalog of maximally-enhanced 7-brane configurations realizing 8d string vacua of maximal non-Abelian rank for all three classes of models, i.e. total rank $(2,18)$, $(2,10)$, and $(2,2)$. We further determine global structure of their non-Abelian subgroup given by $\mathcal{Z}$ and the explicit realization of the fracrtional null junction.

Before we provide the classification we further describe a procedure that allows to incorporate the non-maximally-enhanced cases, with additional $\mathfrak{u}(1)$ factors.

\subsection{Non-maximally-enhanced cases}

In principle, the process of obtaining the global gauge group topology for the non-maximally enhanced cases is equivalent to what was described in the main text: First one obtains the associated brane configuration, with which one has access to the discrete quotients $\mathcal{Z}$ via the fractional null junctions as in \ref{eq:juncZ} as well as $\mathcal{Z}'$ as in \ref{eq:abelian_quotient_from_junctions}. Even though one cannot avoid repeating the computations of $\mathcal{Z}$ and $\mathcal{Z}'$, one fortunately can take a shortcut of finding the corresponding brane configurations (which is technically the most challenging step) by starting from the maximally-enhanced setups and suitably splitting the brane stacks.

Here we stress that, given a single non-Abelian brane stack, all of the natural brane splittings corresponds to Higgs transition with W-boson vacuum expectation values that decrease the rank by 1. Adjoint Higgsing that preserves the rank (such as $\mathfrak{e}_8 \rightarrow \mathfrak{so}_{16}$), on the other hand, are not guaranteed to admit a realization in a specific brane configuration. Even in brane configurations where such adjoint Higgsings are possible, it would necessarily involve not only the constituent branes in the stack but also some additional branes ($\mathbf{E}_8 \cong \bA^7 \bB \bC^2$ and $\mathbf{D}_8 \cong \bA^8 \bB \bC$ in the example). For this reason, we focus on the W-boson Higgsing, which is guaranteed to have a straightforward brane realizations.

\begin{itemize}

    \item{Splitting $\mathfrak{su}_{k}$:} $\quad$ $\mathfrak{su}_{k} \rightarrow \mathfrak{su}_{k^{\prime}} \oplus \mathfrak{su}_{k-k^{\prime}}$: $\bA^{k} \rightarrow \bA^{k' } + \bA^{k - k'}$
    
    \item{Splitting $\mathfrak{sp}_l$:} $\quad$ $\mathfrak{sp}_{l} \rightarrow \mathfrak{su}_{l^{\prime}} \oplus \mathfrak{sp}_{l-l^{\prime}}$: $\bA^{l} {\bf O7}^+ \rightarrow \bA^{l' } + \bA^{l - l'} {\bf O7}^+$
    
    \item{Splitting $\mathfrak{so}_{2m}$:} 
    \begin{itemize}
        \item{$\mathfrak{so}_{2m} \rightarrow \mathfrak{su}_{m}$}: $\bA^m \bB \bC \rightarrow \bA^m + \bB \bC$
        \item{$\mathfrak{so}_{2m} \rightarrow 2\mathfrak{su}_2 \oplus \mathfrak{su}_{m-2}$}: $\bA^m \bB \bC \simeq \bA^{m-2} \bN^2 \bC^2 \rightarrow \bA^{m-2} + \bN^2 + \bC^2 $
        \item{$\mathfrak{so}_{2m} \rightarrow \mathfrak{su}_4 \oplus \mathfrak{su}_{m-3}$}: $\bA^m \bB \bC  \simeq \bA^{m-2} \bN^2 \bC^2 \simeq \bA^{m-3} \bC^4 \bX_{[3, 2]} \rightarrow \bA^{m-3} + \bC^4 + \bX_{[3, 2]}$ 
        \item{$\mathfrak{so}_{2m} \rightarrow \mathfrak{so}_{2m'} \oplus \mathfrak{su}_{m - m'}$ ($4 \leq m' \leq m - 1$)}: $\bA^m \bB \bC \rightarrow \bA^{m'}  \bB \bC + \bA^{m - m'}$
    \end{itemize}

    \item{Splitting $\mathfrak{e}_n$:} 
    \begin{itemize}
        \item{$\mathfrak{e}_n \rightarrow \mathfrak{su}_{n}$}: $\bA^{n-1} \bB \bC^2 \simeq \bA^n \bX_{[3, -1]} \bN \rightarrow \bA^n + \bX_{[3, -1]} + \bN$ (see (2.12) of \cite{DeWolfe:1998eu})
        \item{$\mathfrak{e}_n \rightarrow \mathfrak{so}_{2n-2}$}:  $\bA^{n-1} \bB \bC^2 \rightarrow \bA^{n-1} \bB \bC + \bC$
        \item{$\mathfrak{e}_n \rightarrow \mathfrak{su}_2 \oplus \mathfrak{su}_{n-1}$}: $\bA^{n-1} \bB \bC^2 \rightarrow \bA^{n-1} + \bB + \bC^2$
        \item{$\mathfrak{e}_n \rightarrow \mathfrak{su}_2 \oplus \mathfrak{su}_3 \oplus \mathfrak{su}_{n-3} (\simeq \mathfrak{e}_3 \oplus \mathfrak{su}_{n-3})$}: $\bA^{n-1} \bB \bC^2 \rightarrow \bA^{n-3} + \bA^2 \bB \bC^2 \simeq \bA^{n-3} + \bC \bA^2 \bC^2 \simeq \bA^{n-3} + \bN^2 \bC^3 \rightarrow \bA^{n-3} + \bN^2 + \bC^3$
        \item{$\mathfrak{e}_n \rightarrow \mathfrak{su}_5 \oplus \mathfrak{su}_{n-4} (\simeq \mathfrak{e}_4 \oplus \mathfrak{su}_{n-4})$}: $\bA^{n-1} \bB \bC^2 \rightarrow \bA^{n-4} + \bA^3 \bB \bC^2 \simeq \bA^{n-4} + \bX_{[1, 2]} \bC^5 \rightarrow \bA^{n-4} + \bX^{[1, 2]} + \bC^5$
        \item{$\mathfrak{e}_n \rightarrow \mathfrak{so}_{10} \oplus \mathfrak{su}_{n-5} (\simeq \mathfrak{e}_5 \oplus \mathfrak{su}_{n-5})$}: $\bA^{n-1} \bB \bC^2 \rightarrow  \bA^{n-5} + \bA^4 \bB \bC^2 \simeq \bA^{n-5} + \bC^5 \bA \bX_{[1, 2]}$ (see (2.11) of \cite{DeWolfe:1998eu}).
        \item{$\mathfrak{e}_n \rightarrow \mathfrak{e}_{n'} \oplus \mathfrak{su}_{n-n'}\ \  (6 \leq n' \leq n-1)$}: $\bA^{n-1} \bB \bC^2 \rightarrow \bA^{n - n'} + \bA^{n'-1} \bB \bC^2$
    \end{itemize}
\end{itemize}

These splittings matches with the ``substitution rules" as given in Table 2.2 of \cite{shimada2000}.

\newpage 

\begin{landscape}

\subsection{No \texorpdfstring{O7$^+$}{O7+}}

We give all possible brane configurations with rank $(2,18)$ realizing maximally-enhanced non-Abelian gauge algebras. Our list reproduces the mathematical classification of \cite{shimada2000} of the ADE-singularities of elliptically-fibered K3 surfaces. This is an expected result, since the junctions describe the same physics in a type IIB perspective. For each brane configuration, we give not only the non-Abelian fundamental group $\pi_1(G_{\text{nA}}) = \mathcal{Z}$ but also its particular embedding into the center $\pi_1(G_{\text{nA}}) \hookrightarrow Z(\widetilde{G})$ using string junctions, where $\widetilde{G}$ is the simply-connected cover of the no-Abelian gauge algebra $G_{\text{nA}} = \widetilde{G}_{\text{nA}}/\mathcal{Z}$. 

\captionsetup{width=23cm}
\begin{longtable}{|c|c|c||c|c|c|} 
\caption{All 8d maximally-enhanced rank $(2,18)$ brane configurations. For each entry, we present the brane configuration, the non-Abelian gauge algebra $\mathfrak{g}_\text{nA}$ and $\mathcal{Z} = \pi_1(G_{\text{nA}})$, generators of its associated fractional null junctions (FNJ), and the embedding $\pi_1(G_{\text{nA}}) \hookrightarrow Z(\widetilde{G}_{\text{nA}})$ of the fundamental group into the center of the simply-connected cover $\widetilde{G}_{\text{nA}}$ of the non-Abelian symmetry. For the last entry, the values are ordered in alignment with the ``brane configuration" column rather than the $\mathfrak{g}$ column.}
\label{tab:rank20_braneConfigs} \\
\hline
No. &  $\mathfrak{g}$ & $\pi_1(G_{\text{nA}})$                                              &  Brane Config.  & FNJ & \tabincell{l}{$\pi_1(G_{\text{nA}}) \hookrightarrow Z(\widetilde{G}_\text{nA})$ \\ \tiny{(Order via Brane Config.)}}\\ \hline\hline
1   & $6\mathfrak{su}_4$                            &      $\bbZ_4 \times \bbZ_4$       &     $\bA^4 \bB^4 \bN^4 \bX_{[1, 2]}^4 \bC^4 \bX_{[2, 1]}^4$ &   \tabincell{l}{$\boldsymbol\delta^N_{(1,0)}/4$, \\ $\boldsymbol\delta^N_{(0,1)}/4$}    &    \tabincell{l}{$(0, 3, 1, 2, 1, 1),$\\ $(3, 3, 0, 3, 3, 2)$}    \\\hline \hline
2   & $2\mathfrak{su}_2 \oplus 4\mathfrak{su}_5$                     &      $\bbZ_5$                     &   $\bA^5 \bN^5 \bX_{[1,4]}^5 \bX_{[1,2]}^2 \bC^5 \bX_{[3, 1]}^2$  &  $\boldsymbol\delta^N_{(3,1)}/5$     & $(4, 3, 1, 0, 2, 0)$ \\ \hline
3   & $2\mathfrak{su}_3 \oplus 2\mathfrak{su}_4 \oplus 2\mathfrak{su}_5$              &      -                            & $\bA^3 \bN^5 \bX_{[1,4]}^5 \bX_{[1,3]}^3 \bX_{[1,2]}^4 \bC^4$ & - & -\\
\hline
\end{longtable}

\end{landscape}

\begin{landscape}

\subsection{One \texorpdfstring{O7$^+$}{O7+}}

We proceed in this part to give the full list to brane configurations with a single O7$^+$ realizing maximally-enhanced 8d vacua of rank $(2,10)$. This list precisely matches our previous results in 8d CHL strings in Appendix B of \cite{Cvetic:2021sjm}. For each such brane configuration, in addition to giving all the information as provided in the previous table, we also refer to its particular ``uplift'' to rank $(2,18)$, namely the rank $(2,18)$ configuration that one gets by unfreezing the $\bA^n \mathbf{O7}^+$ stack into a $\bA^{n+8} \bB \bC$ stack.

\begin{longtable}{|c|c|c|c||c|c|c|} 
\caption{All 8d maximally-enhanced rank $(2,10)$ brane configurations, in similar convention as the above rank $(2,18)$ catalog.}
\label{tab:rank12_braneConfigs}\\
\hline
\#  &  $\#_{rk\ 20}$ & $\widehat{\mathfrak{g}}$ ($\mathfrak{g}$) & $\pi_1(G_{\text{nA}})$ &  Brane Config.  & FNJ &  $\pi_1(G_{\text{nA}}) \hookrightarrow Z(\widetilde{G}_\text{nA})$   \\ \hline 
1   &320 & ${\mathfrak e}_{8} \oplus \mathfrak{sp}_{2}/(\mathfrak{so}_{20})$    &  0   &  $(\bA^2 {\bf O7}^+)\bC (\bA^7 \bB \bC^2) \bX_{[4, 1]}$  & -  & - \\ \hline
2   &319 & $\mathfrak{e}_8 \oplus \mathfrak{sp}_{1}/(\mathfrak{so}_{18}) \oplus \mathfrak{su}_2$  &  0  & $(\bA {\bf O7}^+) \bC \bX_{[3,1]}^2 (\bX_{[3,1]}^7 \bX_{[13, 4]} \bX_{[7,2]}^2)$ & - & -  \\ \hline
3   &291 & ${\mathfrak e}_{7} \oplus \mathfrak{sp}_{3}/(\mathfrak{so}_{22})$    &  0  & $(\bA^3 {\bf O7}^+) \bC (\bA^6 \bN \bX_{[2,1]}^2) \bX_{[5, 1]}$ & - & -  \\ \hline
4   &290 & ${\mathfrak e}_{7} \oplus \mathfrak{sp}_{2}/(\mathfrak{so}_{20}) \oplus \mathfrak{su}_2$    &  $\bbZ_2$  &  $(\bA^2 {\bf O7}^+) \bB (\bA^6\bB\bC^2)\bX_{[3, 1]}^2$ &  $\boldsymbol\delta^N_{(1,1)}/2$  &  $(1, 1, 0)$ \\ \hline
5   &289 & ${\mathfrak e}_{7} \oplus \mathfrak{sp}_{1}/(\mathfrak{so}_{18}) \oplus \mathfrak{su}_3$    & 0  & $(\bA {\bf O7}^+) \bC \bX_{[3, 1]}^3(\bX_{[3, 1]}^6 \bX_{[13, 4]} \bX_{[7, 2]}^2)$ & - & - \\ \hline
6   &288 & $(\mathfrak{so}_{16} \oplus ) {\mathfrak e}_{7} \oplus \mathfrak{su}_3 \oplus \mathfrak{su}_2$    &  $\bbZ_2$  & $(\bA^6\bB\bC^2) {\bf O7}^+ \bN^2\bC^3$ &  $\boldsymbol\delta^N_{(1,1)}/2$  & $(1, 1, 0)$  \\ \hline
7   &255 & ${\mathfrak e}_{6} \oplus \mathfrak{sp}_{4}/(\mathfrak{so}_{24})$    &  $0$  & $(\bA^4 {\bf O7}^+) (\bA^5 \bC \bX_{[3,1]}^2) \bX_{[6, 1]}$ & - & - \\   \hline
8   &254 & ${\mathfrak e}_{6} \oplus \mathfrak{sp}_{3}/(\mathfrak{so}_{22}) \oplus \mathfrak{su}_2$ & $0$   & $(\bA^5 \bB \bC^2) (\bA^3 {\bf O7}^+) \bX_{[2,-1]} \bN^2  $ & - & - \\ \hline
9   &253 & ${\mathfrak e}_{6} \oplus \mathfrak{sp}_{1}/(\mathfrak{so}_{18}) \oplus \mathfrak{su}_4$  & $0$ & $(\bA {\bf O7}^+) \bC \bX_{[3, 1]}^4 (\bX_{[3, 1]}^5 \bX_{[13, 4]} \bX_{[7, 2]}^2)$ & - & - \\ \hline
10  &252 & ${\mathfrak e}_{6} \oplus \mathfrak{sp}_{1}/(\mathfrak{so}_{18}) \oplus \mathfrak{su}_3 \oplus \mathfrak{su}_2$    &  $0$  & $(\bA^5 \bB \bC^2) (\bA {\bf O7}^+) \bB^2 \bN^3$ & - & - \\ \hline
11  &251 & $(\mathfrak{so}_{16} \oplus ){\mathfrak e}_{6} \oplus \mathfrak{su}_5$    &   $0$  & $(\bA^5 \bB \bC^2) {\bf O7}^+ \bN^5 \bX_{[1,3]}$ & - & - \\ \hline
12  &218 & $\mathfrak{sp}_{10}/(\mathfrak{so}_{36})$    &   $0$  &  $ (\bA^{10} {\bf O7}^+) \bC \bX_{[4, 1]} \bX_{[8,1]} \bX_{[11,1]}$ & - & -  \\ \hline
13  &217 & $\mathfrak{sp}_{9}/(\mathfrak{so}_{34}) \oplus \mathfrak{su}_2$    &   $0$  & $(\bA^9 {\bf O7}^+) \bX_{[5,-1]} \bX_{[2, -1]} \bC^2 \bX_{[3,1]}$ & - & - \\ \hline
14  &216 & $\mathfrak{sp}_{8}/(\mathfrak{so}_{32}) \oplus \mathfrak{su}_3$    &   $\bbZ_2$    &  $ (\bA^8 {\bf O7}^+) \bX_{[-4, 1]} \bN \bX_{[4, 1]} \bX_{[6, 1]}^3 $ & $\boldsymbol\delta^N_{(0,1)}/2$  &  $(1, 0)$  \\ \hline
15  &215 & $\mathfrak{sp}_{8}/(\mathfrak{so}_{32}) \oplus 2\mathfrak{su}_2$     &   $\bbZ_2$     &  $ (\bA^8 {\bf O7}^+) \bX_{[-4, 1]} \bX_{[2, 1]} \bX_{[4, 1]}^2\bX_{[6, 1]}^2 $ & $\boldsymbol\delta^N_{(0,1)}/2$  &  $(1, 0, 0)$ \\ \hline
16  &214 & $\mathfrak{sp}_{7}/(\mathfrak{so}_{30}) \oplus \mathfrak{su}_3 \oplus \mathfrak{su}_2$ &  $0$       &  $(\bA^7 {\bf O7}^+) \bC \bX_{[5,1]}^2 \bX_{[6,1]}^3 \bX_{[9, 1]}$ & - & - \\ \hline
17  &213 & $\mathfrak{sp}_{6}/(\mathfrak{so}_{28}) \oplus \mathfrak{su}_5$    &   $0$  & $(\bA^6 {\bf O7}^+) \bC \bX_{[9,2]} \bX_{[5, 1]}^5 \bX_{[8, 1]}$ & - & - \\ \hline
18  &212 & $\mathfrak{sp}_{6}/(\mathfrak{so}_{28}) \oplus \mathfrak{su}_4 \oplus \mathfrak{su}_2$    &   $\bbZ_2$   &   $ (\bA^6 {\bf O7}^+) \bX_{[-4, 1]} \bX_{[2, 1]} \bX_{[4, 1]}^4\bX_{[9, 2]}^2 $ & $\boldsymbol\delta^N_{(0,1)}/2$  &   $(1, 0, 1)$  \\ \hline
19  &210 & $\mathfrak{sp}_{6}/(\mathfrak{so}_{28}) \oplus 2\mathfrak{su}_3$     &  $0$      & $(\bA^6 {\bf O7}^+) \bX_{[2,-1]} \bN^3 \bC^3 \bX_{[3,1]}$ & - & - \\ \hline
20  &211 & $\mathfrak{sp}_{6}/(\mathfrak{so}_{28}) \oplus \mathfrak{su}_3 \oplus 2\mathfrak{su}_2$  & $\bbZ_2$    &   $ (\bA^6 {\bf O7}^+) \bX_{[-4, 1]} \bC^2 \bX_{[2, 1]}^3\bX_{[4, 1]}^2 $   &  $\boldsymbol\delta^N_{(0,1)}/2$  &  $(1, 1, 0, 0)$ \\ \hline
21  &209 & $\mathfrak{sp}_{5}/(\mathfrak{so}_{26}) \oplus \mathfrak{so}_{10}$     &  $0$      &  $(\bA^5\bB\bC)(\bN^{13}\bA\bX_{[1,-2]})\bX_{[2,15]} \bX_{[1,6]}$  & - & - \\ \hline
22  &208 & $\mathfrak{sp}_{5}/(\mathfrak{so}_{26}) \oplus \mathfrak{su}_6$      &  $0$      & $(\bA^5 {\bf O7}^+) \bB \bX_{[2,3]} \bC^6 \bX_{[3,1]}$ & - & - \\ \hline
23  &207 & $\mathfrak{sp}_{5}/(\mathfrak{so}_{26}) \oplus \mathfrak{su}_5 \oplus \mathfrak{su}_2$  &  $0$     &   $(\bA^5 {\bf O7}^+) \bC (\bX_{[11,2]} \bX_{[6,1]}^5) \bX_{[8, 1]}^2$ & - & - \\ \hline
24  &206 & $\mathfrak{sp}_{4}/(\mathfrak{so}_{24}) \oplus \mathfrak{so}_{12}$     & $\bbZ_2$    &   $ (\bA^4 {\bf O7}^+) (\bX_{[2, -1]}^6 \bX_{[1, -1]} \bX_{[3, -1]}) \bN \bX_{[4, 1]} $   & $\boldsymbol\delta^N_{(0,1)}/2$ & $(1, (1,1))$  \\ \hline
25  &205 & $\mathfrak{sp}_{4}/(\mathfrak{so}_{24}) \oplus \mathfrak{so}_{10} \oplus \mathfrak{su}_2$ & $\bbZ_2$   &  $ (\bA^4 {\bf O7}^+) (\bX_{[2, -1]}^5 \bX_{[1, -1]} \bX_{[3, -1]}) \bN^2 \bX_{[2, 1]} $  &  $\boldsymbol\delta^N_{(0,1)}/2$   &  $(1, 2, 0)$  \\ \hline
26  &204 & $\mathfrak{sp}_{4}/(\mathfrak{so}_{24}) \oplus \mathfrak{su}_5 \oplus 2\mathfrak{su}_2$  & $\bbZ_2$  &   $ (\bA^4 {\bf O7}^+) \bX_{[-4, 1]} \bC^2 \bX_{[2, 1]}^5\bX_{[5, 2]}^2 $ &  $\boldsymbol\delta^N_{(0,1)}/2$  &  $(1, 1, 0, 1)$  \\ \hline
27  &203 & $\mathfrak{sp}_{4}/(\mathfrak{so}_{24}) \oplus \mathfrak{su}_4 \oplus \mathfrak{su}_3 \oplus \mathfrak{su}_2$   & $\bbZ_2$   &    $ (\bA^4 {\bf O7}^+) \bN^2 \bC^4 \bX_{[2, 1]}^3\bX_{[4, 1]} $  & $\boldsymbol\delta^N_{(0,1)}/2$  &   $(1, 0, 2, 0)$ \\ \hline
28  &202 & $\mathfrak{sp}_{4}/(\mathfrak{so}_{24}) \oplus 2\mathfrak{su}_3 \oplus 2\mathfrak{su}_2$   & $\bbZ_2$    &   $ (\bA^4 {\bf O7}^+) \bN^3 \bC^2 \bX_{[5, 1]}^2 \bX_{[6, 1]}^3 $   &  $\boldsymbol\delta^N_{(0,1)}/2$  &  $((1, 0, 1, 1, 0)$  \\ \hline
29  &201 & $\mathfrak{sp}_{3}/(\mathfrak{so}_{22}) \oplus \mathfrak{su}_7 \oplus \mathfrak{su}_2$    &  $0$ & $(\bA^3 {\bf O7}^+) \bC (\bX_{[11,2]} \bX_{[6,1]}^7) \bX_{[13, 2]}^2$ & - & - \\ \hline
30  &200 & $\mathfrak{sp}_{3}/(\mathfrak{so}_{22}) \oplus \mathfrak{su}_6 \oplus \mathfrak{su}_3$    &  $0$ & $(\bA^3 {\bf O7}^+) \bC \bX_{[9,2]} \bX_{[5, 1]}^6 \bX_{[6, 1]}^3$ & - & - \\ \hline
31  &199 & $\mathfrak{sp}_{3}/(\mathfrak{so}_{22}) \oplus \mathfrak{su}_5 \oplus \mathfrak{su}_3 \oplus \mathfrak{su}_2$    &  $0$ & $(\bA^3 {\bf O7}^+)\bX_{[2,3]}\bC^5\bX_{[5, 1]}^2\bX_{[6, 1]}^3$ & -  & - \\ \hline
32  &198 & $\mathfrak{sp}_{3}/(\mathfrak{so}_{22}) \oplus \mathfrak{su}_4 \oplus 2\mathfrak{su}_3$    &  $0$ & $(\bA^3 {\bf O7}^+) (\bN^4 \bC^3 \bX_{[2,1]}^3) \bX_{[5, 1]}$ & - & - \\ \hline
33  &197 & $\mathfrak{sp}_{2}/(\mathfrak{so}_{20}) \oplus \mathfrak{so}_{14} \oplus \mathfrak{su}_2$    &  $\bbZ_2$   &  $ (\bA^2 {\bf O7}^+) \bX_{[3, -1]}^2 \bN (\bX_{[2, 1]}^7 \bX_{[1, 1]} \bX_{[5, 3]})$   & $\boldsymbol\delta^N_{(0,1)}/2$   & $(1, 1, 2)$  \\ \hline
34  &196 & $\mathfrak{sp}_{2}/(\mathfrak{so}_{20}) \oplus \mathfrak{so}_{12} \oplus \mathfrak{su}_3$    &  $\bbZ_2$   &   $ (\bA^2 {\bf O7}^+) (\bX_{[2, -1]}^6 \bX_{[1, -1]} \bX_{[3, -1]}) \bB^3 \bC $   &  $\boldsymbol\delta^N_{(1,1)}/2$  &  $(1, (1, 0), 0) $ \\ \hline
35  &195 & $\mathfrak{sp}_{2}/(\mathfrak{so}_{20}) \oplus \mathfrak{so}_{10} \oplus \mathfrak{su}_3 \oplus \mathfrak{su}_2$    &  $\bbZ_2$   &   $ (\bA^2 {\bf O7}^+) (\bX_{[2, -1]}^5 \bX_{[1, -1]} \bX_{[3, -1]}) \bN^3 \bC^2 $  &  $\boldsymbol\delta^N_{(0,1)}/2$  &  $(1, 2, 0, 1)$ \\ \hline
36  &194 & $\mathfrak{sp}_{2}/(\mathfrak{so}_{20}) \oplus \mathfrak{su}_9$    &  $0$ & $(\bA^2 {\bf O7}^+) \bX_{[4,-1]} \bX_{[1,-2]} \bN^9 \bX_{[1, 3]} $ & - & - \\ \hline
37  &193 & $\mathfrak{sp}_{2}/(\mathfrak{so}_{20}) \oplus \mathfrak{su}_7 \oplus \mathfrak{su}_3$    &  $0$ & $(\bA^2 {\bf O7}^+) (\bX_{[3,4]} \bC^7 \bX_{[2,1]}^3) \bX_{[5, 1]}$ & - & - \\ \hline
38  &192 & $\mathfrak{sp}_{2}/(\mathfrak{so}_{20}) \oplus \mathfrak{su}_6 \oplus \mathfrak{su}_4$    &  $\bbZ_2$   &   $ (\bA^2 {\bf O7}^+) \bB \bX_{[3, 5]} \bC^4 \bX_{[2, 1]}^6 $   & $\boldsymbol\delta^N_{(1,1)}/2$  &   $(1, 0, 3)$  \\ \hline
39  &191 & $\mathfrak{sp}_{2}/(\mathfrak{so}_{20}) \oplus \mathfrak{su}_6 \oplus 3\mathfrak{su}_2$   &  $\bbZ_2 \times \bbZ_2$  &  $ (\bA^2 {\bf O7}^+) \bN^6 \bX_{[1, 5]}^2 \bX_{[1, 3]}^2 \bC^2$  & \tabincell{l}{ $\boldsymbol\delta^N_{(1,1)}/2$, \\ $\boldsymbol\delta^N_{(0,1)}/2$} &  \tabincell{l}{$(1, 3, 0, 0, 0)$,\\ $(1, 0, 1, 1, 1)$}  \\ \hline
40  &189 & $\mathfrak{sp}_{2}/(\mathfrak{so}_{20}) \oplus 2\mathfrak{su}_5$   &  $0$ & $(\bA^2 {\bf O7}^+)\bX_{[2,3]}\bC^5\bX_{[5, 1]}^5\bX_{[16, 3]}$ & - & - \\ \hline
41  &190 & $\mathfrak{sp}_{2}/(\mathfrak{so}_{20}) \oplus \mathfrak{su}_5 \oplus \mathfrak{su}_4 \oplus \mathfrak{su}_2$ & $\bbZ_2$  &    $ (\bA^2 {\bf O7}^+) \bN^4 \bX_{[1,2]}^2 \bC^5 \bX_{[3, 1]} $  &  $\boldsymbol\delta^N_{(1,1)}/2$   &  $(1, 0, 2, 1)$   \\ \hline
42  &188 & $\mathfrak{sp}_{2}/(\mathfrak{so}_{20}) \oplus 2\mathfrak{su}_4 \oplus 2\mathfrak{su}_2$   &   $\bbZ_2 \times \bbZ_2$  &  $ (\bA^2 {\bf O7}^+) \bN^4 \bX_{[1, 3]}^4 \bX_{[2, 5]}^2 \bC^2$  & \tabincell{l}{$\boldsymbol\delta^N_{(1,1)}/2$, \\ $\boldsymbol\delta^N_{(0,1)}/2$}   & \tabincell{l}{$(1, 2, 0, 1, 0)$,\\ $(1, 0, 2, 0, 1)$} \\ \hline
43  &179 & $\mathfrak{sp}_{1}/(\mathfrak{so}_{18}) \oplus \mathfrak{so}_{18}$    &  $0$ & $(\bA {\bf O7}^+) \bB (\bC^9 \bA \bX_{[1, 2]}) \bX_{[3, 2]}$ & - & - \\ \hline
44  &187 & $\mathfrak{sp}_{1}/(\mathfrak{so}_{18}) \oplus \mathfrak{so}_{10} \oplus \mathfrak{su}_5$    &  $0$ &  $(\bA {\bf O7}^+)(\bX_{[2, -1]}^5 \bB \bX_{[3, -1]}) \bN^5 \bX_{[1,2]}$ & - & - \\ \hline
45  &186 & $\mathfrak{sp}_{1}/(\mathfrak{so}_{18}) \oplus \mathfrak{su}_{10}$       &  $0$ &  $(\bA {\bf O7}^+) \bC  (\bX_{[3, 1]}^{10} \bX_{[19, 6]} \bX_{[10, 3]})$ & - & - \\ \hline
46  &185 & $\mathfrak{sp}_{1}/(\mathfrak{so}_{18}) \oplus \mathfrak{su}_9 \oplus \mathfrak{su}_2$       &  $0$ &  $(\bA {\bf O7}^+) \bC \bX_{[3, 1]}^9 \bX_{[13, 4]} \bX_{[7, 2]}^2$ & - & - \\ \hline
47  &184 & $\mathfrak{sp}_{1}/(\mathfrak{so}_{18}) \oplus \mathfrak{su}_8 \oplus 2\mathfrak{su}_2$    & $\bbZ_2$  &    $(\bA {\bf O7}^+) \bN^8 \bX_{[1, 4]} \bX_{[2, 5]}^2 \bC^2$  &  $\boldsymbol\delta^N_{(1,0)}/2$  &   $(0, 4, 1, 1)$ \\ \hline
48  &183 & $\mathfrak{sp}_{1}/(\mathfrak{so}_{18}) \oplus \mathfrak{su}_7 \oplus \mathfrak{su}_3 \oplus \mathfrak{su}_2$   &  $0$ & $(\bA {\bf O7}^+) \bC \bX_{[3, 1]}^7 \bX_{[10, 3]}^2 \bX_{[7,2]}^3$ & - & - \\ \hline
49  &182 & $\mathfrak{sp}_{1}/(\mathfrak{so}_{18}) \oplus \mathfrak{su}_6 \oplus \mathfrak{su}_5$    &  $0$ & $(\bA {\bf O7}^+) \bC (\bX_{[3, 1]}^6 \bX_{[17, 5]} \bX_{[7, 2]}^5)$ & - & - \\ \hline
50  &181 & $\mathfrak{sp}_{1}/(\mathfrak{so}_{18}) \oplus \mathfrak{su}_6 \oplus \mathfrak{su}_4 \oplus \mathfrak{su}_2$  &  $\bbZ_2$ &   $(\bA {\bf O7}^+) \bN^6 \bX_{[1, 5]}^4 \bX_{[2, 9]}^2 \bX_{[1, 2]}$  &  $\boldsymbol\delta^N_{(1,0)}/2$  &  $(0, 3, 2, 1)$  \\ \hline
51  &180 & $\mathfrak{sp}_{1}/(\mathfrak{so}_{18}) \oplus \mathfrak{su}_5 \oplus 2\mathfrak{su}_3 \oplus \mathfrak{su}_2$   &  $0$ &  $(\bA {\bf O7}^+) \bX_{[1,2]}^5 \bX_{[5,9]}^3 \bX_{[3, 5]}^2 \bC^3$ & - & - \\ \hline
52  &169 & $(\mathfrak{so}_{16} \oplus ) \mathfrak{so}_{16} \oplus 2\mathfrak{su}_2$   &  $\bbZ_2 \times \bbZ_2$  &  ${\bf O7}^+ (\bX_{[2, -1]}^8 \bX_{[1,-1]} \bX_{[3, -1]}) \bB^2 \bC^2$   &  \tabincell{l}{$\boldsymbol\delta^N_{(0,1)}/2$,\\ $\boldsymbol\delta^N_{(1,1)}/2$} &  \tabincell{l}{$((1, 1), 1, 1)$,\\ $((0, 1), 0, 0 )$}  \\ \hline
53  &178 & $(\mathfrak{so}_{16} \oplus ) \mathfrak{so}_{12} \oplus \mathfrak{su}_4 \oplus \mathfrak{su}_2$  &  $\bbZ_2 \times \bbZ_2$  &    ${\bf O7}^+ (\bX_{[2, -1]}^6 \bX_{[1,-1]} \bX_{[3, -1]}) \bB^4 \bN^2$ &  \tabincell{l}{$\boldsymbol\delta^N_{(0,1)}/2$, \\ $\boldsymbol\delta^N_{(1,1)}/2$} & \tabincell{l}{$((1, 1), 2, 0)$, \\ $((1, 0), 0, 1)$ }   \\ \hline
54  &177 & $(\mathfrak{so}_{16} \oplus ) 2\mathfrak{so}_{10}$    &     $\bbZ_2$  &  ${\bf O7}^+ (\bX_{[2, -1]}^5 \bX_{[1,-1]} \bX_{[3, -1]}) (\bN^5 \bC \bB)$  &  $\boldsymbol\delta^N_{(0,1)}/2$ &  $(2, 2)$ \\ \hline
55  &176 & $(\mathfrak{so}_{16} \oplus ) \mathfrak{su}_{10} \oplus \mathfrak{su}_2$  &     $\bbZ_2$  &  ${\bf O7}^+ \bX_{[1,2]} \bC^{10} \bX_{[3, 2]} \bX^2_{[3, 1]} $    &  $\boldsymbol\delta^N_{(1,0)}/2$  &  $(5, 1)$  \\ \hline
56  &175 & $(\mathfrak{so}_{16} \oplus ) \mathfrak{su}_8 \oplus \mathfrak{su}_3 \oplus \mathfrak{su}_2$  &     $\bbZ_2$   &   ${\bf O7}^+ \bB^2 \bX_{[1, -3]} \bN^8 \bC^2$   & $\boldsymbol\delta^N_{(1,1)}/2$   & $(0, 4, 0)$  \\ \hline
57  &174 & $(\mathfrak{so}_{16} \oplus ) \mathfrak{su}_7 \oplus 2\mathfrak{su}_3$       &  $0$ &  ${\bf O7}^+ \bB^3 \bN^3 \bC^7 \bX_{[3,2]}$ & - & - \\ \hline
58  &171 & $(\mathfrak{so}_{16} \oplus ) 2\mathfrak{su}_6$  &     $\bbZ_2$   &    ${\bf O7}^+ \bB^6 \bX_{[1, -2]} \bC^6 \bX_{[3, 2]}$   & $\boldsymbol\delta^N_{(1,0)}/2$    &  $(3, 3)$  \\ \hline
59  &173 & $(\mathfrak{so}_{16} \oplus ) \mathfrak{su}_6 \oplus \mathfrak{su}_5 \oplus \mathfrak{su}_2$  &     $\bbZ_2$   &    ${\bf O7}^+ \bB^6 \bX_{[2, -3]} \bN^5 \bC^2$   & $\boldsymbol\delta^N_{(0,1)}/2$   &  $(3, 0, 1)$  \\ \hline
60  &172 & $(\mathfrak{so}_{16} \oplus ) \mathfrak{su}_6 \oplus \mathfrak{su}_4 \oplus 2\mathfrak{su}_2$  &  $\bbZ_2 \times \bbZ_2$ &  ${\bf O7}^+ \bB^4 \bX_{[1, -2]}^2 \bN^6 \bC^2$   &  \tabincell{l}{$\boldsymbol\delta^N_{(0,1)}/2$,\\ $\boldsymbol\delta^N_{(1,1)}/2$}  &  \tabincell{l}{$(2, 1, 0, 1)$,\\  $(0, 1, 3, 0)$}   \\ \hline
61  &170 & $(\mathfrak{so}_{16} \oplus ) 2\mathfrak{su}_4 \oplus 2\mathfrak{su}_3$  &  $\bbZ_2$  &  ${\bf O7}^+ \bB^3 \bN^4 \bX_{[1, 2]}^4 \bC^3$ & $\boldsymbol\delta^N_{(0,1)}/2$   &   $(0, 2, 2, 0)$ \\ \hline
\end{longtable}  

\subsection{Two \texorpdfstring{O7$^+$}{O7+}'s}
\label{subapdx:rank4_catalog}

Finally, we give all six vacua in the rank $(2,2)$ branch via brane configurations, this time no longer restricting to maximally-enhanced cases. This is the first time that the global structure of such string vacua without a heterotic or CHL description has been computed.

\renewcommand{\arraystretch}{1.4}

\begin{table}[H]
    \centering
    \begin{tabular}{|c|c|c||c|c|c|c|} \hline
        $\#$  &  $\#_{\text{rk} 12}$   &  $\#_{\text{rk} 20}$   & $(\mathfrak{g}, Z)_{(2,18)}$ & Brane Config. & FNJ & $\pi_1(G_{\text{nA}}) \hookrightarrow Z(\widetilde{G}_\text{nA})$ \\ \hline \hline
        1  &  -  &   -   &   $(2\mathfrak{so}_{16}, \bbZ_2)$  & ${\bf O7}^+ \bB {\bf O7}^+ \bX_{[3, -1]} \bB \bC$ &- &- \\ \hline
        2  &  -  &     -   &   $(2\mathfrak{so}_{16}, -)$  & ${\bf O7}^+ \bB\bC {\bf O7}^+_{[[1,1],[0,1]]} \bA\bX_{[1, 2]}$ &- &-  \\ \hline
        3  &  -  &   -   &   $(2\mathfrak{so}_{16} \oplus \mathfrak{su}_2, \mathbb{Z}_2)$  & ${\bf O7}^+ \bB {\bf O7}^+ \bB\bC^2$ &- &- \\ \hline
        4  &  52 &   169   &   $(2\mathfrak{so}_{16} \oplus 2\mathfrak{su}_2, \mathbb{Z}_2 \times \mathbb{Z}_2)$  & ${\bf O7}^+ {\bf O7}^+_{[[2, -1],[1,0]]} \bB^2 \bC^2$ & $\boldsymbol\delta^N_{(1,1)}/2$ & (1,1) \\ \hline
        5  &  -  &     -   &   $(\mathfrak{so}_{18} \oplus \mathfrak{so}_{16}, -)$  & $\bA {\bf O7}^+ \bB (\bC\ {\bf O7}^+_{[[1, 1],[0,1]]}) \bX_{[3, 2]}$ &- &- \\ \hline
        6  &  43 &   179   &   $(2\mathfrak{so}_{18}, -)$  & $(\bA\ {\bf O7}^+)\bB (\bC\  {\bf O7}^+_{[[1, 1],[0,1]]}) \bX_{[3 ,2]}$ &- &- \\ \hline
    \end{tabular}
    \caption{All 6 vacua of 8d rank $(2,2)$ with two O7$^+$-planes, not limited to maximally-enhanced vacua. We give their rank $(2,18)$ and rank $(2,10)$ uplift for the two maximally-enhanced vacua. Here the two ${\bf O7}^+$ could be mutually non-locally, and in this case the subscript $[[p, q],[r, s]]$ (with $ps - qr = 1$) stands for the $SL(2, \bbZ)$ transformation that ones need to transform a standard ${\bf O7}^+$ into a ${\bf O7}^+_{[[p, q],[r, s]]}$.}
    \label{tab:rank4}
\end{table}

\renewcommand{\arraystretch}{1}

\end{landscape}

\bibliographystyle{JHEP.bst}
\bibliography{references.bib}

\end{document}